\documentclass[twocolumn,fleqn]{svjour2}    
\usepackage{graphicx}
\usepackage{graphicx}
\usepackage{dcolumn}
\usepackage{bm}
\usepackage{subfigure}
\usepackage{amsmath}
\usepackage{graphicx}

\journalname{Astrophysics and Space Science}
\begin{document}

\title{Self-Similar Dynamics of a Magnetized Polytropic Gas
}


\author{Wei-Gang Wang        \and
        Yu-Qing Lou
}


\institute{W.-G. Wang \at
              Physics Department and Tsinghua Center
for Astrophysics (THCA), Tsinghua University, Beijing 100084,
China;
          \email{wwg03@mails.tsinghua.edu.cn; weigwang@stanford.edu}
          \and
           Y.-Q. Lou \at
           1. Physics Department and the Tsinghua Center for
Astrophysics (THCA), Tsinghua University, Beijing 100084, China;\\
2. Department of Astronomy and Astrophysics, The University
of Chicago, 5640 S. Ellis Avenue, Chicago, IL 60637 USA;\\
3. National Astronomical Observatories, Chinese Academy
of Sciences, A20, Datun Road, Beijing 100012, China.\\
\email{louyq@tsinghua.edu.cn; lou@oddjob.uchicago.edu} }

\date{Received: date / Accepted: date}

\maketitle

\begin{abstract}
In broad astrophysical contexts of large-scale gravitational
collapses and outflows and as a basis for various further
astrophysical applications, we formulate and investigate a
theoretical problem of self-similar magnetohydrodynamics (MHD) for
a non-rotating polytropic gas of quasi-spherical symmetry
permeated by a completely random magnetic field.
Within this framework, we derive two coupled nonlinear MHD
ordinary differential equations (ODEs), examine properties of the
magnetosonic critical curve, obtain various asymptotic and global
semi-complete similarity MHD solutions, and qualify the
applicability of our results. Unique to a magnetized gas cloud, a
novel asymptotic MHD solution for a collapsing core is
established. Physically, the similarity MHD inflow towards the
central dense core proceeds in characteristic manners before the
gas material eventually encounters a strong radiating MHD shock
upon impact onto the central compact object.
Sufficiently far away from the central core region enshrouded by
such an MHD shock, we derive regular asymptotic behaviours. We
study asymptotic solution behaviours in the vicinity of the
magnetosonic critical curve and determine smooth MHD
eigensolutions across this curve. Numerically, we construct global
semi-complete similarity MHD solutions that cross the magnetosonic
critical curve zero, one, and two times. For comparison,
counterpart solutions in the case of an isothermal unmagnetized
and magnetized gas flows are demonstrated in the present MHD
framework at nearly isothermal and weakly magnetized conditions.
For a polytropic index $\gamma=1.25$ or a strong magnetic field,
different solution behaviours emerge. With a strong magnetic field,
there exist semi-complete similarity solutions crossing the
magnetosonic critical curve only once, and the MHD counterpart of
expansion-wave collapse solution disappears. Also in the
polytropic case of $\gamma=1.25$, we no longer observe the trend
in the speed-density phase diagram of finding infinitely many
matches to establish global MHD solutions that cross the
magnetosonic critical curve twice.
\end{abstract}
\section[]{Introduction}\label{intro}
\renewcommand{\topfraction}{0.98}

The self-similar gas dynamics in spherical symmetry involving
self-gravity and thermal gas pressure has been studied over past
several decades with complementary perspectives and various
applications. In astrophysical and cosmological contexts, Larson
(1969) and Penston (1969a, b) independently studied self-similar
flow solutions in a self-gravitating gas. For modelling star
formation processes in a molecular cloud and in contrast to the
earlier results (Larson, 1969; Penston, 1969a, b), Shu (1977)
explored self-similar collapse behaviours of an isothermal gas and
obtained the central free-fall asymptotic solution and the static
solution at large radii initially.
Shu (1977) focused on the expansion-wave collapse solution (EWCS)
by numerically joining the inner free-fall collapse solution with
the outer static isothermal sphere solution (i.e., the outer part
of a singular isothermal sphere). In different notations, Hunter
(1977) constructed complete isothermal self-similar solutions
crossing the isothermal sonic critical line once by matching
solutions in the speed-density phase diagram. The variation trend
of a spiral pattern in the speed-density phase diagram suggests
that there may exist infinitely many discrete solutions. Hunter
succeeded in connecting the two parts ($t<0$ and $t>0$ with $t$
being the time) of flows smoothly. In the same model framework,
Whitworth and Summers (1985) identified and distinguished such
sonic critical points as nodal and saddle points through a
systematic analysis of the behaviours in the neighborhood of the
isothermal sonic critical line, noted the numerical stability
issue of integration directions in the vicinity of the sonic
critical line, and suggested two-dimensional continua of solutions
with weak discontinuities across the sonic critical line. Hunter
(1986) promptly examined their continua of solutions and pointed
out the weak discontinuity in their solutions (Whitworth and
Summers, 1985), suggesting that these solutions may be unphysical
for being unstable (see also Lazarus 1981 for more details).
Moreover, Hunter (1986) proposed to use as many expansion terms as
possible for a numerical integration away from nodes along the
sonic critical line. Lou and Shen (2004) emphasized that the EWCS
solution (Shu 1977) being static sufficiently far away only
represents a special limiting case of a more general class of
constant speed solutions at large $x$ and constructed global
semi-complete solutions for envelope expansion with core collapse
(EECC), connecting the inner free-fall asymptotic solutions with
asymptotic flow solutions at large $x$ using the similar matching
procedure of Hunter (1977). In particular, Lou and Shen (2004)
constructed EECC solutions crossing the isothermal sonic critical
line twice with radial similarity oscillations in the subsonic
regime and with the divergent free-fall asymptotic behaviour in the
limit of small $x$. We note in passing that steady or self-similar
accretions of dark matter under self-gravity might be relevant in
understanding the formation of a few recently reported
supermassive black holes (SMBHs) in the early Universe (Hu et al.,
2005). Lou (2005) outlined a two-fluid similarity dynamics which
is fairly similar to the isothermal model mentioned above, to
model the gravitational coupling between a dark matter halo and a
hot interstellar gas medium. Furthermore, by incorporating a
random magnetic field into the model analysis (Lou, 2005), it is
possible to set up a model framework to further examine
synchrotron radio emissions and magnetic Sunyaev-Zel'dovich effect
in galaxy clusters (Hu and Lou, 2004). Also, the stability problem
has been tackled (Ori and Piran, 1988; Hanawa and Matsumoto, 1999,
2000; Hanawa and Makayama, 1997) for the Hunter type isothermal
solutions. However, because of technical issues, the free-fall
solutions (Shu, 1977) remain to be analyzed in this aspect. Semelin
et al. (2001) took into account of viscosity to examine the flow
stability problem.

Physically, we imagine that shocks occur naturally under various
astrophysical flow situations (see, e.g., Kennel and Coroniti 1984
for a model of the Crab Nebula involving MHD pulsar wind shocks
and Bagchi et al. 2006 for observational evidence of MHD galaxy
cluster wind shocks). In contexts of star formation, self-similar
isothermal shock flows have been investigated and applied to
astrophysical systems such as Bok globules and the so-called
`champagne flows' in H$_{\mbox{II}}$ regions (Shu et al., 2002;
Tsai and Hsu, 1995; Shen and Lou, 2004; Bian and Lou, 2005). We
briefly touch upon the subject of self-similar MHD shocks here,
because our MHD results (Yu and Lou, 2005) are candidate solutions
for both the upstream and downstream regions across an MHD shock.
The self-similar polytropic MHD shock flows will be presented
elsewhere in more details (see, e.g., Yu et al., 2006 for
self-similar isothermal MHD shock flows and Lou and Wang, 2007).

In polytropic MHD collapses and outflows, the polytropic index
$\gamma$ represents a gross simplification from complicated
physical processes including possible nuclear reactions, energy
transport, neutrino transport and electron capture and so forth
in various contexts (see, e.g., Bouquet et al., 1985 and Yahil,
1983). In parallel with earlier results under the isothermal
approximation, the polytropic treatment has also been pursued
in various astrophysical contexts including supernovae. Goldreich
and Weber (1980) studied the $\gamma=4/3$ case with a focus on
the homologous collapse of the inner core of a progenitor star
prior to the emergence of a rebound shock (e.g., Lou and Wang,
2006, 2007). Yahil (1983) noted limitations of Goldreich-Weber model,
studied the polytropic gas dynamics within the polytropic index range
of $6/5\leq\gamma\leq 4/3$, and discussed possible applications to
the pre-catastrophe as well as post-catastrophe phases separately.
A few years earlier, Cheng (1978) investigated the polytropic
hydrodynamics with $1\leq\gamma\leq 5/3$, focussing on the initial
distribution of mass density $\rho$ with a radial scaling of
$\rho\sim r^{\alpha}$. For a polytropic gas, Cheng (1978) derived
the inner free-fall solution and generalized the isothermal EWCS
to the polytropic EWCS. Bouquet et al. (1985) introduced a dual
space of parameters as well as the systems I and II for the
polytropic gas dynamics with $1\leq \gamma\leq 4/3$, and analyzed
properties of the sonic critical curve in details. Suto and Silk
(1988) performed a similarity transformation for the polytropic
gas dynamics and, from the resulting nonlinear ordinary
differential equations (ODEs), obtained regular and asymptotic
solutions as well as numerical solutions with $n=1$ and
$\gamma=2-n$ for $1<\gamma<4/3$. By taking into account of the
total radiative emissivity from the gas in the form of
$\epsilon\sim\rho^{\alpha}T^{4+\beta}$, Boily and Lynden-Bell
(1995) replaced the polytropic equation of state with an energy
equation and discussed physical, mathematical and numerical
properties of such radiative self-similar gas dynamics. McLaughlin
and Pudritz (1997) considered the limiting case of the so-called
logotropic gas dynamics also involving sonic critical points and
obtained expansion-wave collapse solutions in their analysis. In
the context of star formation, Fatuzzo et al. (2004) studied a
combination or a transition of an initial `equation of state'
$P\propto\rho^{\Gamma}$ and a later dynamic equation of state
$P\propto\rho^{\gamma}$, and noted that in the special case of
$\Gamma=1$, asymptotic solutions may have constant flow speed at
far away regions, analogous to the earlier results of Lou and Shen
(2004) and Shen and Lou (2004). Fatuzzo et al. (2004) carried out
their analysis for the case of $\gamma<1$ without involving the
sonic critical curve.

Magnetic field can be extremely important in many astrophysical
processes on different scales and in particular, for star
formation activities at various stages (e.g., Shu et al., 1987).
In mostly neutral gas medium, such as molecular clouds and cores
etc., the magnetic field can only couple to the gas medium if MHD
wave frequencies are less than the ion-neutral collision frequency,
corresponding to a lower limit on the wavelength that MHD waves
need in order to propagate in such a magnetized cloud (e.g., Myers,
1998). At the end of stellar evolution, magnetic fields are observed to
exist in various stellar systems such as the well-known Crab Nebula
(e.g., Woltjer, 1957, 1958a, b; Kennel and Coroniti, 1984a, b;
Wilson et al., 1985; Lou, 1993; Wolf et al., 2003). Chiueh and Chou
(1994) discussed the gravitational collapse of an isothermal
magnetized gas cloud, including the magnetic pressure force term
in the radial momentum equation together with the magnetic
induction equation. They assumed a randomly distributed magnetic
field such that a quasi-spherical symmetry is sustained during the
MHD similarity evolution of a gas cloud on large scales. Magnetic
tension force was ignored in their formulation.\protect\footnote{The
magnetic induction equation in Chiueh and Chou (1994) involves typos,
both incorrect in its original form and inconsistent with their
later nonlinear ODEs after the similarity MHD transformation.} Yu and
Lou (2005) approached the same physical problem yet with a different
formulation and discussed MHD consequences of a random magnetic field.
Using the frozen-in condition on magnetic field, Yu and Lou (2005)
managed to reduce the three apparently coupled nonlinear MHD equations
to two key coupled nonlinear MHD ODEs in an equivalent manner. Yu et
al. (2006) further explored various self-similar isothermal MHD shocks
under the approximation of large-scale quasi-spherical symmetry.

Self-similar gas dynamics for stellar collapse problems under
self-gravity and thermal pressure have been studied extensively
from various
perspectives. In this paper, we construct similarity MHD solutions
to explore nonlinear effects of a random magnetic field. In
general, MHD similarity collapses and outflows evolve nonlinearly
by simply changing certain profile scalings of the enclosed mass,
gas mass density, radial flow speed, and mean transverse magnetic
field energy density without changing their shapes. Hydrodynamic
simulations have pointed to possible similarity evolutions as
all sorts of transients peter out in time (see, e.g., Bodenheimer
and Sweigart, 1968 and Foster and Chevalier, 1993). The well-known
example is the Sedov-Taylor similarity blast waves resulting from
a point explosion (Sedov, 1959; Landau and Lifshitz, 1959; Barenblatt
and Zel'dovich, 1972). Self-similar flow solutions have been
explored in different geometries (Fillmore and Goldreich, 1984;
Hennebelle, 2003; Terebey et al., 1984; Inutsuka and Miyama, 1992;
Shadmehri, 2005; Krasnopolsky and K\"onigl, 2002; Shen and Lou,
2006) and we here work with the quasi-spherical geometry mainly
by a more phenomenological consideration. For the example of the
Crab Nebula projected onto the plane of the sky, a quasi-spherical
morphology (more or less elliptical in reality) has been sustained
on large scales. Another example is the Cassiopeia A supernova
remnant (resulting from a type II supernova explosion presumably)
which, projected onto the plane of sky, appears more or less round
with a central neutron star manifested as a bright X-ray point.
There are also examples of more or less round planetary nebula
systems where magnetic fields, be they weak or strong, are also
likely involved. We invoke these morphological examples involving
magnetic fields to justify an MHD collapse and expansion problem
with a quasi-spherical symmetry on large scales as a first
approximation. We also assume that small-scale deviations from
the quasi-spherical symmetry is relatively insignificant in
large-scale MHD, i.e., small-scale transverse flow components are
random and may be neglected to simplify the mathematical treatment.
Since magnetic field strengths can be significant in various
astrophysical systems (Yu and Lou, 2005), we should take into
account of the MHD influence in the evolution of a magnetized
gas cloud or a magnetized star (Lou, 1993, 1994) as well as
MHD gas systems on much larger scales.

For a random magnetic field in a cloud, we envision a simple `ball
of thread' scenario in a vast spatial volume of gas medium. A
magnetic field line follows the `thread' meandering within a thin
spherical `layer' in space in a random manner. In the strict
sense, there is always a random weak radial magnetic field
component such that random magnetic field lines in adjacent
`layers' are actually connected throughout in space. By taking a
large-scale ensemble average of such a magnetized gas system, we
are then left with `layers' of random magnetic field components
transverse to the radial direction. Having gone thus far in our
idealization, we would admit that the magnetic fields in the Crab
Nebula, the SNR Cas A as well as several round planetary nebulae
may not be fully represented by of ``ball of thread" scenario. What
we have been trying to emphasize is the large-scale quasi-spherical
geometry of magnetized astrophysical systems rather than detailed
magnetic field configurations. We note also that, in our model,
the ``ball of thread" scenario is mainly for the transverse magnetic
field effect on average, while the MHD effect of a weak radial
magnetic field may be negligible. As a matter of fact, we will still
need further observational information to infer whether our magnetic
field configuration can roughly describe some round-shaped
morphologies of astrophysical systems.

In reference to the recent isothermal self-similar MHD analysis
(Yu and Lou, 2005), we show in this paper that an {\it isothermal}
similarity MHD treatment can be naturally extended to a magnetized
{\it polytropic} gas in a systematic manner. Parallel to the
self-similar transformation for relevant variables (Suto and Silk,
1988) with an additional transformation for the transverse
magnetic field, we derive three apparently coupled nonlinear MHD
ODEs, as in the case of an isothermal magnetofluid (Chiueh and Chou,
1994). The major technical difference in our polytropic MHD
formalism is that these three ODEs can be readily reduced to two
key ODEs of MHD by invoking the frozen-in condition on magnetic
field (Yu et al., 2006; Lou \& Wang, 2007). This frozen-in
condition\footnote{By combining conservations of mass and
magnetic flux, we can readily derive equation (\ref{wang15})
or in dimensional form $<B_t^2>/(\rho^2r^2)=$consant. }
(\ref{wang15}) naturally leads to an integration constant $h$
denoting physically the ratio of the magnetic energy density to
the self-gravitational energy density and significantly reduces
the complexity of analyzing the nonlinear similarity MHD problem.
Although only a change of equation of state is made in our current
formulation as compared to the isothermal treatment (Yu and Lou,
2005), several distinct differences arise. For example, the
magnetosonic critical curve now shows qualitatively different
asymptotic behaviours as compared to an isothermal gas, both in
unmagnetized and magnetized cases (Lou and Wang, 2006, 2007). By
increasing the polytropic index $\gamma$, we find qualitative
differences in reference to the case of a smaller $\gamma$. Most
importantly, we found a novel asymptotic nonlinear MHD solution
near the central core or at later time and constructed
semi-complete MHD similarity solutions using this asymptotic
solution. We have also discovered the so-called `quasi-static'
asymptotic polytropic MHD solution behaviours
(see Lou and Wang, 2006 for polytropic hydrodynamic asymptotic
solutions). We here focus on the MHD case and provide a
description in Appendix \ref{asystat}. A more detailed analysis
and astrophysical applications of this MHD asymptotic solution
can be found in Lou and Wang (2007).

Motivated by potentially wide astrophysical applications, the main
purpose of this paper is to present possible similarity solutions
from the nonlinear MHD ODEs, distinguish the asymptotic behaviours
of different types including the eigensolutions across the
magnetosonic critical curve, and construct global semi-complete
solutions numerically. Our analyses and results here serve as the
theoretical basis for further specific astrophysical MHD
applications.
We provide the background information in Section \ref{intro} as an
introduction. Section \ref{mhdformulation} contains the basic MHD
formulation of the problem and section \ref{math} presents the
mathematical analysis. Section \ref{numerical} mainly describes
numerical results, including the magnetosonic critical curves,
similarity MHD solutions without crossing the magnetosonic
critical curve, similarity solutions crossing the magnetosonic
critical curve once and twice. In both analytical and numerical
analyses, we focus on differences between the cases with or
without magnetic field and between weak and strong magnetic field.
We also compare the case in which $\gamma$ is almost unity to that
in which $\gamma$ is larger than one, and further discuss
differences between a nearly isothermal polytropic gas and an
exact isothermal case.

\section[]{Similarity MHD Flows}\label{mhdformulation}

In this section, the basic MHD formulation of the similarity
problem is presented and the approximation of quasi-spherical
symmetry is discussed (Appendix \ref{QSphere}).

\subsection{MHD Formulation of the Problem}\label{form}

Under the assumptions of a random magnetic field on smaller
scales, the approximation of quasi-spherical symmetry and the
ideal MHD treatment, the dynamics of a polytropic magnetized gas
in spherical polar coordinates $(r,\theta,\phi)$ is described by
the following equations:
\begin{equation}\label{wang1}
\frac{\partial \rho}{\partial t}
+\frac{1}{r^{2}}\frac{\partial}{\partial r}(r^{2}\rho u)=0\ ,
\end{equation}
\begin{equation}\label{wang2}
\frac{\partial M}{\partial t}+u\frac{\partial M}{\partial r}=0\ ,
\end{equation}
\begin{equation}\label{wang3}
\frac{\partial M}{\partial r}=4\pi r^{2}\rho\ ,
\end{equation}
\begin{eqnarray}\label{wang4}
\rho\bigg(\frac{\partial u}{\partial t} +u\frac{\partial
u}{\partial r}\bigg)&=&-\frac{\partial p}{\partial r}
-\frac{GM\rho}{r^{2}}\nonumber\\
&&-\frac{\partial}{\partial r}\bigg(\frac{
<B_{t}^{2}>}{8\pi}\bigg)-\frac{<B_{t}^{2}>} {4\pi r}\ ,
\end{eqnarray}
\begin{eqnarray}\label{wang5}
\frac{\partial}{\partial t}\big(r^{2}<B_{t}^{2}>\big)&+&
u\frac{\partial}{\partial r} \big(r^{2}<B_{t}^{2}>\big)\nonumber\\
&+&2r^{2}<B_{t}^{2}> \frac{\partial u}{\partial r}=0\ ,
\end{eqnarray}
\begin{equation}\label{wang132}
p=\kappa\rho^{\gamma}\ ,
\end{equation}
where $G=6.67\times 10^{-8}\hbox{ g}^{-1}\hbox{ cm}^{-3}
\hbox{ s}^{-2}$ is the gravitational constant, $\rho(r,t)$ is the
gas mass density, $M(r,t)$ is the enclosed gas mass within radius
$r$ at time $t$, $u(r,t)$ is the bulk radial flow speed,
and $<B_t^2>$ is the mean square of the random transverse magnetic
field $\vec B_t$ proportional to the magnetic energy density
associated with the random transverse magnetic field. In the
conventional polytropic equation of state\footnote{The condition
of specific entropy conservation along streamlines would be more
general and will be considered in a separate paper.}
(\ref{wang132}), the coefficient $\kappa$ remains
constant globally, independent of both $r$ and $t$.
The case of $\gamma=1$ is only a special case corresponding to an
isothermal magnetized gas (Yu and Lou, 2005; Yu et al., 2006). The
Poisson equation relating the mass density and the gravitational
potential is automatically satisfied under the quasi-spherical
symmetry. In the above equations, the radial momentum equation
(\ref{wang4}) involves the magnetic pressure and tension forces on
the right-hand side (RHS), and equation (\ref{wang5}) is derived
from the magnetic induction equation along with certain
simplifications (Yu and Lou, 2005); these two equations will be
further discussed and analyzed in the next subsection. Compared
to the work of Chiueh and Chou (1994), the formulation here is
different by keeping the magnetic tension force and by dealing
with a conventional polytropic gas. For the problem outlined
above, the reader may consult relevant references (Shu 1977;
Suto and Silk, 1988; Chiueh and Chou, 1994; Lou and Shen, 2004;
Bian and Lou, 2005; Yu and Lou, 2005; Yu et al., 2006; Lou \&
Wang, 2006, 2007). We focus on the semi-complete solution
space $0<t<+\infty$ rather than the complete solution space as
introduced by Hunter (1977).\footnote{By the time-reversal
invariance, the correspondence between a complete solution (Hunter,
1977) and semi-complete solutions has been shown explicitly in Lou
and Shen (2004) by concrete examples.}

For a conventional polytropic gas with $\gamma=2-n$ where $\gamma$
and $n$ are defined by equations (\ref{wang12}) and
(\ref{wang19}), respectively (see subsection \ref{simi}), a
combination of equations (\ref{wang1})$-$(\ref{wang5}) together
with equation (\ref{wang12}) leads to the MHD energy conservation
equation as
\begin{eqnarray}\label{wang129}
\frac{\partial}{\partial t}\bigg [\frac{\rho
u^2}{2}&+&\frac{p}{(\gamma-1)} -\frac{1}{8\pi
G}\left(\frac{\partial\Phi}{\partial r}\right)^2
+\frac{<B_t^2>}{(8\pi )}\bigg ]\nonumber\\
&+&\frac{1}{r^2}\frac{\partial}{\partial r}
\bigg\{r^2u\rho\bigg[\frac{u^2}{2} +\frac{\gamma
p}{(\gamma-1)\rho}
\bigg]\bigg\}\nonumber\\
&+&\frac{1}{r^2}\frac{\partial}{\partial r}
\bigg(r^2u\rho\Phi+\frac{r^2\Phi}{4\pi G}\frac{\partial^2
\Phi}{\partial r\partial t}\bigg)\nonumber\\
&+&\frac{1}{r^2}\frac{\partial}{\partial r}\bigg(\frac{r^2
u}{4\pi}<B_t^2>\bigg)=0\ ,
\end{eqnarray}
%
where
$-\partial\Phi/\partial r=-GM/r^2$ and $\Phi$ is the gravitational
potential (Fan and Lou, 1999). This MHD energy conservation equation
reduces to the isothermal cases (Lou and Shen, 2004; Yu and Lou, 2005)
by taking the L'H$\hat{\mbox{o}}$pital rule with respect to $\gamma$
in the limit of $\gamma\rightarrow 1$. The magnetic energy density
and Poynting flux density associated with $<B_t^2>$ can be readily
identified in the MHD energy conservation equation (\ref{wang129}).
With the quasi-spherical symmetry,
the divergence term containing $\Phi$ in MHD energy conservation
equation (\ref{wang129}) vanishes (Lou and Shen, 2004).

\subsection{Comments on the MHD Formalism}\label{validity}

We here briefly comment on the basic MHD formulation, the
quasi-spherical symmetry, the three-dimensional random flow
fluctuations on small scales and the physical basis for the
magnetic force density and the magnetic induction equation.

Our MHD model describes a self-gravitating gas cloud embedded with
a magnetic field presumed to be random and tangled on small scales.
On large scales, the gas mass density, the gas thermal temperature,
the thermal pressure and the entropy are all taken to be
quasi-spherically symmetric. The magnetic field distribution is
presumed to be completely random in space such that in a small
volume (an infinitesimal volume $r^2\text dr\text d\theta \text
d\phi$) the magnetic field is effectively represented by the mean
square averages of $<B_r^2>$ and $<B_t^2>$, proportional to the
radial and transverse magnetic energy densities respectively. In
terms of the magnetic pressure and tension forces for the
large-scale MHD, $<B_t^2>$ plays the dominant role on the dynamics
of a magnetized gas cloud as compared to $<B_r^2>$. As the magnetic
field is randomly distributed with a quasi-spherical symmetry, the
bulk gas flow velocity remains grossly spherically symmetric and
can be characterized by the bulk mean radial flow speed $v_r$; the
transverse component of the flow velocity $v_\theta$ and $v_\phi$
are relatively small and may be neglected in the first approximation.
These transverse components should be part of Alfv\'enic fluctuations
corresponding to magnetic field fluctuations about the mean
configuration; thus the more random the magnetic field
fluctuations are, the better the approximation becomes.

The physical concept of a quasi-spherical symmetry for a magnetized
gas cloud or a magnetized progenitor star is only valid for MHD
processes of sufficiently large scales. Here, `large scales' are
obviously in contrast to `small scales' on which magnetic fields
are presumed completely random locally in our MHD model framework.
In the strict sense, an exact spherical symmetry is impossible due
to the very nature of a magnetic field. However, a quasi-spherical
symmetry may be sustained for large-scale MHD processes. We invoke
the projected quasi-elliptical shape of the Crab Nebula and the
projected more or less round remnant of the Cassiopeia A supernova
as empirical supports for this notion of quasi-spherical symmetry.
Although it is not yet obvious that the actual magnetic field
can be largely approximated by our `ball of thread' scenario, from
the morphology of these systems we suggest a globally random magnetic
field distribution as a plausible yet tractable starting point. In
this scenario, small-scale random flow velocities are ignored as
compared to systematic radial flows. Qualitatively speaking, this
perspective is justifiable when a random magnetic field is weak. In
our model analysis, we sometimes do encounter situations of strong
magnetic fields especially for accretions towards a central compact
object. In such a case, one should really view our asymptotic MHD
solutions as indicating a gross trend of variation that is bound
to be destroyed by non-spherical and transient MHD processes
sufficiently close to the central compact object. Upon impacting
onto a central compact object, we expect the emergence of a strong
radiating MHD shock traveling outward slowly in a self-similar
manner (Shen and Lou, 2004; Yu and Lou, 2005; Yu et al., 2006).
Practically, we can apply our large-scale self-similar MHD
solutions of quasi-spherical symmetry outside this radiating MHD
shock. Within this quasi-spherical MHD shock, the core gravity can
be strong enough to more or less hold on the strongly magnetized
plasma. By the na\"ive solar analogy, we readily imagine that
sporadic violent `flares' or `coronal mass ejections' may erupt
from the strongly magnetized central core region and can even
break into the `self-similar' and `quasi-spherical' domain of
magnetized accreting flows.

Given local random magnetic fields on small scales in a gas
medium, three-dimensional MHD flows on small scales are naturally
expected because of the unbalanced magnetic tension force here and
there. Except for very special self-similar radial flow situations
of zero transverse flows yet with a three-dimensional magnetic
field (see Low 1992), we do generally expect small-scale
three-dimensional random flow fluctuations associated with the
large-scale mean radial flow. By our assumption of a randomly
tangled magnetic field, such flow fluctuations are more or less
confined or trapped locally and advected by the mean radial MHD
flow on large scales. In short, we do not expect mean flows
transverse to the radial direction on large scales in our scenario.
By intuition, we expect transverse flows caused by the random
magnetic tension force, yet such flows will remain locally confined
due to the local random magnetic field and hence be small as
compared with the bulk quasi-spherical radial flow speed. As
suggested by Zel'dovich and Novikov (1971), an isotropic magnetic
pressure is expected from a completely random magnetic field on
small scales. We follow this basic concept and also include the
radial magnetic tension force which is non-negligible in our
formulation. Physically, one may view such small-scale random flow
fluctuations as turbulence and for simplicity, we have ignored
the effects of the effective turbulent pressure, viscosity and
resistivity etc. (see, e.g., Lou and Rosner, 1986) in this
formulation. In our model framework, if the MHD turbulence will
attribute to random fields and fluctuations on smaller scales,
the turbulence scale itself should be small enough such that
the largest turbulence scale is small compared to the overall
quasi-spherical geometry.

\section[]{Model Analysis}\label{math}

With the basic ideal MHD model qualified and the formulation
established, we now perform the analytical and numerical
analyses in order. In the next section, we present the
results of numerical exploration.

\subsection{Self-Similar MHD Processes in a \\
Magnetized Polytropic Gas Cloud}\label{simi}

To seek self-similar solutions to the MHD equations, we introduce
an independent similarity dimensionless variable $x$ and presume
that dependent physical variables are given by the following
similarity forms accordingly\footnote{In terms of the self-similar
transformation, the magnetic field term here distinguishes ours
from that of Suto and Silk (1988).}:
\begin{equation}\label{wang6}
r=ax, u=bv, \rho=c\alpha, p=d\beta, M=em, <B_{t}^{2}>=fw\ ,
\end{equation}
where the six scaling factors $a(t)$ through $f(t)$ are functions
of time $t$ only and are defined by
\begin{eqnarray}\label{wang19}
&&a\equiv k^{1/2}t^{n}\ ,\qquad b\equiv k^{1/2}t^{n-1}\ , \qquad
c\equiv\frac{1}{4\pi Gt^{2}}\ ,\qquad\nonumber\\
&&d\equiv\frac{kt^{2n-4}}{4\pi G}\ ,\quad
e\equiv\frac{k^{3/2}t^{3n-2}}{(3n-2)G}\ ,\quad
f\equiv\frac{kt^{2n-4}}{G}\ .
\end{eqnarray}
Here $k$ and $n$ are two constant parameters. As functions of $x$
only, $v(x)$, $\alpha(x)$, $\beta(x)$, $m(x)$, and $w(x)$ are the
reduced forms of radial flow speed, gas mass density, gas
pressure, enclosed gas mass, and magnetic energy density
(associated with the averaged random transverse magnetic field),
respectively. With this self-similar MHD transformation, equations
(\ref{wang2}) and (\ref{wang3}) lead to an algebraic expression
for $m(x)$ in terms of $\alpha (x)$ and $v(x)$, viz.
\begin{equation}\label{wang8}
m=\alpha x^{2}(nx-v)\ ,
\end{equation}
and an ODE for $\alpha (x)$ and $v(x)$
\begin{equation}\label{wang9}
(nx-v)\alpha'-\alpha v'=-2\frac{(x-v)}{x}\alpha\ ,
\end{equation}
where the prime $'$ denotes the differentiation with respect to
$x$. Relation (\ref{wang8}) leads to the important inequality
\begin{equation}
nx-v>0\
\end{equation}
for a positive gas mass density as noted repeatedly in figure
displays presently. In our later analyses, this inequality is a
key constraint on choosing relevant physical solutions. From
equation (\ref{wang5}), one obtains
\begin{equation}\label{wang10}
(nx-v)w'-2wv'=\frac{2v-(4-2n)x}{x}w\
\end{equation}
for the reduced dependent variables $w(x)$ and $v(x)$. This
constraint for the reduced magnetic energy density is fairly
similar to equation (\ref{wang9}). By equation (\ref{wang4}),
one obtains the reduced radial momentum equation
\begin{equation}\label{wang11}
\frac{\beta'}{\alpha}-(nx-v)v'+\frac{w'}{2\alpha}
=-(n-1)v-\frac{w}{x\alpha}-\frac{nx-v}{3n-2}\alpha\ .
\end{equation}
For a generalized polytropic equation of state with a polytropic
index $\gamma$ (Suto and Silk, 1988), we simply have
\begin{equation}\label{wang12}
\beta=\alpha^{\gamma},\qquad p =k(4\pi G)^{\gamma-1}
t^{2(n+\gamma-2)}\rho^{\gamma}=\kappa\rho^{\gamma}\ ,
\end{equation}
where $\kappa$ may be time dependent in general. For $n=2-\gamma$,
we have a constant $\kappa=k(4\pi G)^{\gamma-1}$ and equation
(\ref{wang12}) is an equation of state for a conventional
polytropic gas.

A combination of equations (\ref{wang11})
and (\ref{wang12}) leads to
\begin{eqnarray}\label{wang13}
&&\gamma\alpha^{\gamma-2}\alpha'-(nx-v)v'+\frac{w'}{2\alpha}\nonumber\\
&&=-(n-1)v-\frac{w}{x\alpha} -\frac{(nx-v)}{(3n-2)}\alpha\ ,
\end{eqnarray}
and equations (\ref{wang9}), (\ref{wang10}) and (\ref{wang13}) are
the three MHD similarity ODEs describing polytropic magnetized gas
flows with a quasi-spherical symmetry. The three nonlinear MHD
ODEs are similar to those of Chiueh and Chou (1994) with the key
differences in the adopted equation of state and in keeping the
magnetic tension force term (see equations \ref{wang1} to \ref{wang5}).
By taking relevant limits as necessary checks, these equations are
consistent with those of Shu (1977), Suto and Silk (1988), Lou and
Shen (2004), Yu and Lou (2005), Yu et al. (2006), Lou and Wang
(2006, 2007) as expected.

The Alfv\'en speed in this formulation is defined by
\begin{equation}\label{wang116}
v_A\equiv\left(\frac{<B_t^2>}{4\pi\rho}\right)^{1/2}\ ,
\end{equation}
and the sound speed $s$ in the polytropic gas determined by the
equation of state with $\gamma=2-n$ (i.e., the polytropic state
equation in the usual sense) is simply
\begin{equation}\label{wang117}
s\equiv\bigg(\frac{\partial p}{\partial\rho}\bigg)^{1/2}\ .
\end{equation}
Thus the ratio of the Alfv\'en wave speed
to the polytropic sound speed becomes
\begin{equation}\label{wang118}
\frac{v_A}{s}=\bigg(\frac{w}{\gamma\alpha^\gamma}\bigg)^{1/2}\ ,
\end{equation}
consistent with the isothermal $\gamma=1$ case
(Yu and Lou, 2005; Yu et al., 2006).

\subsection{Reduction of Nonlinear MHD ODEs }

One can readily reduce the three coupled nonlinear MHD
ODEs to two. From equation (\ref{wang9}), one obtains
\[
\frac{\alpha'}{\alpha}=\frac{v'-2(x-v)/x}{nx-v}\ ,
\]
and from equation (\ref{wang10}), one gets
\[
\frac{w'}{w}=\frac{2v'+2[2v-(4-2n)x]/x}{nx-v}\ .
\]
The above two equations lead to a differential relation
\begin{equation}\label{wang14}
\frac{w'}{w}=\frac{2\alpha'}{\alpha}+\frac{2}{x}\ ,
\end{equation}
which immediately gives a simple integral of
\begin{equation}\label{wang15}
w=h\alpha^{2}x^{2}\ ,
\end{equation}
where $h$ is an integration constant providing a measure for the
magnetic field strength. Integral (\ref{wang15}) gives a new
parameter $h$ of a magnetized gas cloud besides $n$ and $\gamma$,
and reduces the three coupled nonlinear MHD ODEs to two. Expressed
explicitly in physical quantities, $h$ is
\[
h=\frac{<B_t^2>}{16\pi^2G\rho^2r^2}\ ,
\]
representing the ratio of the magnetic energy density to the
self-gravitational energy density.
This simplification reduces tremendously complications in
numerical MHD exploration and physically represents the frozen-in
condition on magnetic field (Yu and Lou, 2005; Yu et al., 2006).
The procedure of numerically constructing global MHD solutions
and matching the solutions across the magnetosonic critical curve
can be carried out similar to that of Lou and Shen (2004). This
reduction of three coupled nonlinear MHD ODEs to two is parallel
to the isothermal case (Yu and Lou, 2005; Yu et al., 2006).

Substitution of equation (\ref{wang15}) into equation
(\ref{wang13}) gives
\begin{eqnarray}\label{wang16}
(\gamma \alpha^{\gamma-2}&+&hx^{2})\alpha'-(nx-v)v'\nonumber\\
&=&-(n-1)v -\bigg(2hx+\frac{nx-v}{3n-2}\bigg)\alpha\ .
\end{eqnarray}
Equations (\ref{wang9}) and (\ref{wang16}) together lead to the
two coupled nonlinear MHD ODEs in the forms of
\begin{eqnarray}\label{wang17}
\alpha'=\alpha^{2}\bigg[(n\!\!&-&\!\!1)v
+\bigg(2hx+\frac {nx-v}{3n-2}\bigg)\alpha\nonumber\\
-\frac {2(x-v)(nx-v)}{x}\bigg]
\!\!\!\!&\bigg/&\!\!\!\!\bigg[\alpha (nx-v)^{2}
-\gamma\alpha^{\gamma}-h\alpha^{2}x^{2}\bigg]\ ,
\nonumber\\
\end{eqnarray}
\[
v'=\bigg\{(n-1)\big[\alpha v(nx-v)+2h\alpha^{2}x^{2}\big]
+\frac{(nx-v)^{2}}{(3n-2)}\alpha^2
\]
\begin{equation}\label{wang18}
\quad-2\gamma \alpha^{\gamma}\frac{(x-v)}{x}\bigg\}
\bigg/\bigg[\alpha (nx-v)^{2}-\gamma\alpha^{\gamma}
-h\alpha^{2}x^{2}\bigg]\ .
\end{equation}
We now introduce simplifying notations as follows
\begin{eqnarray}\label{wang77}
&&A(\alpha,v,x)\equiv\alpha^2\bigg[(n-1)v
\qquad\nonumber\\
&&\quad+\bigg(2hx+\frac{nx-v}{3n-2}\bigg)\alpha
-2\frac{(x-v)(nx-v)}{x}\bigg]\ ,\qquad\quad
\end{eqnarray}
\begin{eqnarray}\label{wang78}
&&V(\alpha,v,x)\equiv(n-1)\big[\alpha
v(nx-v)+2h\alpha^2x^2\big]\qquad\qquad\nonumber\\
&&\qquad+\frac{(nx-v)^2}{(3n-2)}\alpha^2
-2\gamma\alpha^{\gamma}\frac{(x-v)}{x}\ ,
\end{eqnarray}
\begin{equation}\label{wang79}
X(\alpha,v,x)\equiv\alpha(nx-v)^2
-\gamma\alpha^\gamma-h\alpha^2x^2\ ,
\end{equation}
to transform equations (\ref{wang17}) and (\ref{wang18}) into
\begin{equation}\label{wang80}
\frac{\text{d}x}{X(\alpha,v,x)}=\frac{\text{d}\alpha}{A(\alpha,v,x)}
=\frac{\text{d}v}{V(\alpha,v,x)}=\text{d}\xi\ ,
\end{equation}
where $\xi$ is a new dependent variable.

The two coupled nonlinear MHD ODEs (\ref{wang17}) and (\ref{wang18})
are analyzed to determine the magnetosonic critical curve and
asymptotic solution behaviours near the magnetosonic critical curve,
and can be integrated numerically using the standard fourth-order
Runge-Kutta scheme (e.g., Press et al., 1986). Along with the
similarity MHD transformation, these two coupled nonlinear MHD ODEs
describe an important subset of MHD solutions to the original
nonlinear partial differential MHD equations (\ref{wang1})$-$(\ref{wang5}).

Using the above simplification, the ratio of Alfv\'en
wave speed $v_A$ to the gas sound speed $s$ becomes
\begin{equation}\label{wang119}
{v_A}/{s}=(h\alpha^{2-\gamma}x^2/\gamma )^{1/2}\
\end{equation}
for $\gamma+n=2$.

\subsection{Singular Surface and
Magnetosonic Critical Curve}\label{crdet}

Given parameters $n$, $\gamma$ and $h$ in MHD ODEs (\ref{wang17})
and (\ref{wang18}), there exists a characteristic surface in the
$(v,\alpha,x)$ space on which the denominators on the RHSs of both
equations vanish (Whitworth and Summers, 1985). Physically, we have
averaged over small-scale MHD fluctuations in our formulation.
Therefore, the magnetosonic critical point or curve should correspond
to a layer of a thickness comparable the mean scale of MHD fluctuations.
In other words, our model analysis relates to an averaged condition
in the actual MHD flow. Mathematically, this singular surface is
determined by computing $v$ from specific $x$ and $\alpha$ in the
ranges of $0<x<+\infty$ and $0<\alpha<+\infty$, namely,
\begin{equation}\label{wang65}
v=nx\mp (\gamma\alpha^{\gamma-1}+h\alpha x^2)^{1/2}\ ,
\end{equation}
in which one should pick up the upper minus sign
in order to satisfy the physical constraint of $nx>v$ and
thus to ensure $m(x)>0$. The solutions of the two coupled ODEs
(\ref{wang17}) and (\ref{wang18}) cannot cross the singular
surface unless they cross it at points along the so-called
magnetosonic critical curve where both the numerators and
denominators vanish simultaneously. Along this magnetosonic
critical curve, the derivatives of the dependent variables $v(x)$
and $\alpha(x)$ can be calculated from equations (\ref{wang17})
and (\ref{wang18}) using the L'H$\hat{\mbox{o}}$pital rule.
Mathematically, this magnetosonic critical curve is defined
by the following pair of equations
\begin{equation}\label{wang110}
(nx-v)^2=\gamma\alpha^{\gamma-1}+h\alpha x^2\
\end{equation}
and
\begin{equation}\label{wang39}
(n-1)v+\bigg(2hx+\frac{nx-v}{3n-2}\bigg)\alpha
-\frac{2(x-v)(nx-v)}{x}=0 ;
\end{equation}
these two equations immediately give
\[
v=nx\mp (\gamma\alpha^{\gamma-1}+h\alpha x^2)^{1/2}\ ,
\]
\begin{eqnarray}\label{wang40}
\pm\bigg[n-1&+&\frac{\alpha}{(3n-2)}\bigg]
(\gamma\alpha^{\gamma-1}+h\alpha x^2)^{1/2}\qquad\qquad\nonumber\\
&&\qquad\ =\frac{2\gamma}{x}\alpha^{\gamma-1}-n(n-1)x\ ,
\end{eqnarray}
with the latter leading to a quadratic equation of $x^{2}$ as
\begin{equation}\label{wang41}
A_1x^{4}+B_1x^{2}+C_1=0\ ,
\end{equation}
where the coefficients $A_1$, $B_1$ and $C_1$ are defined by
\[
A_1\equiv\bigg[n-1+\frac{\alpha}{(3n-2)}
\bigg]^{2}h\alpha-n^{2}(n-1)^{2}\ ,
\]
\[
B_1\equiv\gamma\alpha^{(\gamma-1)}\left\lbrace\bigg[n-1
+\frac{\alpha}{(3n-2)}\bigg]^{2}+4n(n-1)\right\rbrace\ ,
\]
\begin{equation}\label{wang42}
C_1\equiv -4\gamma^{2}\alpha^{(2\gamma-2)}\ .
\end{equation}

If one substitutes equations (\ref{wang39}) and (\ref{wang40})
into equation (\ref{wang18}), the numerator vanishes. By the
physical constraint of $m(x)>0$, the lower minus sign in equation
(\ref{wang40}) will be ignored, even though mathematically, it may
represent a new branch of the critical curve should this branch do
exist. The above expressions of the magnetosonic critical curve
appear far more complicated than the isothermal case (Shu, 1977;
Lou and Shen, 2004; Yu and Lou, 2005; Yu et al., 2006; Lou and Gao,
2006), as a result of a polytropic gas under the influence of a
random magnetic field characterized by a constant $h$ and a
reduced magnetic energy density $w(x)$. In reference to earlier
results of determining $v(x)$ and $\alpha(x)$ in terms of $x$
along the critical curve, the most straightforward procedure
one can take in the current polytropic MHD problem is to first
determine $x$ from a given $\alpha$ and then obtain the
corresponding $v$. This is somewhat unusual in determining the
magnetosonic critical curve for a given sequence of $\alpha$
values. The additional constraints for the magnetosonic critical
curve in the semi-complete space are $x>0$ and $\alpha>0$ besides
equation (\ref{wang41}). Physically, we are interested in the
parameter regime of $nx-v>0$ such that $m(x)>0$. It is obvious
that we always have $C_1<0$ in definition (\ref{wang42}). For
different $\alpha$ values and depending on the values of
coefficients $A_1$, $B_1$ and $C_1$, there are three possible
cases listed below.

Case I: subcase (i) of both $A_1<0$ and $B_1<0$ or subcase
(ii) a negative determinant $B_1^{2}-4A_1C_1<0$; there is
then no positive root for $x^2$ satisfying quadratic equation
(\ref{wang41}) and therefore there is no point along the
magnetosonic critical curve corresponding to such a range
of $\alpha$ values.

Case II: with $A_1<0$, $B_1>0$ and a non-negative determinant
$B_1^{2}-4A_1C_1\geq 0$, there are two positive roots for $x^2$
satisfying quadratic equation (\ref{wang41}); there are thus two
points on the magnetosonic critical curve corresponding to such a
range of $\alpha$ values.

Case III: with $A_1>0$, there is only one positive root for $x^2$
satisfying quadratic equation (\ref{wang41}) and there is thus one
point on the magnetosonic critical curve corresponding to such a
range of $\alpha$ values.

Once an $x$ value is determined for a given $\alpha>0$ by the
above procedure, we readily obtain the corresponding $v$ value.
According to equation (\ref{wang40}) for
\begin{equation}\label{wang62}
\frac{2\gamma\alpha^{\gamma-1}/x-n(n-1)x}
{[n-1+\alpha/(3n-2)]}>0\ ,
\end{equation}
we should obviously pick up the upper sign (i.e., the $'+'$ sign
in the second relation and $'-'$ sign in the first relation) in
equation (\ref{wang40}); otherwise, we should pick up the other
sign accordingly. As the physical constraint $m(x)>0$ requires
that $nx-v>0$, inequality (\ref{wang62}) sets the criterion for
a physical solution.

This sequence of determining the magnetosonic critical curve
appears more involved than those in the isothermal case without
a random magnetic field (Shu, 1977; Lou and Shen, 2004).

\subsection{Asymptotic and Global Similarity Solutions}\label{asym}

In order to specify initial or boundary conditions for numerical
integrations, one needs to derive asymptotic similarity MHD
solutions. These asymptotic MHD solutions also carry their
physical implications. It is also possible to derive some regular
solutions from the coupled nonlinear MHD ODEs for physical
interpretations and for reference of numerical results. In general,
we have found the MHD counterparts of the isothermal asymptotic
solutions (Lou and Shen, 2004), and, in particular, we have derived
a novel MHD asymptotic solution as well as a new class of
asymptotic behaviours. We also examine the corresponding ratio of
the Alfv\'en wave speed to the sound speed and the dominant force
among the gravity force, the thermal pressure gradient force, the
magnetic pressure gradient force and the magnetic tension force
for each MHD asymptotic solution. We also analyze the behaviour
of crossing the magnetosonic critical curve for MHD solutions.

\subsubsection{Asymptotic Solutions in the Limit of Large $x$}

As $x$ approaches infinity in equations (\ref{wang17}) and
(\ref{wang18}) and with finite $\alpha(x)$ and $v(x)$, one obtains
the following pair of MHD ODEs to leading orders of large $x$
\begin{equation}\label{wang20}
\alpha'=-\frac{2\alpha}{nx}\ ,
\end{equation}
\begin{equation}\label{wang21}
v'=\frac{(n-1)v}{nx}+\bigg[\frac{1}{3n-2}
+\frac{2h(n-1)}{n^{2}}\bigg]\alpha
-\frac{2\gamma\alpha^{\gamma-1}}{n^{2}x^{2}}\ .
\end{equation}
We solve these two asymptotic MHD ODEs to obtain
\begin{equation}\label{wang22}
\alpha=A_0x^{-{2}/{n}}\ ,
\end{equation}
\begin{eqnarray}\label{wang23}
&&v=B_0x^{1-1/n}-\bigg[\frac{n}{3n-2}
+\frac{2h(n-1)}{n}\bigg]A_0x^{1-2/n}\nonumber\\
&&\qquad +\frac{2\gamma A_0^{\gamma-1}}
{n[2(n+\gamma)-3]}x^{{(2-2\gamma-n)}/{n}}\ ,\qquad\qquad
\end{eqnarray}
where $A_0$ and $B_0$ are two constants of integration. With
similarity MHD transformation (\ref{wang6}) and (\ref{wang19}),
this asymptotic MHD solution becomes
\begin{equation}
\rho\equiv c\alpha=cA_0 x^{- 2/n} =\frac{k^{1/n}A_0r^{-2/n}}{4\pi
G}\ ,
\end{equation}
\begin{equation}\label{wang114}
u\equiv bv=bB_0x^{{(n-1)}/{n}}=B_0k^{{1}/{(2n)}}r^{1-1/n}\ ,
\end{equation}
to the leading order of large $x$, indicating that the gas mass
density and radial flow speed profiles are both independent of
time $t$ at large $r$. For $n=1$, expression (\ref{wang114})
represents a constant radial flow speed at very large $r$ as
emphasized earlier (Shen and Lou, 2004; Lou and Shen, 2004) in an
isothermal gas. For $\alpha(x)$ and $v(x)$ to be non-increasing at
large $x$, this type of asymptotic MHD solutions requires that
\begin{equation}\label{wang24}
\max[0,\ 2(1-\gamma)]\leq n\leq 1\ .
\end{equation}

\noindent Note that for $h$ greater than the critical value $h_c$,
i.e.,
\begin{equation}\label{wang111}
h>h_c\equiv\frac{n^2}{2(1-n)(3n-2)}\ ,
\end{equation}
the coefficient of $x^{{(n-2)}/{n}}$ term in equation
(\ref{wang23}) becomes positive, while for $h<h_c$ this
coefficient is negative. The presence of this critical value $h_c$
for $h$ is a consequence of the polytropic and magnetic nature of
our MHD problem, which will be further discussed. The
corresponding reduced mean magnetic energy density $w$ at large
$x$ is
\begin{equation}\label{wang56}
w=A_0^2 hx^{2-{4}/{n}}\ ,
\end{equation}
which goes to zero as $x\rightarrow +\infty$. With similarity MHD
transformations (\ref{wang6}) and (\ref{wang19}), we have
\begin{equation}\label{wang115}
<B_t^2>\equiv fw=\frac{A_0^2h}{G}k^{2/n}r^{2-4/n}\
\end{equation}
in dimensional form, indicating that the magnetic field does not
change with time $t$ at large $r$ in a magnetized gas cloud. For a
usual polytropic gas with $\gamma+n=2$, the corresponding ratio of
the Alfv\'en wave speed $v_A$ to the sound speed $s$ for large $x$
remains constant
\begin{equation}\label{wang120}
\frac{v_A}{s}=\bigg(\frac{h}{\gamma}A_0^{2-\gamma}\bigg)^{1/2}
\end{equation}
at large $x$. As $x$ approaches infinity, the denominators of both
equations (\ref{wang17}) and (\ref{wang18}) approach $n^2\alpha
x^2$ for this series of solutions and these solutions will not
encounter the singular surface $X(\alpha,v,x)=0$.
In the regime of large $x$ and for $\gamma=2-n$, the dominant
forces are both magnetic pressure and gas pressure, and the
magnetic pressure force is stronger than the magnetic tension
force in magnitude with a ratio of $(2/n)-1$. The magnetic
pressure gradient force points radially outward.

\subsubsection{Asymptotic Solutions in the Limit of Small $x$}

With the assumptions of $v^{2}\gg\alpha^{\gamma-1}+h\alpha x^{2}$
and of $\alpha^{\gamma-2}\ll xv$ as $x$ approaches zero in MHD
ODEs (\ref{wang17}) and (\ref{wang18}), we derive the following
pair of asymptotic MHD equations to leading orders of small $x$,
viz.
\begin{equation}\label{wang25}
\frac{\text{d}\alpha}{\text{d}x}=-\frac{2\alpha}{x}
-\frac{\alpha^{2}}{(3n-2)v}\ ,
\end{equation}
\begin{equation}\label{wang26}
\frac{\text{d}v}{\text{d}x}=\frac{\alpha}{(3n-2)}\ ,
\end{equation}
which, by a direct integration, lead to two integrals
\begin{equation}\label{wang27}
\alpha(x)=\bigg[\frac{(3n-2)m(0)}{2x^{3}}\bigg]^{1/2}\ ,
\end{equation}
\begin{equation}\label{wang28}
v(x)=-\bigg[\frac{2m(0)}{(3n-2)x}\bigg]^{1/2}\ ,
\end{equation}
where $m(0)$ is an integration constant representing the core mass
at the centre. For this family of asymptotic solutions at small
$x$, both assumptions stated at the beginning of this subsection
are satisfied when
\begin{equation}\label{wang29}
\gamma<\frac{5}{3}\ \qquad\text{ and }\qquad\ n>\frac{2}{3}\ .
\end{equation}
The corresponding reduced magnetic energy density $w$ is
\begin{equation}\label{wang70}
w=\frac{(3n-2)hm(0)}{2x}\ ,
\end{equation}
which diverges as $x\rightarrow 0^{+}$. For a conventional
polytropic gas with $\gamma=2-n$, the corresponding ratio of the
Alfv\'en wave speed to the sound speed becomes
\begin{equation}\label{wang121}
\frac{v_A}{s}=\bigg\lbrace\frac h\gamma
\left[\frac{(3n-2)m(0)}{2}\right]^{{(2-\gamma)}/{2}}
x^{(3\gamma/2-1)}\bigg\rbrace^{1/2}\ .
\end{equation}
Since we take $\gamma>1$, this speed ratio approaches zero as
$x\rightarrow 0^{+}$. These asymptotic similarity MHD solutions
are the free-fall solutions entrained with a random magnetic
field. The $x-$dependence in this limiting behaviour is related to
the value of $n$. As $x$ approaches zero, the denominators of both
equations (\ref{wang17}) and (\ref{wang18}) approach $\alpha v^2$
for this series of asymptotic MHD solutions and these MHD
solutions will not reach the singular surface. Under the
assumptions above, one further obtains higher-order terms
for the asymptotic similarity MHD solutions
\begin{eqnarray}\label{wang74}
&&\alpha(x)=\bigg[\frac{(3n-2)m(0)}{2x^3}\bigg]^{1/2}\nonumber\\
&&\qquad+\frac{(3n-2)}{(\gamma-1)}\frac{\gamma x}
{2m(0)}\bigg[\frac{(3n-2)m(0)}{2x^3}\bigg]^{{\gamma}/{2}}\ ,\qquad
\end{eqnarray}
\begin{eqnarray}\label{wang75}
&&v(x)=-\bigg[\frac{2m(0)}{(3n-2)x}\bigg]^{1/2}\qquad\nonumber\\
&&\qquad+\frac{1}{(\gamma-1)}\frac{\gamma x^2}{m(0)}
\bigg[\frac{(3n-2)m(0)}{2x^3}\bigg]^{{\gamma}/{2}}\ ,\qquad
\end{eqnarray}
with the corresponding reduced magnetic energy density
\begin{eqnarray}\label{wang76}
&&w(x)=\frac{(3n-2)hm(0)}{2x}+h\gamma\bigg[\frac{(3n-2)m(0)}
{2x^3}\bigg]^{\gamma/2}\nonumber\\
&&\qquad\times\frac{(3n-2)}{(\gamma-1)}
\bigg[\frac{(3n-2)x^3}{2m(0)}\bigg]^{1/2}\ .
\end{eqnarray}
In comparison with the isothermal analysis (Whitworth and Summers,
1985), this solution appears somewhat different. This is mainly
because Whitworth and Summers (1985) considered the isothermal case
of $\gamma=1$; in that case the denominators of the second-order
terms vanish, and one should take the L'H$\hat{\mbox{o}}$pital
rule with respect to $\gamma$ in order to derive the proper form
of expansion solution, that is, an $x\ln x$ form for the
second-order term. For this central free-fall asymptotic solution,
the leading force is the gravity force, and the magnetic tension
force is twice the magnetic pressure force in magnitude. The
magnetic pressure gradient force points radially outward.
\begin{figure}
\includegraphics[width=3.3in,bb=100 270 480 570]{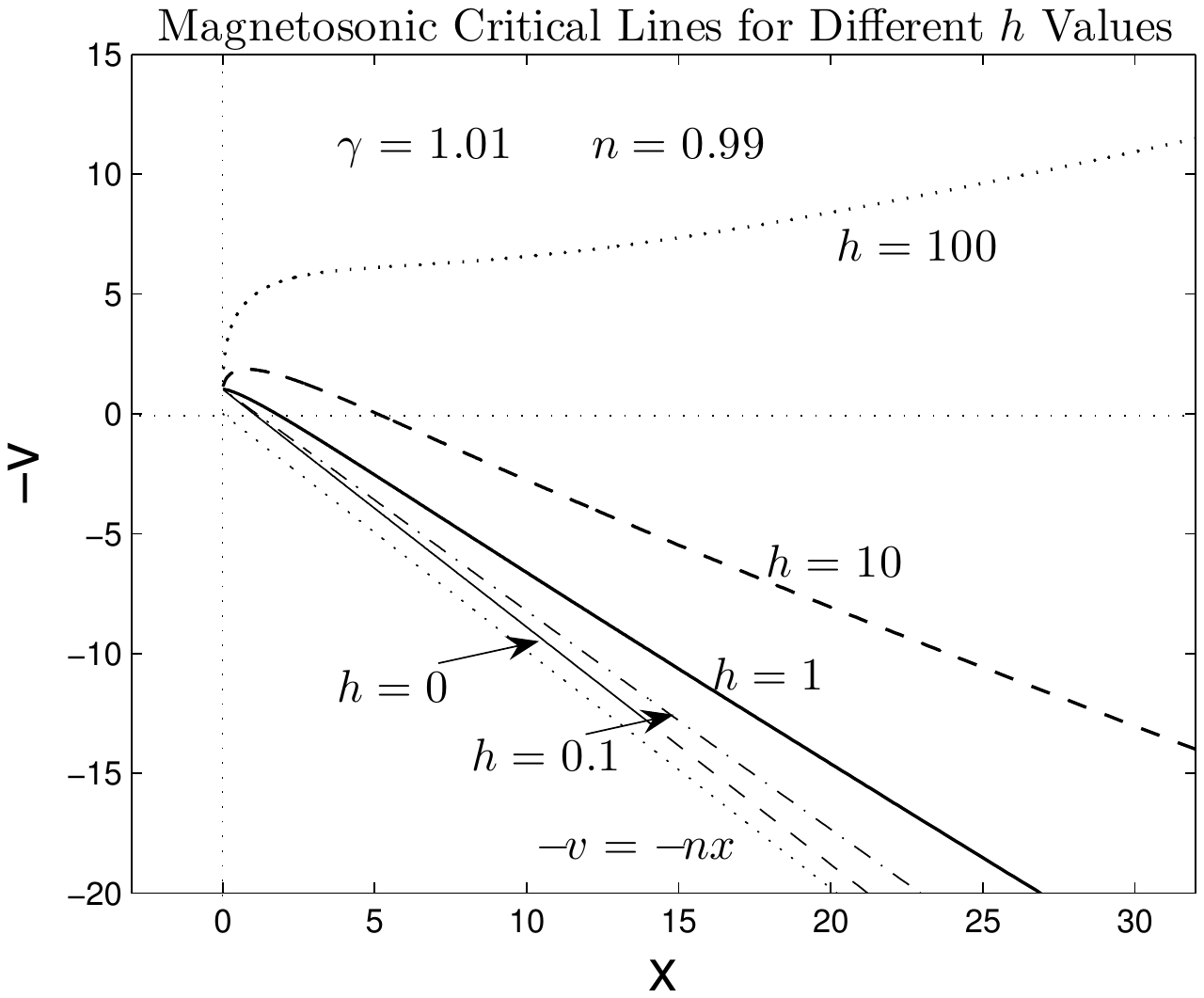}
\caption{The magnetosonic critical curves for different $h$ values
with given parameters $n=0.99$ and $\gamma=1.01$ ($\gamma=2-n$ is
imposed for a usual polytropic gas). The straight dotted line
passing through the origin represents $nx-v=0$; physical MHD flow
solutions with positive mass should be to the upper right of this
straight line. As $h$ increases, the average slope
$\text{d}(-v)/\text{d}x$ of an individual magnetosonic critical
curve increases from negative to positive. The two segments for
the case of $h=0$ with two different line types (i.e., light solid
and dashed lines) indicate that this curve consists of two
portions corresponding to two different roots of equation
(\ref{wang41}). Too compact to be seen here, the diverging
behaviours at small $x$ for all the magnetosonic curves are
shown in Fig. \ref{critical101s}. } \label{critical101}
\end{figure}

\subsubsection{Novel Magnetic Solutions in the Limit of Small $x$}

\begin{figure}
\includegraphics[width=3.3in,bb=100 270 480 570]{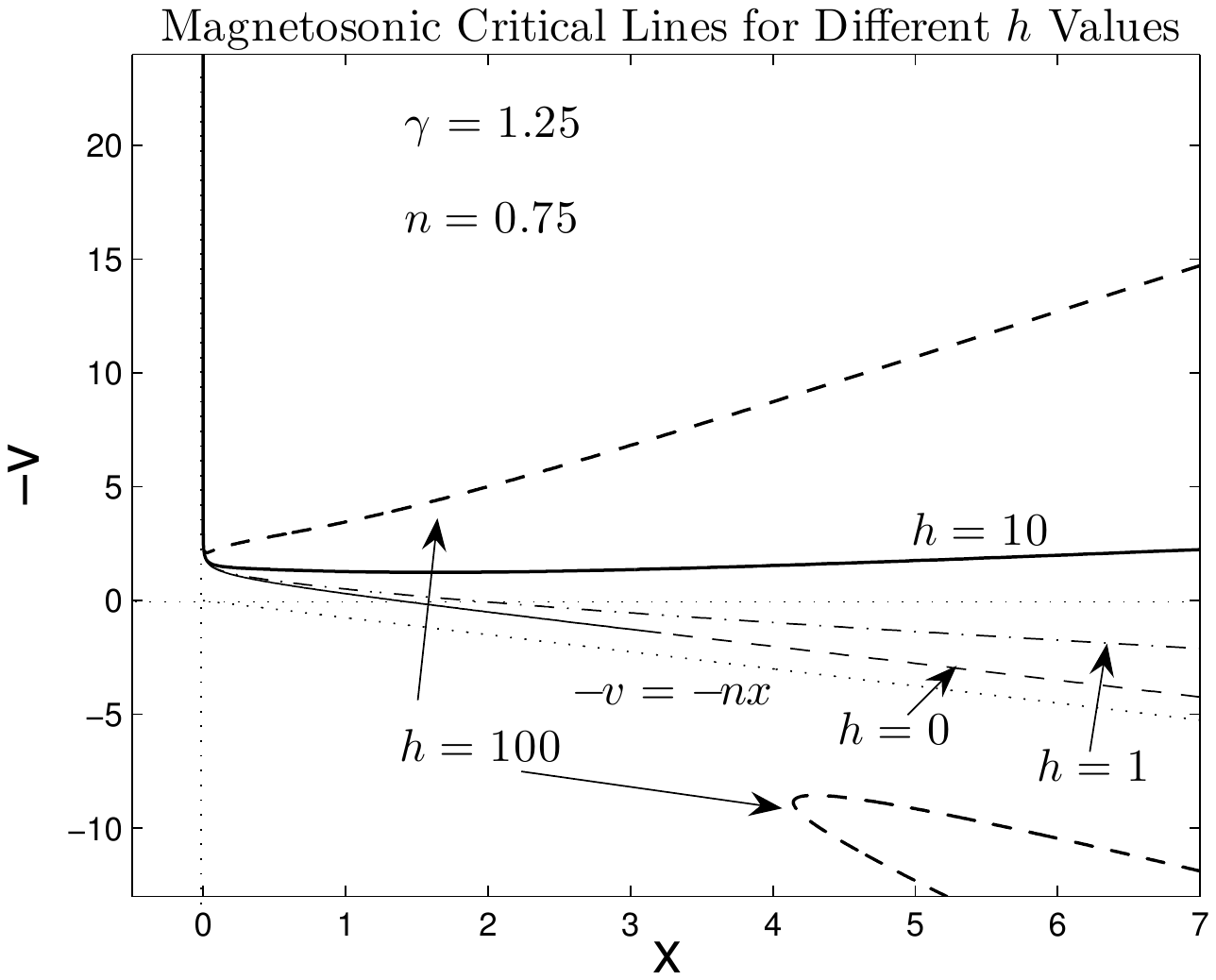}
\caption{The magnetosonic critical curves for different values of
$h$ with given parameters $n=0.75$ and $\gamma=1.25$ ($n=2-\gamma$
for a usual polytropic gas). The light dotted straight line
$nx-v=0$ passes through the origin; the two perpendicular light
dotted lines represent the abscissa and ordinate axes,
respectively. Similar to variation trends shown in Fig.
\ref{critical101}, as $h$ increases (i.e., $0,\ 1,\ 10,\ 100$),
the average slope $\text{d}(-v)/\text{d}x$ of an individual
magnetosonic critical curve increases from negative to positive,
and the two segments (i.e., the light solid and dashed lines) for
the case of $h=0$ with two different line types correspond to the
two sensible roots of equation (\ref{wang41}). The magnetosonic
critical curve with $h=100$ (heavy dashed lines) has two branches,
as indicated in the figure and the lower branch beneath $nx-v=0$.
}\label{critical125}
\end{figure}

\begin{figure}
\includegraphics[width=3.3in,bb=100 270 480 570]
{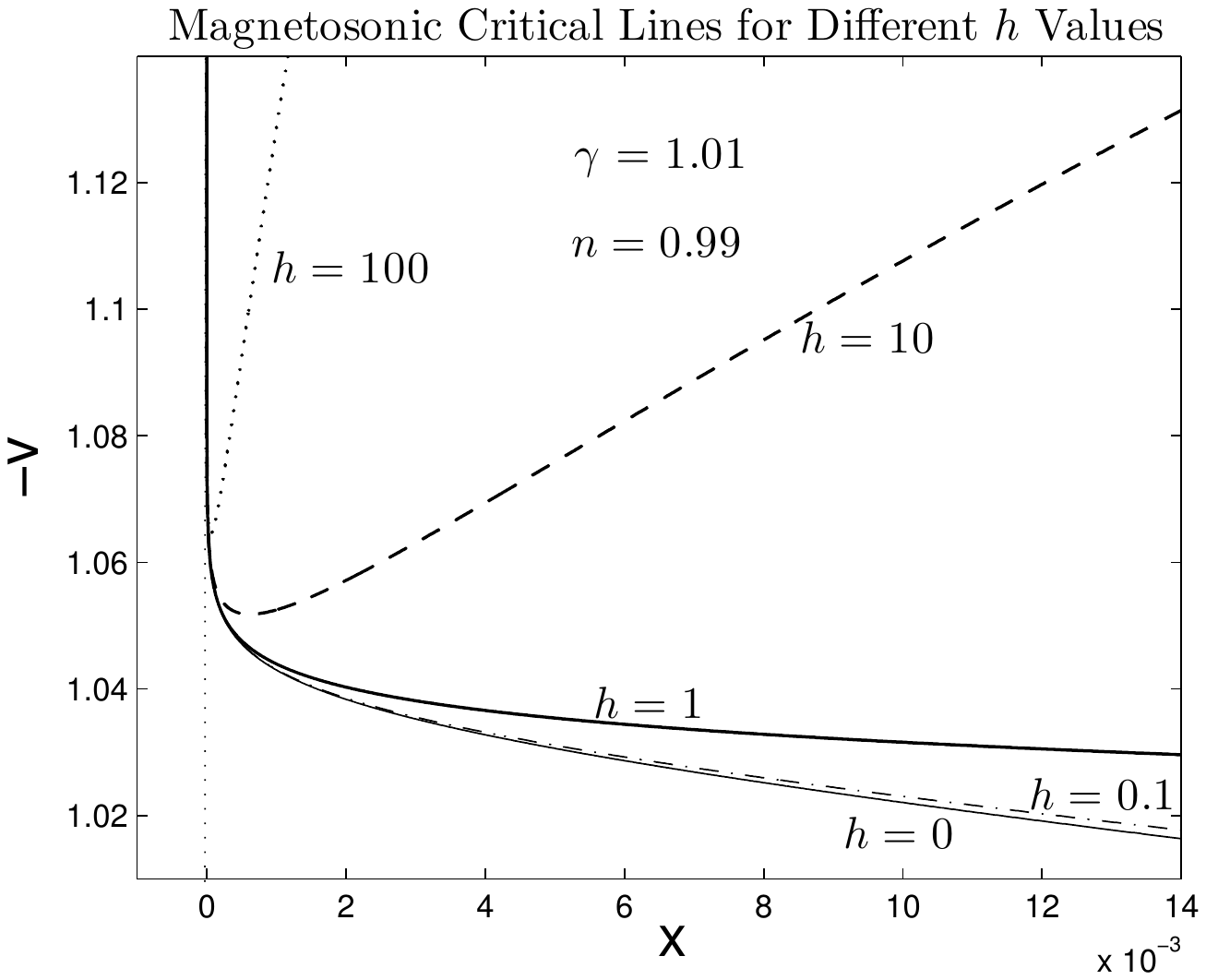}\caption{Enlarged version of Fig.
\ref{critical101} with the same parameters yet for much smaller
$x$, intended to clearly show the diverging behaviours of the
magnetosonic critical curves as $x\rightarrow 0^{+}$. The
horizontal $x$ axis is in a numerical scale of $10^{-3}$.
}\label{critical101s}
\end{figure}

\begin{figure}
\includegraphics[width=3.3in,bb=100 270 480 570]
{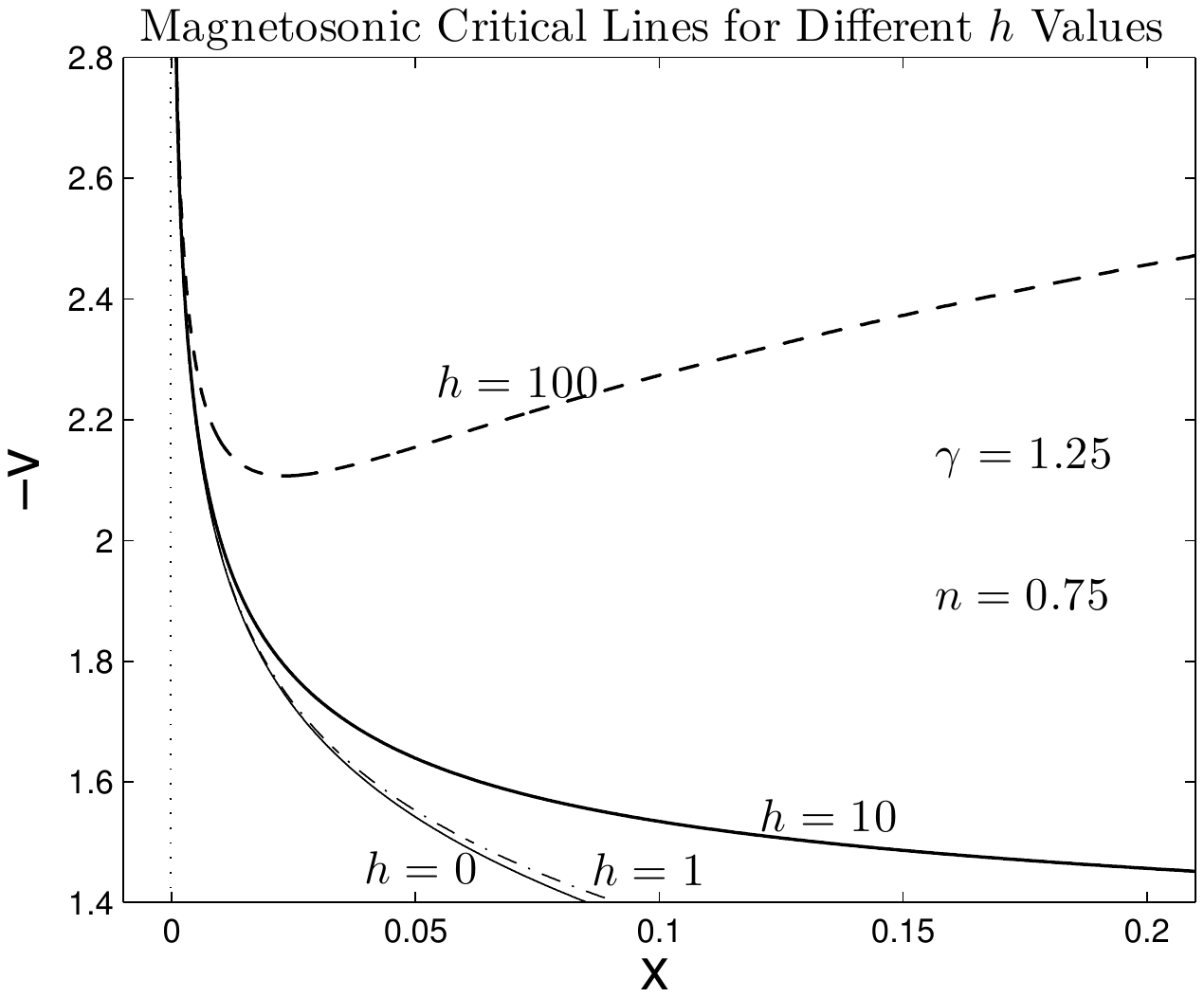} \caption{Enlarged version of Figure
\ref{critical125} with the same parameters yet for much smaller
$x$, intended to clearly show the diverging behaviours of the
magnetosonic critical curves in the limit of $x\rightarrow 0^{+}$.
}\label{critical125s}
\end{figure}

For a magnetized gas cloud, it is possible to derive another MHD
asymptotic series solution in which $v={\cal O}(x)$ and
$\alpha\rightarrow +\infty$ as $x\rightarrow 0^+$. With the
assumption of $v={\cal O}(x)$ and $\alpha^{\gamma}\ll\alpha^2x^2$
in equation (\ref{wang18}), we have $v'$ approaching a constant
and obtain
\begin{equation}\label{wang66}
v=-Cx\ ,
\end{equation}
where $C$ is an integration constant. Substituting expression
(\ref{wang66}) into equations (\ref{wang17}) and (\ref{wang18}),
we obtain
\begin{equation}\label{wang67}
\alpha'=-\frac{\big[2h+(n+C)/(3n-2)\big]\alpha}{hx}\ ,
\end{equation}
\begin{equation}\label{wang68}
v'=-2(n-1)-\frac{(n+C)^2}{h(3n-2)}\ .
\end{equation}
By equation (\ref{wang66}), this in turn requires
\begin{equation}\label{wang69}
2(n-1)+\frac{(n+C)^2}{h(3n-2)}=C\
\end{equation}
for consistency. Once this quadratic equation (\ref{wang69}) for
$C$ has at least one positive root of $C$, we obtain at least one
possible asymptotic MHD solution in the form of
\begin{equation}\label{wang71}
\alpha(x)=Dx^{-2-(n+C)/[(3n-2)h]}\ ,
\end{equation}
\begin{equation}\label{wang72}
v(x)=-Cx\ ,
\end{equation}
where $D$ is yet another integration constant. Quadratic
equation (\ref{wang69}) for $C$ can be readily solved to give
\begin{equation}\label{wang125}
C=-n+\frac{(3n-2)}{2}\left[h\pm (h^2-4h)^{1/2}\right]\
\end{equation}
and the requirement of $nx-v>0$ is satisfied for both roots of
$C$. This immediately requires $h>4$ (i.e., a sufficiently strong
magnetic field) for a valid asymptotic MHD solution of this kind.
In short, this asymptotic solution for small $x$ is described by
\begin{equation}\label{wang126}
v(x)=\bigg\lbrace n-\frac{(3n-2)}{2} \left[h\pm
(h^2-4h)^{1/2}\right]\bigg\rbrace x\ ,
\end{equation}
\begin{equation}\label{wang127}
\alpha(x)=Dx^{-5/2\mp\sqrt{h^2-4h}/(2h)}\ ,
\end{equation}
\begin{equation}\label{wang73}
w(x)=hD^2 x^{-3\mp\sqrt{h^2-4h}/h}\ ,
\end{equation}
and the corresponding reduced enclosed mass $m(x)$
\begin{equation}\label{wang128}
m(x)=\frac{(3n-2)}{2}D\left[h\pm (h^2-4h)^{1/2}\right]
x^{1/2\mp\sqrt{h^2-4h}/(2h)}\ .
\end{equation}
The requirement on $h$ is discussed below. For
$\alpha^\gamma\ll\alpha^2 x^2$, we obtain the following inequality
\begin{equation}\label{wang131}
\gamma<\frac{6h\pm2(h^2-4h)^{1/2}}{5h\pm(h^2-4h)^{1/2}}\ .
\end{equation}
For the upper plus signs, if $\gamma<{6}/{5}$, this requirement
is automatically satisfied, while if $\gamma>{6}/{5}$, this
requirement means that $h>4/[1-(5\gamma-6)^2/(2-\gamma)^2]$.
For the lower minus signs, if $\gamma<6/5$, this requirement
means that $4<h<4/[1-(6-5\gamma)^2/(2-\gamma)^2]$, while if
$\gamma>6/5$, this condition cannot be met, i.e. there does not
exist such asymptotic solutions with the lower negative signs. In
the absence of magnetic field with $h=0$, this form of asymptotic
solution disappears completely. This is a brand-new asymptotic MHD
solution in a magnetized gas cloud, and global semi-complete
solutions matching this asymptotic solution can be constructed
numerically (see Figures \ref{Once101_10}, \ref{Once101_100_1}
and \ref{Once125_10_1} for specific examples). Physically, this
asymptotic MHD solution describes a much compressed accreting
nucleus where the magnetic pressure $<B_t^2>/(8\pi)$ becomes much
stronger than the thermal gas pressure $p$ to oppose the
gravitational collapse such that the reduced radial inflow speed
$v(x)$ approaches zero linearly with $x$ as $x\rightarrow 0^{+}$.
Physically, we anticipate that a very strong random magnetic field
confined to a sufficiently small spatial volume would certainly
destroy the quasi-spherical symmetry at some point and drive
random flows. We further expect violent and sporadic magnetic
activities to destroy the similarity evolution. In spite of all
these, we count on the gravity of accreted core materials to more
or less control a central sphere. In other words, sufficiently far
away from this central magnetized sphere of influence, we may
ignore feedbacks of central activities and apply our self-similar
MHD inflow solutions. The scenario envisioned here essentially
parallels that of a spherical symmetric central inflow without
magnetic field. Ultimately, there must be a central object to
confront radial inflows and thus destroy the similarity flow
evolution. A self-similar flow solution is only valid on large
scales and outside a certain sphere surrounding the core. As $x$
approaches zero, the denominators of both equations (\ref{wang17})
and (\ref{wang18}) for this asymptotic series approach
$-h\alpha^2x^2$ and these MHD solutions do not encounter the
magnetosonic singular surface. For this asymptotic MHD solution,
the magnetic pressure, tension and gravity forces are in the same
order of magnitude, all overpowering the thermal gas pressure
force, and the magnetic pressure force is the strongest. Including
one more term in the series expansion, this novel magnetic
asymptotic similarity solution appears as
\begin{eqnarray}\label{wang136}
&&v(x)=\bigg\lbrace n-\frac{(3n-2)}{2}\left[h\pm
(h^2-4h)^{1/2}\right]\bigg\rbrace x
\nonumber\\
&&
+\gamma D^{\gamma-2}
\frac{(3n-2)\big[3h-2\pm3(h^2-4h)^{1/2}\big]}
{(6-5\gamma)h\mp\gamma(h^2-4h)^{1/2}}\nonumber\\
&&\times
x^{(2-\gamma)\big[5/2\pm\sqrt{h^2-4h}/(2h)\big]-1}+\cdots\ ,
\end{eqnarray}
\begin{eqnarray}\label{wang137}
&&\alpha(x)=Dx^{-5/2\mp\sqrt{h^2-4h}/(2h)}+\gamma
D^{\gamma-1}\nonumber\\&&\times\bigg[6h-4\pm6(h^2-4h)^{1/2}
+\big[5\pm(h^
2-4h)^{1/2}/h\big]\nonumber\\&&\times\big[(6-5\gamma)h\mp
(h^2-4h)^{1/2}\big]\bigg]\nonumber\\&&
\times\bigg[\big[(6-5\gamma)h\mp\gamma(h^2-4h)^{1/2}
\big]\nonumber\\&&
\big[(6-5\gamma)h\pm(2-\gamma)(h^2-4h)^{1/2}\big]\bigg]^{-1}\nonumber\\
&&\times x^{(\gamma-1)\big[-5/2\mp
\sqrt{h^2-4h}/(2h)\big]-2}+\cdots\ .
\end{eqnarray}

\begin{figure}
\includegraphics[width=3.3in,bb=100 270 480 570]
{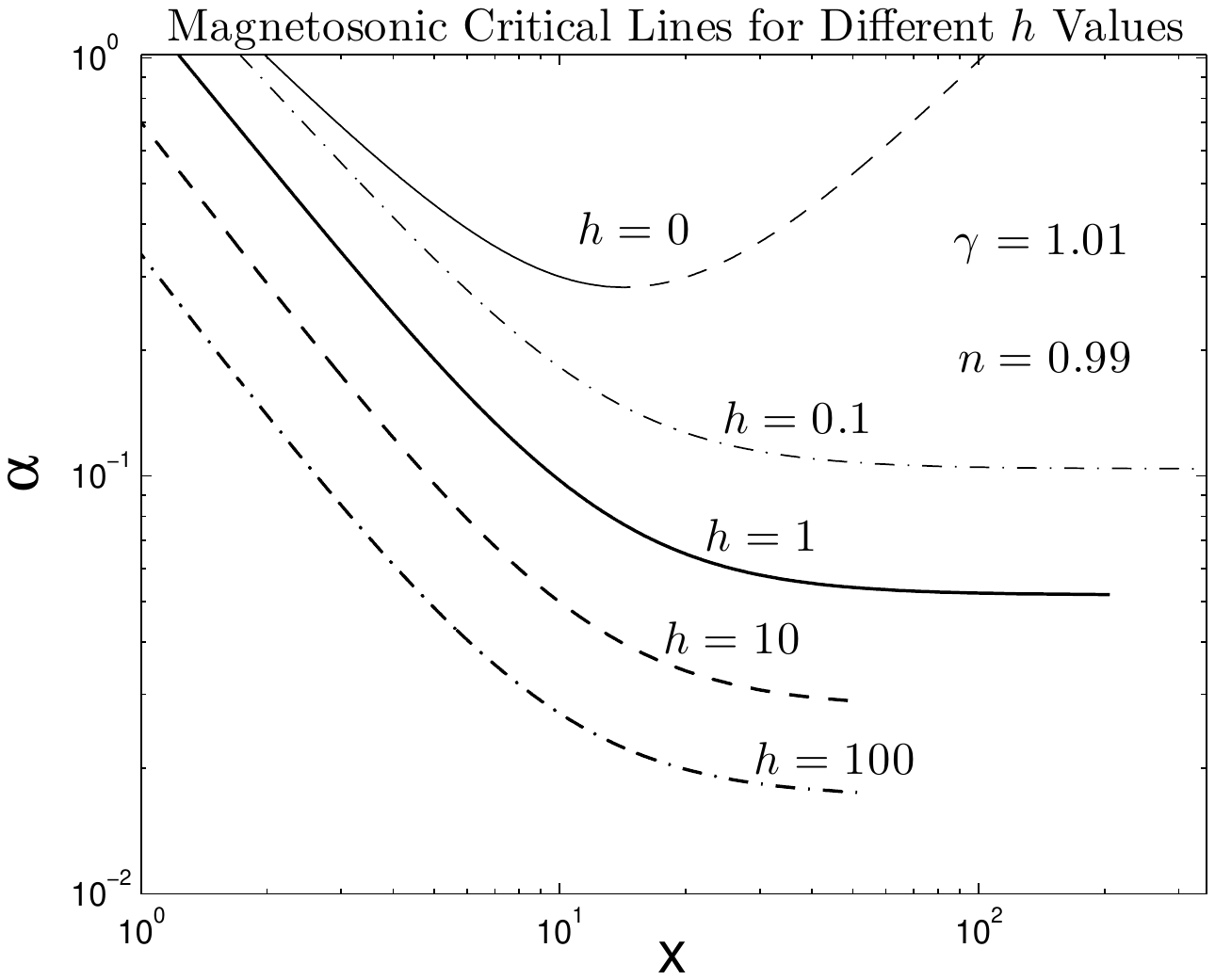}\caption{Magnetosonic critical curves of
$\alpha$ versus $x$ displayed in logarithmic scales for different
values of $h$ with given $\gamma=1.01$ and $n=0.99$ as shown in
Figure \ref{critical101}. }\label{criticalalpha101}
\end{figure}

\subsubsection{A Singular Global Magnetostatic Solution }

\begin{figure}
\includegraphics[width=3.3in,bb=100 270 480 570]
{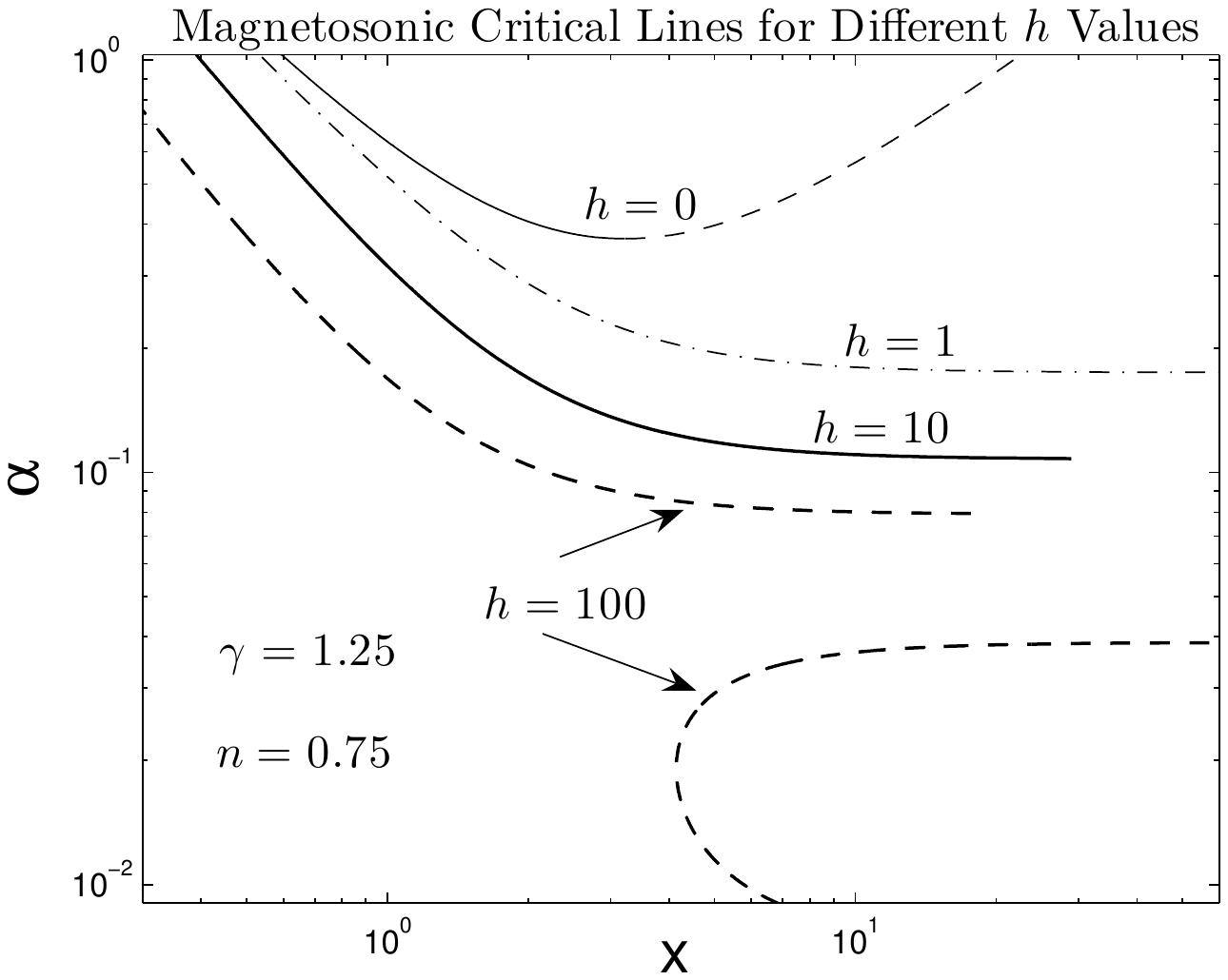}\caption{Magnetosonic critical curves of
$\alpha$ versus $x$ displayed in logarithmic scales for different
values of $h$ with given $\gamma=1.25$ and $n=0.75$ as shown in
Figure \ref{critical125}. }\label{criticalalpha125}
\end{figure}

For a constant $v$, equation (\ref{wang9}) reduces to
\begin{equation}\label{wang30}
\quad (nx-v)\alpha'=-2\frac{(x-v)}{x}\alpha\ ,
\end{equation}
which can be readily integrated for $\alpha(x)$ in the form of
\begin{equation}\label{wang31}
\qquad \alpha(x)=cx^{-2}(nx-v)^{2-2/n}\ ,
\end{equation}
where $c$ is an integration constant. Along with equation
(\ref{wang16}), there exists a special singular global
magnetostatic solution such that $v=0$ when $n+\gamma=2$,
namely
\[
v=0\ ,\qquad\alpha=\bigg[\frac{(2-\gamma)^{2}}
{2\gamma(4-3\gamma)}+\frac{(1-\gamma)} {\gamma}h\bigg]^{\frac{1}
{\gamma-2}}x^{\frac{-2}{2-\gamma}}\ ,
\]
\begin{equation}\label{wang32}
m=(2-\gamma)\bigg[\frac{(2-\gamma)^{2}} {2\gamma
(4-3\gamma)}+\frac{(1-\gamma)}
{\gamma}h\bigg]^{\frac{1}{\gamma-2}} x^{3-\frac{2}{2-\gamma}}\ .
\end{equation}
The reduced magnetic energy density $w(x)$ is
\begin{equation}\label{wang57}
w(x)=h\bigg[ \frac{(2-\gamma)^{2}}{2\gamma(4-3\gamma)}
+\frac{(1-\gamma)}{\gamma}h\bigg]^{{2}/{(\gamma-2)}}
x^{2-{4}/{(2-\gamma)}}\ ,
\end{equation}
which diverges as $x\rightarrow 0^{+}$ and vanishes as
$x\rightarrow +\infty$. Note that for $h\geq h_c$, this global
similarity MHD solution does not exist. When $\gamma=2-n$ for a
usual polytropic gas, the corresponding ratio of the Alfv\'en wave
speed to the sound speed becomes a constant
\begin{equation}\label{wang122}
\frac{v_A}{s}=\bigg(\frac{h}{\gamma}\bigg)^{1/2}
\bigg[\frac{(2-\gamma)^{2}}{2\gamma(4-3\gamma)}
+\frac{(1-\gamma)}{\gamma}h\bigg]^{-1/2}\ .
\end{equation}
This global similarity MHD solution reduces to that of Suto and
Silk (1988) for $h=0$ or $w=0$ as expected. This is a new singular
polytropic magnetostatic solution, constructed in the similar
manner as has been done before (Shu, 1977; Suto and Silk, 1988;
Lou and Shen, 2004). When $h<h_c$, this solution encounters the
magnetosonic singular surface at the point
\begin{eqnarray}\label{wang135}
x=&&\bigg[\frac{2n}{2h+n/(3n-2)}\bigg]^{{(\gamma-2)}/{2}}\nonumber\\
&&\times\bigg[\frac{(2-\gamma)^{2}}{2\gamma (4-3\gamma)}
+\frac{(1-\gamma)}{\gamma}h\bigg]^{-1/2}\ ,
\end{eqnarray}
and this point is the intersection of the $v=0$ surface and the
magnetosonic critical curve; when $h>h_c$, this point does not
exist. More precisely, one readily finds that for a fixed
$\gamma$, this point moves to infinity as $h$ approaches
$h_c^{-}$. Expression (\ref{wang135}) can also be used to
determine the location of the `kink point' of the mEWCSs. In this
solution, the four forces, viz. thermal gas pressure, magnetic
pressure, magnetic tension and gravity forces are in the same
order, and the magnetic pressure force is stronger than the
magnetic tension force with a ratio of $-1+2/(2-\gamma)$, with the
magnetic pressure gradient force pointing radially outward. In
other words, the random magnetic field is not force-free. In the
isothermal limit of $\gamma=1$, this ratio becomes unity and the
random magnetic field is essentially quasi-force-free (Yu and
Lou, 2005; Yu et al., 2006).

\begin{figure}
\includegraphics[width=3.3in,bb=100 270 480 570]{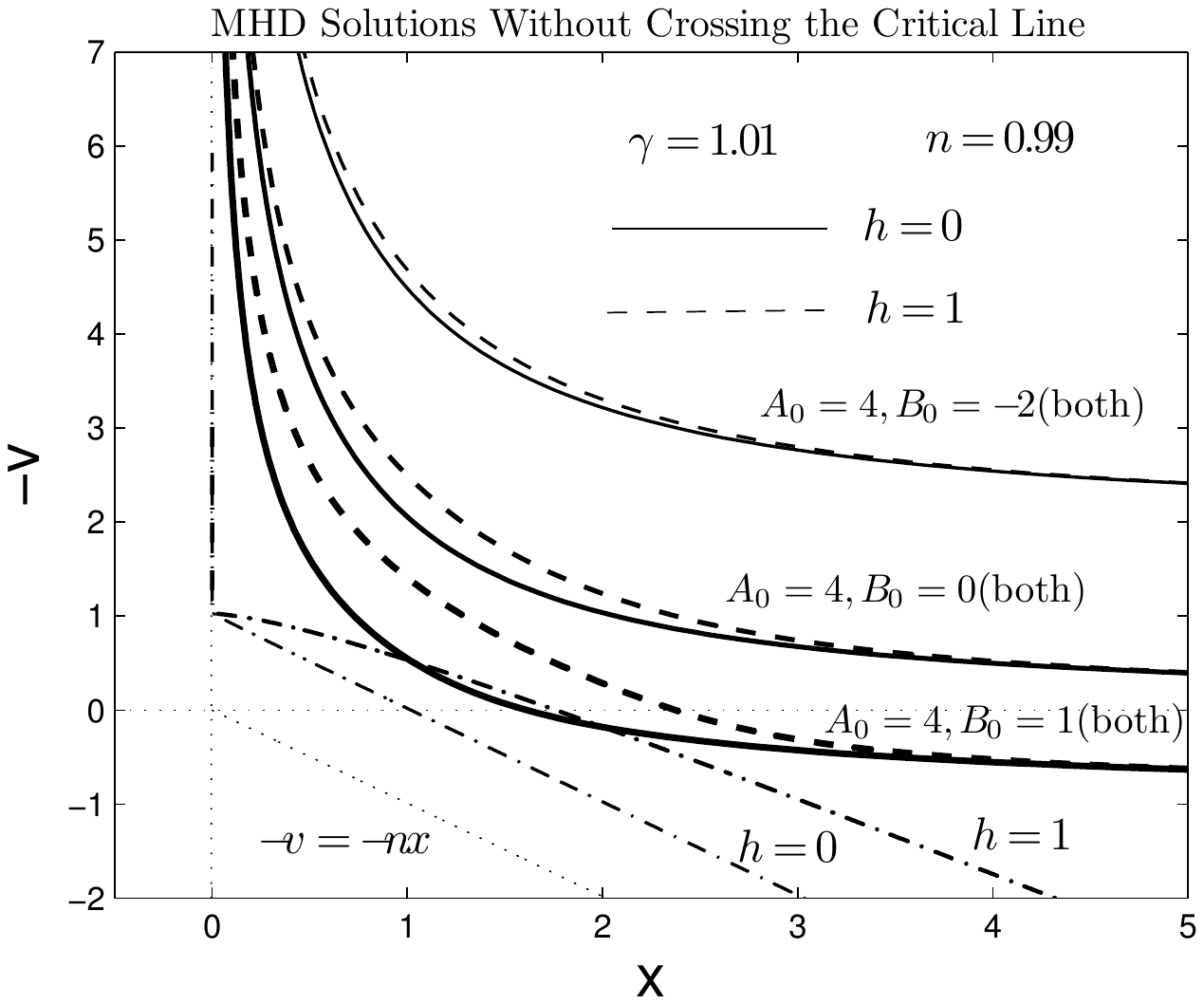}
\caption{Global numerical similarity MHD solutions in the
semi-complete space without crossing the magnetosonic critical
curve can be readily constructed using asymptotic solutions
(\ref{wang22}) and (\ref{wang23}) to start at a sufficiently large
$x$ (e.g., $x=100$), with the two parameters $\gamma=1.01$ and
$n=0.99$ being specified for a conventional polytropic gas. The
two perpendicular dotted lines denote the abscissa and ordinate
axes, respectively, and the straight dotted line $-v=-nx$ is the
demarcation to the lower left of which solutions become unphysical
for a negative enclosed gas mass. The two dash-dotted curves are
the sonic ($h=0$) and magnetosonic ($h=1$) critical curves. The
solid ($h=0$) and dashed ($h=1$) curves are similarity MHD flow
solutions with $h=0$ and $h=1$, respectively; for the first pair
solutions with light linetypes, we have both $A_0=4$ and $B_0=-2$;
for the second pair solutions with heavy linetypes, we have both
$A_0=4$ and $B_0=0$; and for the third pair solutions with
extremely heavy linetypes, we have both $A_0=4$ and $B_0=1$ as
examples of illustrations for similarity MHD solutions without
crossing the critical curves and with outward flow speeds at
far-away region. }\label{numer101_1}
\end{figure}

Note that this solution can also serve as an asymptotic
quasi-static condition, with $v$ approaching zero faster than
$\sim {\cal O}(x)$ and $\alpha\simeq x^{-2/n}$. These behaviours
are primarily caused by the polytropic equation of state with the
magnetic field playing the role of modification. This type of
asymptotic solutions are referred to as `quasi-static' MHD
asymptotic solutions, which can be further sub-divided into two
types - type I `quasi-static' MHD asymptotic solutions without
oscillations and type II `quasi-static' MHD asymptotic solutions
with oscillatory behaviours. A detailed analysis of this asymptotic
solution in a polytropic gas without magnetic field has been given
by Lou and Wang (2006). We here
show such MHD solutions in our figure illustrations (type II
`quasi-static' asymptotic solutions in Figures \ref{Once125_0_1}
and \ref{Once125_2_1}). An analysis of these newly found MHD
asymptotic solutions is contained in Appendix \ref{asystat}.

\begin{figure}
\includegraphics[width=3.3in,bb=100 270 480 570]{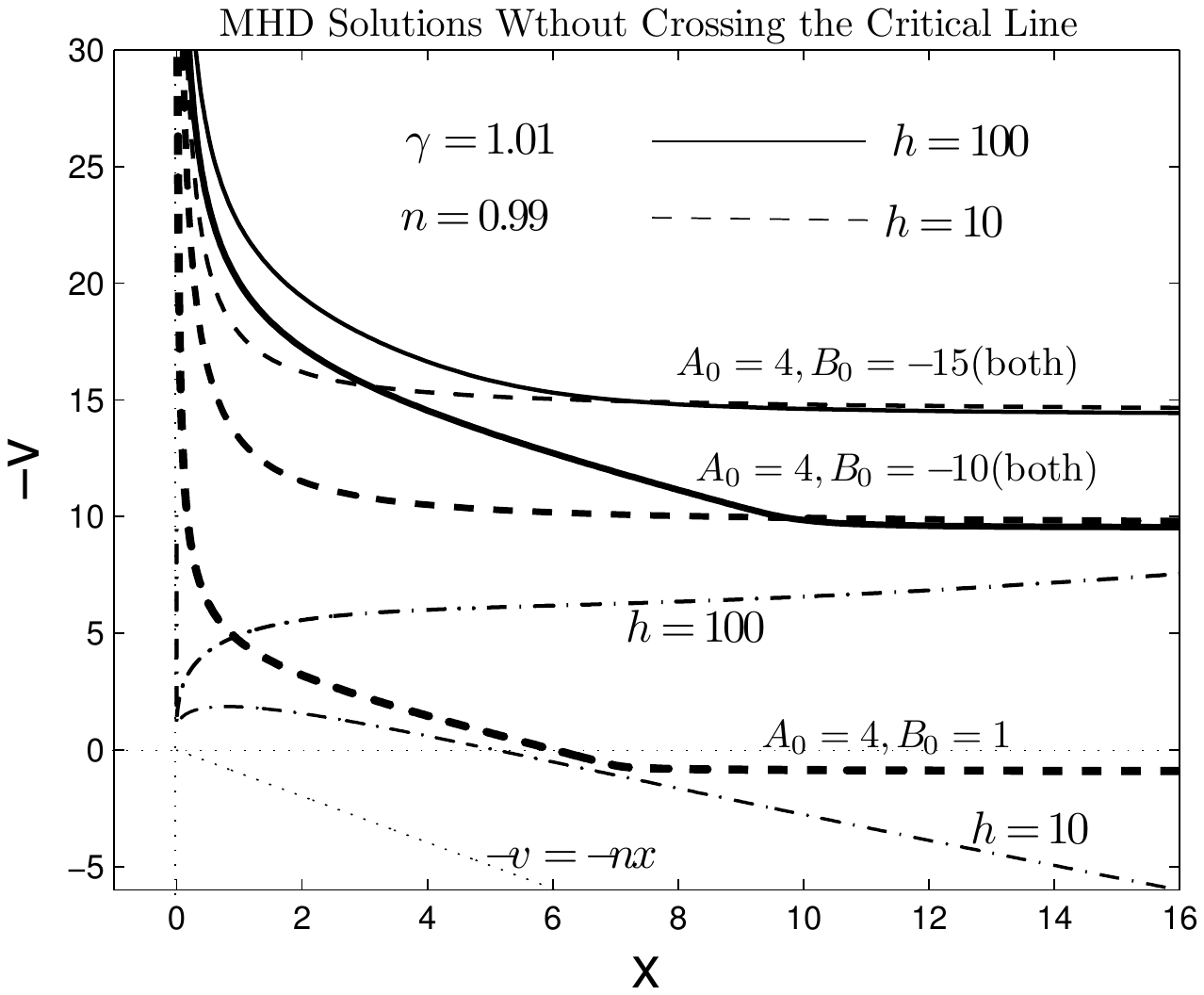}
\caption{The numerical similarity MHD flow solutions without
crossing the magnetosonic critical curve can be readily
constructed using asymptotic solutions (\ref{wang22}) and
(\ref{wang23}) to start at a sufficiently large $x$ (e.g., $x=100$
here), with $\gamma=1.01$ and $n=0.99$ being specified for a
conventional polytropic gas. The two perpendicular dotted lines
denote the abscissa and ordinate axes, respectively, and the
straight dotted line $-v=-nx$ is the demarcation to the lower left
of which solutions become unphysical for a negative enclosed gas
mass. The two dash-dotted lines are the magnetosonic critical
curves for $h=10$ and $h=100$, respectively. The solid and dashed
curves are MHD solutions with $h=100$ and $h=10$, respectively;
for the pair solutions with light linetypes, we have both $A_0=4$
and $B_0=-15$, while for the other pair solutions with heavy
linetypes, we have both $A_0=4$ and $B_0=-10$. The dashed curve
with $A_0=4$ and $B_0=1$ for $h=10$ (with an extremely heavy
linetype) is an example of MHD similarity solutions without
crossing the magnetosonic critical curve and with an outward flow
speed in the far-away region. }\label{numer101_2}
\end{figure}

\subsubsection{A Global MHD Expansion Solution}

For a constant $\alpha$ in equation (\ref{wang9}), we obtain
\begin{equation}\label{wang33} v'={2(x-v)}/{x}\ ,
\end{equation}
which can be readily integrated to attain
\begin{equation}\label{wang34}
v= {2}x/{3}+cx^{2}\ ,
\end{equation}
where $c$ is an integration constant. By equation (\ref{wang16}),
we should set $c=0$ and thus obtain a global MHD solution
\begin{equation}\label{wang35}
v=\frac{2x}{3}\ ,\ \ \ \alpha=\frac{2}{3(6h+1)}\ , \ \ \
m=\frac{2(n-2/3)}{3(6h+1)}x^{3}
\end{equation}
accordingly; the reduced magnetic energy density $w$ is
\begin{equation}\label{wang60}
w(x)=\frac{4hx^2}{9(6h+1)^2}\ ,
\end{equation}
which increases with $x$ quadratically along with the increase of
the radial flow speed $v$ linearly in $x$. For a polytropic gas
with $\gamma=2-n$, the ratio of the Alfv\'en wave speed to the
sound speed becomes
\begin{equation}\label{wang123}
\frac{v_A}{s}=\bigg(\frac{h}{\gamma}\bigg)^{1/2}
\left[\frac{2}{3(6h+1)}\right]^{(2-\gamma)/2}x\ .
\end{equation}
This solution reaches the singular surface at
\[
x=\bigg[\frac{\gamma(4/3)^\gamma
/(18h+3)^\gamma}{2(n-2/3)^2/(18h+3) -4h/(18h+3)^2}\bigg]^{1/2}
\]
on the magnetosonic critical curve.

\subsubsection{Asymptotic MHD Expansion\\
\qquad\quad Solutions in the Limit of Large $x$}

Solutions (\ref{wang35}) and (\ref{wang60}) can also be regarded
as an asymptotic solution as $x$ approaches infinity. To leading
orders, the asymptotic MHD solution can be written as
\[
v(x)=\frac{2}{3}x+v_0\ ,
\]
\begin{eqnarray}\label{wang133}
&&\alpha=\frac{2}{3(6h+1)}\qquad\nonumber\\
&&-\frac{9(3n+1)(3n-2)(6h+1)^2-6(6h+1)}
{6(n-3/2)^2(3n-2)(6h+1)-2h(3n-2)}\frac{v_0}{x}\ ,\qquad
\end{eqnarray}
where $v_0$ is an integration constant. In this case, the
denominators of both equations (\ref{wang17}) and (\ref{wang18})
approach
\[
\frac{2}{3(6h+1)}\bigg[\bigg(n-\frac 2 3\bigg)^2
-\frac{2h}{3(6h+1)}\bigg]x^2
\]
which does not encounter the magnetosonic critical curve, unless
under extremely rare situations.

\subsubsection{Regular Similarity MHD Solutions for Small $x$}

We may assume an asymptotic MHD series solution as
\[
\alpha(x)=\alpha_{\ast}+\alpha_{1}x+\alpha_{2}x^{2}+\cdots,
\]
\begin{equation}\label{wang36}
v(x)=v_{0}+v_{1}x+v_{2}x^{2}+\cdots\ ,
\end{equation}
as $x$ approaches zero. Substitution of solution (\ref{wang36})
into equations (\ref{wang9}) and (\ref{wang16}) leads to
\begin{equation}\label{wang37}
v(x)=\frac{2}{3}x-\frac{\alpha_{\ast}^{(1-\gamma)}}
{15\gamma}\bigg[(6h+1)\alpha_{\ast}-\frac{2}{3}\bigg]
\bigg(n-\frac{2}{3}\bigg)x^{3}+\cdots\ ,
\end{equation}
\begin{equation}\label{wang38}
\alpha(x)=\alpha_{\ast}-\frac{\alpha_{\ast}^{(2-\gamma)}}{6\gamma}
\bigg[(6h+1)\alpha_{\ast}-\frac{2}{3}\bigg]x^{2}+\cdots\ ,
\end{equation}
\begin{equation}\label{wang61}
w(x)=h\alpha_{\ast}^2x^2-2h
\frac{\alpha_{\ast}^{(2-\gamma)}}{6\gamma}
\bigg[(6h+1)\alpha_{\ast}-\frac{2}{3}\bigg]x^{4} +\cdots\ .
\end{equation}
For a usual polytropic gas with $\gamma=2-n$ and for very small
$x$, the corresponding ratio of the Alfv\'en wave speed $v_A$ to
the sound speed $s$ becomes
\begin{equation}\label{wang124}
{v_A}/{s}=(h\alpha_\ast^{2-\gamma}/\gamma)^{1/2}x\ ,
\end{equation}
which vanishes as $x\rightarrow 0^{+}$. As $x$ approaches zero,
the denominators of both equations (\ref{wang17}) and
(\ref{wang18}) for this series expansion approach
$-\gamma\alpha^\gamma$ and these solutions do not encounter the
magnetosonic singular surface. For this asymptotic solution, the
four forces including the thermal pressure, magnetic pressure,
magnetic tension and gravity forces are in the same order, and the
magnetic pressure and tension forces tend to be the same in
magnitude. The magnetic pressure gradient force points radially
outward.

\subsection{Asymptotic Behaviours of Critical Curves}\label{crasy}

Asymptotic behaviours of magnetosonic critical curves are important
from a global perspective. We summarize below the major asymptotic
behaviours of critical curves with or without magnetic field,
respectively.

Case I for $h\neq 0$ in the presence of magnetic field.

(i) The limiting regime of $\alpha\rightarrow +\infty$. According
to quadratic equation (\ref{wang41}), the asymptotic quadratic
equation of $x^2$ for the magnetosonic critical curve is
\begin{equation}\label{wang43}
\frac{h\alpha^{3}}{(3n-2)^{2}}x^4
+\frac{\gamma\alpha^{\gamma+1}}{(3n-2)^{2}}x^{2}
-4\gamma^{2}\alpha^{2\gamma-2}=0\ ,
\end{equation}
which has only one positive root of small $x$
\begin{equation}\label{wang44}
x^{2}\cong 4(3n-2)^{2}\alpha^{\gamma-3}\ ,
\end{equation}
and along with equation (\ref{wang40}), one obtains
correspondingly a diverging radial inflow speed
\begin{equation}\label{wang45}
v\cong -\sqrt{\gamma}\alpha^{(\gamma-1)/2}\ .
\end{equation}
As $\alpha$ approaches positive infinity, $x$ approaches zero and
$v$ approaches $-\infty$. This appears to be the case also for a
purely hydrodynamic case with $h=0$ (Lou and Wang, 2006) and was not
discussed in Suto and Silk (1988). Thus, the magnetosonic critical
curve does not intersect the $v-$axis as compared to the
isothermal unmagnetized cases (Shu, 1977; Lou and Shen, 2004) where
the sonic critical line is the straight line $-v=1-x$. For a
magnetized isothermal gas (Yu and Lou, 2005; Yu et al., 2006), the
magnetosonic critical lines are curves intersecting the $v-$axis.

\begin{figure}
\includegraphics[width=3.3in,bb=100 270 480 570]{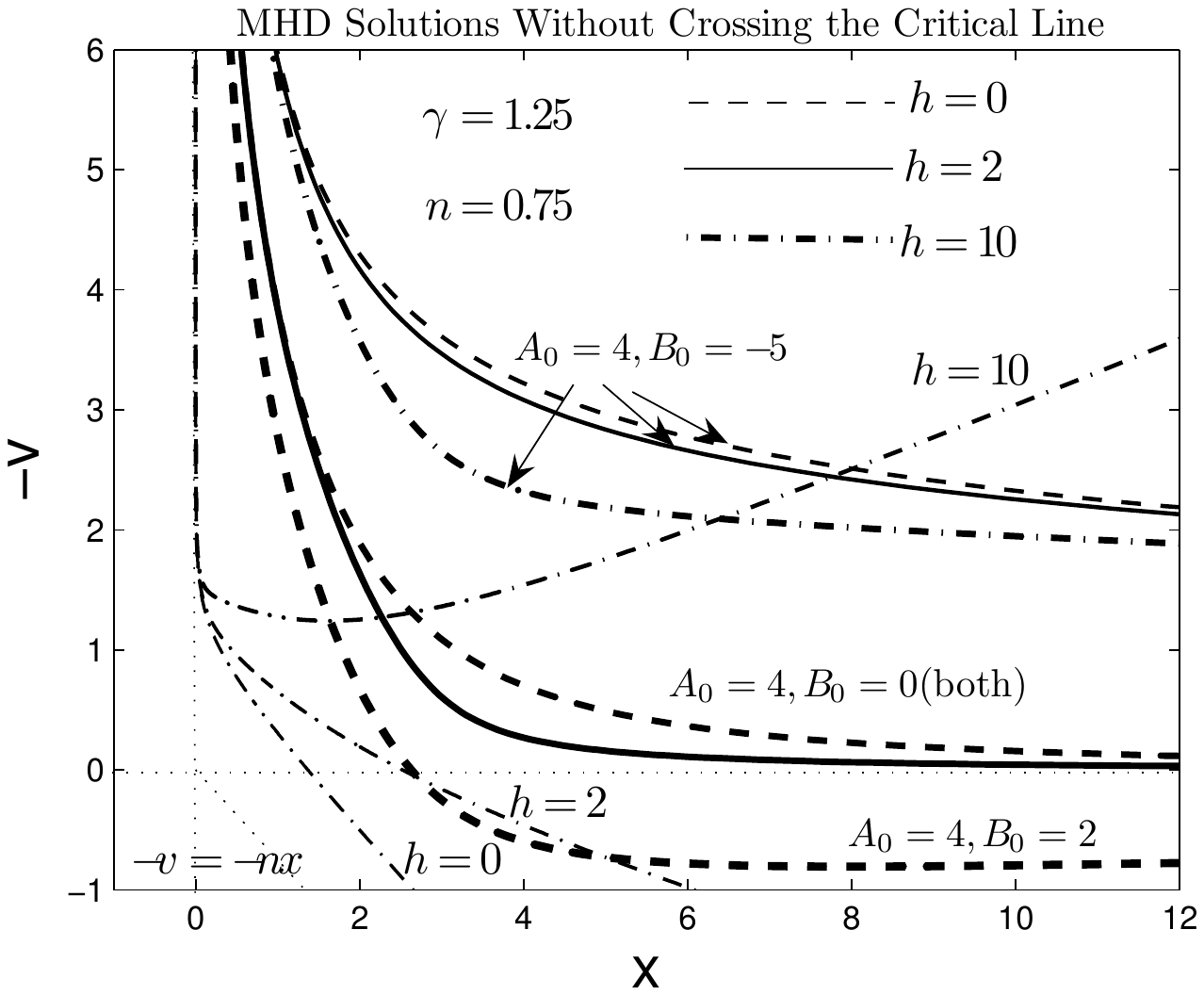}
\caption{Similarity MHD flow solutions without crossing the
magnetosonic critical curve with the values of $\gamma=1.25$ and
$n=0.75$ being specified for a usual polytropic gas. The
perpendicular dotted lines denote the abscissa and ordinate axes,
respectively, and the straight dotted line $-v=-nx$ is a
demarcation to the lower left of which the solutions become
unphysical for a negative enclosed mass. The light dash-dotted
curves are the magnetosonic critical curves for $h=0,\ 2,\ 10$.
The dashed and solid curves are global solutions with $h=0$ and
$h=2$, respectively (heavy linetypes: $A_0=4$ and $B_0=0$; light
linetypes: $A_0=4$ and $B_0=-5$). The heavy dash-dotted curve is
the solution with $h=10$, $A_0=4$ and $B_0=-5$. The dashed curve
with $A_0=4$ and $B_0=2$ for $h=0$ (the extremely heavy linetype)
is an example of hydrodynamic similarity solutions without
crossing the sonic critical curve and with an outward flow
speed in the far-away region. }\label{numer125_2}
\end{figure}

(ii) The limiting case of $x\rightarrow +\infty$. According to
equation (\ref{wang40}), we obtain the following condition for
$\alpha$
\begin{equation}\label{wang46}
\pm\bigg[n-1+\frac{\alpha}{(3n-2)}\bigg](h\alpha)^{1/2}=-n(n-1)\ ,
\end{equation}
and correspondingly, the behaviour of $v$ versus $x$ as
\begin{equation}\label{wang47}
v\cong [n\mp (h\alpha)^{1/2}]x\ .
\end{equation}
This means that as $x$ approaches infinity, $v$ varies linearly
with $x$ and $\alpha$ approaches a certain constant value
determined by equation (\ref{wang46}). This linear behaviour of $v$
in $x$ is qualitatively similar to the isothermal case (Shu, 1977;
Lou and Shen, 2004), where the sonic critical line is a straight
line throughout the entire semi-complete space; we note that for
the isothermal and non-magnetized case, $\alpha$ approaches zero
as the sonic critical line approaches infinity. For a magnetized
gas cloud, the slope of this linear relation now depends on $h$
parameter. By the constraint of $m(x)>0$, the lower signs in
solutions (\ref{wang46}) and (\ref{wang47}) are unphysical, yet
mathematically these equations may describe asymptotic behaviours
of a new branch of critical line.

The special case of $h=h_c$ with $h_c$ defined by equation
(\ref{wang111}) should be noted with interest. In this case, the
corresponding value of $\alpha$ as $x\rightarrow +\infty$ becomes
\begin{equation}\label{wang112}
\alpha=2(1-n)(3n-2)\ ,
\end{equation}
giving rise to $(h\alpha)^{1/2}=n$ and thus the vanishing of the
leading order term of $v$ by the upper minus sign in equation
(\ref{wang47}). This hints that the magnetosonic critical curve
tends to be parallel to the $x-$axis as $x$ approaches infinity.
Naturally for $(h\alpha)^{1/2}>n$ as $x\rightarrow +\infty$, one
infers that $h>h_c$ and vice versa. This reveals an interesting
trend of variation that as $h$ exceeds the critical value $h_c$,
the magnetosonic critical curve will head up (i.e., increase
eventually) towards the first quadrant (see subsection \ref{numcri})
and $h=h_c$ marks the marginal case for the asymptotic behaviour of
the magnetosonic critical curve at large $x$.

Case II for $h=0$ in the absence of magnetic field.

As $\alpha\rightarrow +\infty$, the asymptotic
quadratic equation for $x^2$ now becomes
\begin{equation}\label{wang48}
n^2(n-1)^2x^4-\frac{\gamma\alpha^{\gamma+1}x^2}
{(3n-2)^2}+4\gamma^{2}\alpha^{2\gamma-2}=0\ ,
\end{equation}
[see equation (\ref{wang43}) in parallel for the
different coefficient of the first term $x^4$]
which has two positive roots for $x^2$.

(i) For the small $x$ regime of
\begin{equation}\label{wang49}
x^2\cong 4\gamma(3n-2)^{2}\alpha^{\gamma-3}\ ,
\end{equation}
one obtains a diverging radial inflow speed of
\begin{equation}\label{wang50}
v\cong -\sqrt{\gamma}\alpha^{(\gamma-1)/2}\ .
\end{equation}
As $x$ approaches zero, $\alpha$ approaches positive infinity
while $v$ approaches negative infinity. Again, the sonic critical
curve does not intersect the $v-$axis in contrast to the
isothermal case (Shu, 1977; Lou and Shen, 2004). Compared with the
isothermal case, this sonic critical curve behaviour appears to be
a unique feature for a polytropic gas (Lou and Wang, 2006, 2007),
not realized before (Suto and Silk, 1988). For an isothermal
magnetized cloud (Yu and Lou, 2005; Yu et al., 2006), the
magnetosonic critical curve involves a cubic equation in terms
of $v$; the physical portion of the curve still intersects the
$v-$axis (see figures 1 and 2 of Yu and Lou, 2005).

(ii) For the large $x$ regime of
\begin{equation}\label{wang51}
x^2\cong\frac{\gamma \alpha^{\gamma+1}}
{n^{2}(n-1)^{2}(3n-2)^{2}}\
\end{equation}
and accordingly
\begin{equation}\label{wang52}
v\cong nx\ ,
\end{equation}
we have the variation trend for the magnetosonic critical curve
such that as $x$ approaches infinity, $\alpha$ approaches infinity
and $v$ increases with $x$ linearly. This feature differs from the
isothermal case (Shu, 1977; Lou and Shen, 2004); in the isothermal
case with $n=1$, the above asymptotic behaviour is invalid although
the asymptotic behaviour (\ref{wang52}) seems valid.

\subsection{Series Expansions near the\\
$\qquad$Magnetosonic Critical Curve}\label{behav}

\begin{figure}
\includegraphics[width=3.3in,bb=100 270 480 570]{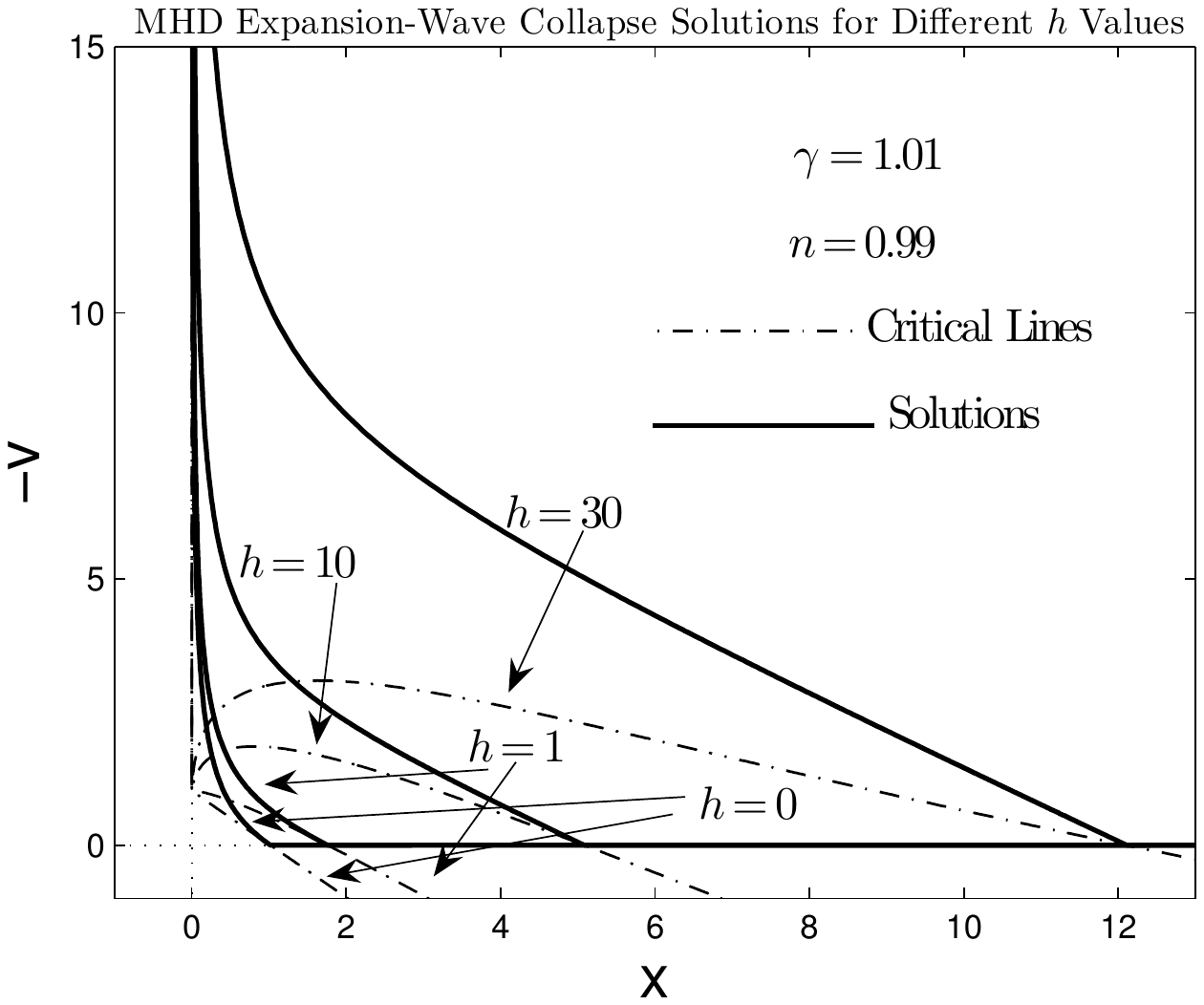}
\caption{Examples of semi-complete MHD expansion wave collapse
solutions (mEWCSs) and the corresponding magnetosonic critical
curves for different $h$ values with $B_0=0$, $\gamma=1.01$ and
$n=0.99$ being specified for a usual polytropic gas. The two
perpendicular dotted straight lines are the abscissa and ordinate
axes, respectively, and the dash-dotted curves are the
magnetosonic critical curves. The heavy solid curves are the
mEWCSs. The corresponding values of $A_0$ for the cases of $h=0$,
$1$, $10$ and $30$ are $2.0133$, $2.0544$, $2.5163$ and $5.004$,
respectively. } \label{numer101_3}
\end{figure}

Generally, smooth analytical solutions cannot go across the
singular surface as noted in Section \ref{crdet} unless they cross
the magnetosonic critical curve. Along the magnetosonic critical
curve, one cannot calculate the derivatives of the variables
directly from nonlinear MHD ODEs (\ref{wang17}) and
(\ref{wang18}). From equation (\ref{wang9}), we use
\begin{equation}\label{wang53}
\alpha'=\frac{\alpha v'-2\alpha(x-v)/x}{(nx-v)}\
\end{equation}
to compute $\alpha'$ once $v'$ is known. Applying the
L'H$\hat{\mbox{o}}$pital rule to equation (\ref{wang17}), we
immediately obtain the following quadratic equation in terms of
$v'$
\begin{equation}\label{wang54}
A_{2}(v')^{2}+B_{2}v'+C_{2}=0\
\end{equation}
along the magnetosonic critical curve, where the three
coefficients $A_2$, $B_2$ and $C_2$ are defined explicitly by
\[
A_{2}\equiv 1+\frac{(\gamma^2\alpha^{\gamma-1} +2h\alpha
x^2)}{(\gamma\alpha^{\gamma-1}+h\alpha x^2)}\ ,
\]
\begin{eqnarray*}
&&B_{2}\equiv n+1-4\frac{(\gamma^{2}\alpha^{\gamma-1}+2h\alpha
x^{2})}{(\gamma\alpha^{\gamma-1}+h\alpha x^{2})}\nonumber\\
&&+4\bigg[\frac{(\gamma^2\alpha^{\gamma-1}+2h\alpha x^2)}
{(\gamma\alpha^{\gamma-1}+h\alpha x^2)}-1\bigg]\frac{v}{x}
+\frac{4h\alpha x}{(nx-v)}\ ,
\end{eqnarray*}
\begin{eqnarray}\label{wang55}
&&C_{2}\equiv 2\bigg[\frac{2(\gamma^2\alpha^{\gamma-1}+2h\alpha
x^2)}{(\gamma\alpha^{\gamma-1}+h\alpha
x^2)}-1\bigg]\frac{v^{2}}{x^{2}}\nonumber\\
&&+2\bigg[\frac{\alpha}{(3n-2)}-2n
-4\frac{(\gamma^2\alpha^{\gamma-1}+2h\alpha x^2)}
{(\gamma\alpha^{\gamma-1}+h\alpha
x^2)}+4\bigg]\frac{v}{x}\nonumber\\
&&+\frac{(n-2)\alpha}{(3n-2)}
+2\bigg[n+2\frac{(\gamma^2\alpha^{\gamma-1}+2h\alpha
x^2)}{(\gamma\alpha^{\gamma-1}+h\alpha x^2)}-2\bigg]\nonumber\\
&&+2h\alpha\bigg[1-\frac{4(x-v)}{(nx-v)}\bigg]\ .
\end{eqnarray}
By setting $h=0$, equations (\ref{wang54}) and (\ref{wang55})
reduce to the hydrodynamic results (Suto and Silk, 1988) as a
necessary requirement; by further letting $\gamma\rightarrow 1$,
these two equations reduce to those of the isothermal case (Shu,
1977). For $h\neq 0$ and letting $\gamma\rightarrow 1$, these two
equations are equivalent to those of the isothermal MHD case (Yu
and Lou, 2005; Yu et al., 2006). One can determine eigensolution
behaviours by Taylor series expansions near the magnetosonic
critical curve using equations (\ref{wang54}) and (\ref{wang55}).
By quadratic equation (\ref{wang54}), one obtains two types of
eigensolutions across the magnetosonic critical curve.

\begin{figure}
\includegraphics[width=3.3in,bb=100 270 480 570]{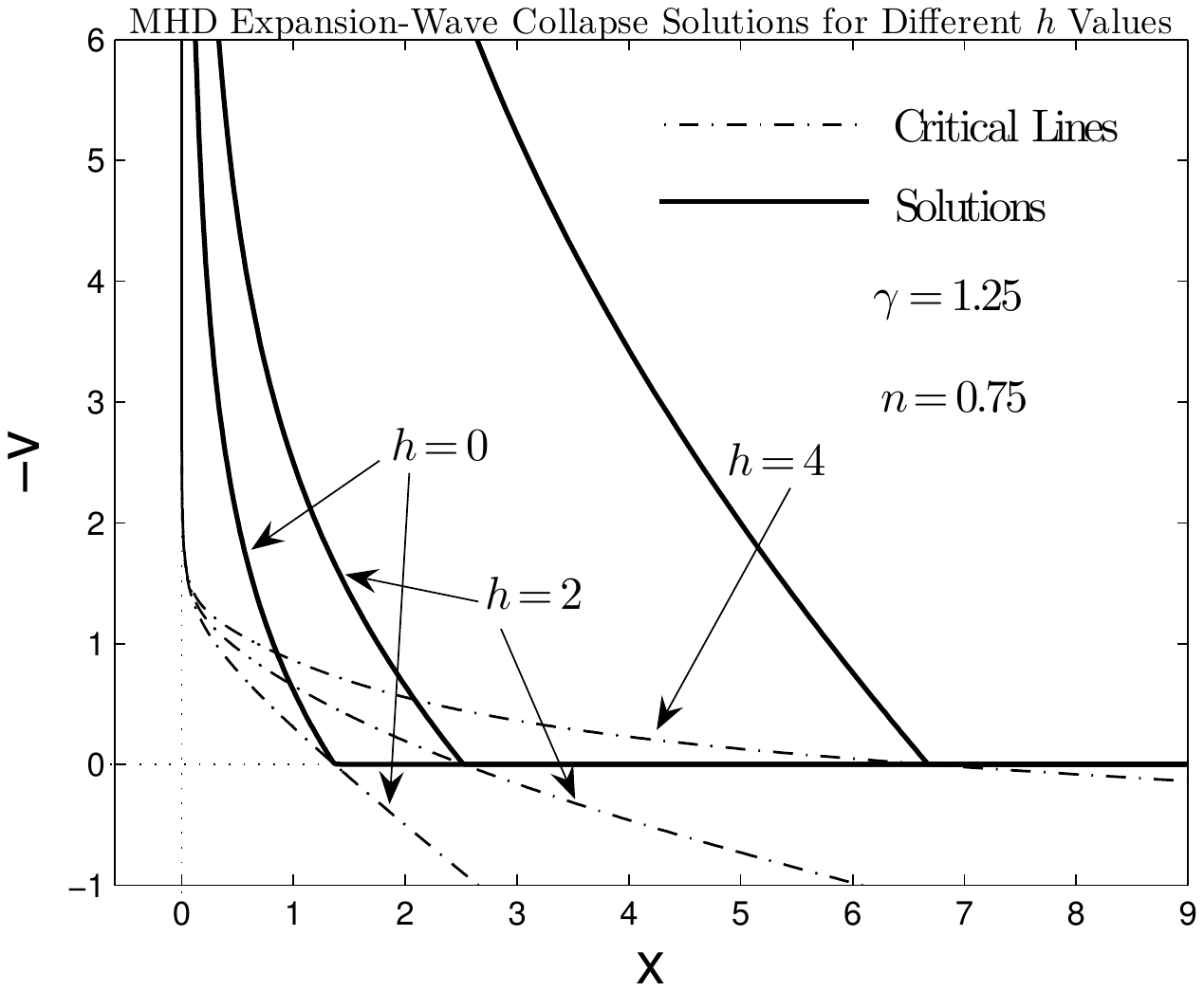}
\caption{Examples of semi-complete mEWCSs and the corresponding
magnetosonic critical curves for different $h$ values with
$B_0=0$, $\gamma=1.25$ and $n=0.75$ being specified for a usual
polytropic gas. The two perpendicular dotted straight lines are
the abscissa and ordinate axes, respectively, and the dash-dotted
curves are the magnetosonic critical curves. The heavy solid
curves represent mEWCSs. The corresponding values of $A_0$
parameter for the cases of $h=0$, $2$ and $4$ are $1.151$,
$2.520$ and $21.55$, respectively. }\label{numer125_3}
\end{figure}

\begin{figure}
\includegraphics[width=3.3in,bb=100 270 480 570]{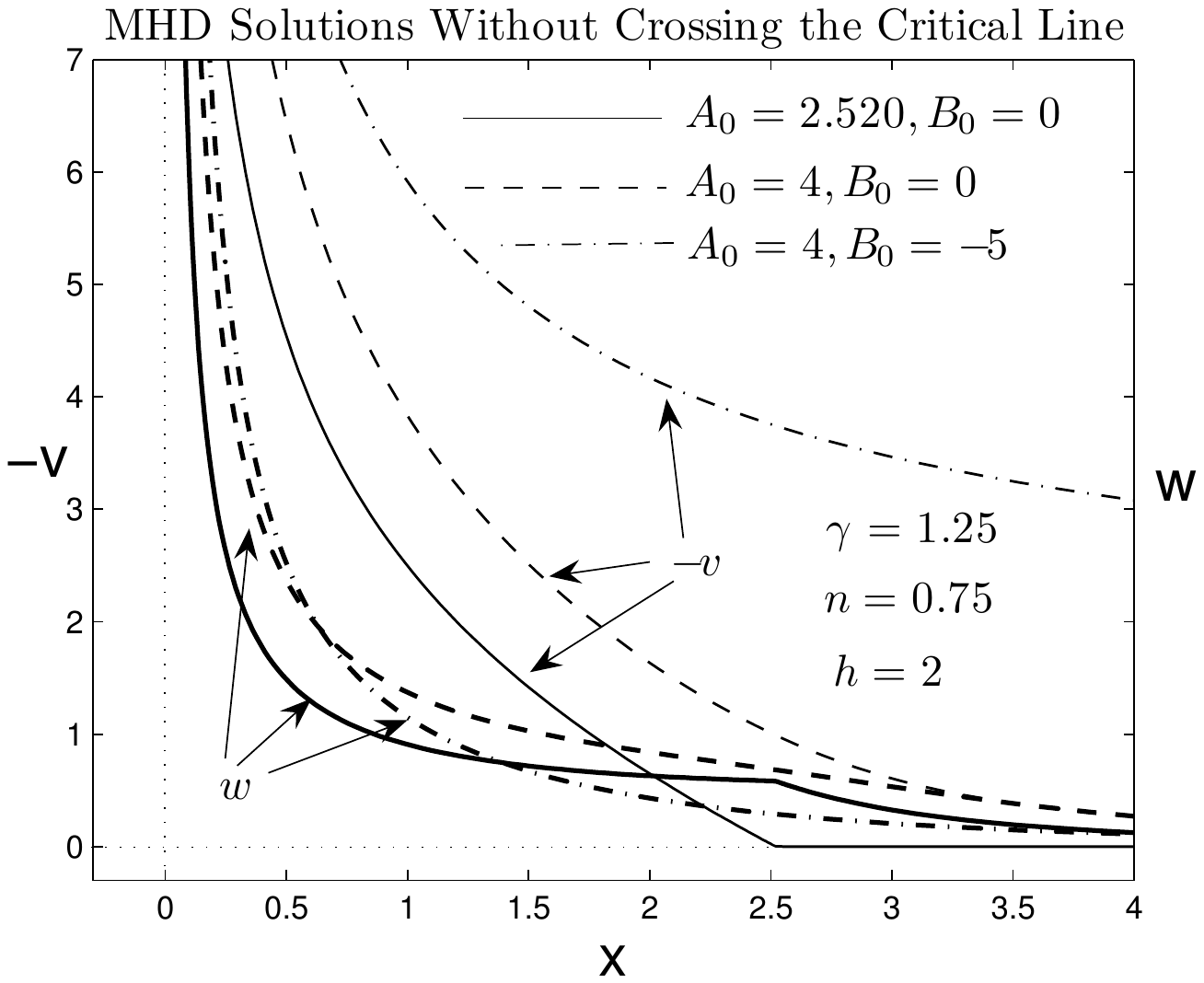}
\caption{Examples of semi-complete similarity MHD flow solutions
without crossing the magnetosonic critical curve. The reduced
magnetic energy density $w$ associated with a random transverse
magnetic field is displayed in curves (right ordinate) using heavy
linetypes and the corresponding $-v$ is in curves (left ordinate)
using light linetypes. The $-v$ curves are the same as those in
Figure \ref{numer125_2}. The case of $A_0=2.520$ and $B_0=0$
(solid curve) is an mEWCS, while the case of $A_0=4$ and $B_0=0$
(dashed curve; not an mEWCS) is a global similarity MHD flow.
}\label{numer125m}
\end{figure}

For the unmagnetized isothermal case, Whitworth and Summers (1985)
noticed that there are different behaviours near the vicinity of
the sonic critical line, depending on the topological structure
of paths through a sonic point. This topological structure is
investigated by the eigenvalues of the following matrix
\begin{equation}\label{wang81}
\left(\begin{matrix} A_\alpha & A_v & A_x \cr V_\alpha & V_v & V_x
\cr X_\alpha & X_v & X_x \cr \end{matrix}\right)
\end{equation}
where $A$, $V$ and $X$ are the three functions defined by
equations (\ref{wang77}), (\ref{wang78}) and (\ref{wang79}),
respectively, and $A_\alpha$ denotes a partial differentiation of
$A$ with respect to $\alpha$, taking $\alpha$, $v$ and $x$ as
three independent variables; other symbols follow the same
notational convention by inference. The explicit expressions of
the partial differentiations contained in the above matrix are
summarized in Appendix \ref{partial}. The characteristic equation
for the matrix (\ref{wang81}) is simply
\begin{equation}\label{wang91}
\left|\begin{matrix}
 \lambda-A_\alpha & -A_v & -A_x \cr
 -V_\alpha & \lambda-V_v & -V_x \cr
 -X_\alpha & -X_v & \lambda-X_x \cr\end{matrix}\right|=0\ ,
\end{equation}
and the signs of the $\lambda$ roots determine the behaviours in
the vicinity of the magnetosonic critical curve. Equation
(\ref{wang91}) is equivalent to
\begin{eqnarray}\label{wang92}
&&\lambda^3-\big(A_\alpha+V_v+X_x\big)\lambda^2 +\big(A_\alpha
V_v+A_\alpha X_x+V_v X_x\nonumber\\
&&-A_x X_\alpha-A_v V_\alpha-X_vV_x\big)\lambda -X_{\alpha}A_v
V_x-A_\alpha V_v X_x\nonumber\\
&&+X_\alpha V_v A_x+X_x V_\alpha A_v+A_\alpha V_x X_v-V_{\alpha}
A_x X_v=0\ ,\nonumber\\
\end{eqnarray}
which can be cast into a succinct form of
\begin{equation}\label{wang93}
\lambda^3+B_3\lambda^2+C_3\lambda+D_3=0\ ,
\end{equation}
with apparent definitions for the three coefficients $B_3$, $C_3$
and $D_3$ by referring to equation (\ref{wang92}). Because the
magnetosonic critical curve is continuous, this characteristic
equation must have one zero root, i.e., $D_3=0$, corresponding to
a path which stays on the magnetosonic curve (see Whitworth and
Summers 1985 for a reference and Appendix \ref{proofd3} for a
proof). The other two eigenvalues for $\lambda$ of matrix
(\ref{wang81}) are then given by quadratic equation
\begin{equation}\label{wang94}
\lambda^2+B_3\lambda+C_3=0\ .
\end{equation}
The behaviours in the vicinity of the magnetosonic critical curve
depend upon the signs of the two $\lambda$ roots of quadratic
equation (\ref{wang94}).

Case I. For a negative determinant $B_3^2-4C_3<0$, we have a
spiral or a centre case (e.g., Jordan and Smith, 1977). In this
case, the solutions do not have a one-to-one correspondence to $x$
and are thus regarded as unphysical. For a polytropic magnetized
gas, such points may exist (see Appendix \ref{vicinity}).

Case II. For a positive determinant $B_3^2-4C_3>0$ with $C_3<0$,
we have a saddle point along the magnetosonic critical curve with
two eigensolutions determined by equations (\ref{wang53}),
(\ref{wang54}) and (\ref{wang55}). Numerical integrations away
from these saddle points tend to be stable.

Case III. For a positive determinant $B_3^2-4C_3>0$ with $C_3>0$,
we have a nodal point along the magnetosonic critical curve with
infinitely many solutions crossing the magnetosonic critical
curve. As noted by Hunter (1986), among these solutions only the
two eigensolutions are analytical, while others involve weak
discontinuities or weak shocks (e.g., Boily and Lynden-Bell, 1995)
and might be unstable (e.g., Lazarus, 1981). Although only
integrations towards nodal points are stable, to pick out the
analytic eigensolutions among the solutions having weak
discontinuities, we have integrated outward from these points
using second-order derivatives. Explicit expressions of the
relevant second-order derivatives are summarized in Appendix
\ref{second}.

Case IV. For a vanishing determinant $B_3^2-4C_3=0$, we have
inflection nodal points. If $C_3=0$, we have degenerate points
along the magnetosonic critical curve.

\section{Global Similarity Solutions}\label{numerical}

\begin{table}
\center \caption{The corresponding $m(0)$ values for similarity
MHD flow solutions without crossing the magnetosonic critical
curve for a polytropic gas with $\gamma=1.01$ and $n=0.99$ and
relatively small values of $h$ parameter. The two coefficients
$A_0$ and $B_0$ are specified in asymptotic MHD solutions
(\ref{wang22}) and (\ref{wang23}). Parameter $m(0)$ is directly
computed from equation (\ref{wang8}). Here, the $m(0)$ value is
computed after integrating from large $x$ values such as $x=100$
to small $x$ values. }\label{m0_101_1} \vskip 0.2cm
\begin{tabular}{ccccc}\hline
$h$ & $0$ & $0$ & $1$ & $1$ \\
$A_0$, $B_0$ &  $4,\ 0$ & $4,\ -2$ & $4,\ 0$ & $4,\ -2$ \\ \hline
$m(0)$ & $5.513$ & $12.34$ & $5.912$ & $12.48$   \\
\hline
\end{tabular}
\end{table}

\begin{table}
\center\caption{The corresponding $m(0)$ values for similarity MHD
solutions without crossing the magnetosonic critical curve for a
polytropic gas with $\gamma=1.01$ and $n=0.99$ and relatively
large values of $h$ parameter. The two coefficients $A_0$ and
$B_0$ are specified in asymptotic MHD solutions (\ref{wang22}) and
(\ref{wang23}). Parameter $m(0)$ is directly computed from
equation (\ref{wang8}). Here, the $m(0)$ value is computed after
numerically integrating from large $x$ values such as $x=100$ to
small $x$ values. }\label{m0_101_2} \vskip 0.2cm
\begin{tabular}{ccccc}\hline
$h$         & $10$     & $10$     & $100$    & $100$\\
$A_0$, $B_0$& $4,\ -10$& $4,\ -15$& $4,\ -10$& $4,\ -15$\\ \hline
$m(0)$      & $40.28$  & $57.59$  & $44.58$  & $59.09$  \\
\hline
\end{tabular}
\end{table}

With compatible initial and boundary conditions together with a
proper treatment of the magnetosonic critical curve, the two
coupled nonlinear MHD ODEs for self-similar collapses and flows
can be integrated numerically. We have explored MHD solutions
numerically, including the properties of the magnetosonic critical
curve and $\alpha-v-x$ solutions of the MHD ODEs. In Suto and Silk
(1988), both cases of $n=1$ and $n=2-\gamma$ were considered;
here, we focus on the case of a usual polytropic gas with
$n=2-\gamma$. In contrast to the case of $\gamma<1$ considered by
Fatuzzo et al. (2004), we are mainly concerned with the range of
$1<\gamma<4/3$. We intend to find semi-complete solutions valid in
the range of $0<x< +\infty$, and discuss how such MHD solutions
can be constructed through numerical integrations.

\subsection{Magnetosonic Critical Curves}\label{numcri}

The magnetosonic critical curves for different parameters can be
systematically searched by numerical means, and for the completion
of a magnetosonic critical curve, one needs the relevant
analytical results summarized in subsections \ref{crdet} and
\ref{crasy}. We have extensively explored the behaviours of
magnetosonic critical curves for specified values of $n$, $\gamma$
and $h$ parameters, and present the main results below. More
details can be found in Appendix \ref{infocritical}.

\begin{figure}
\includegraphics[width=3.3in,bb=100 270 480 570]{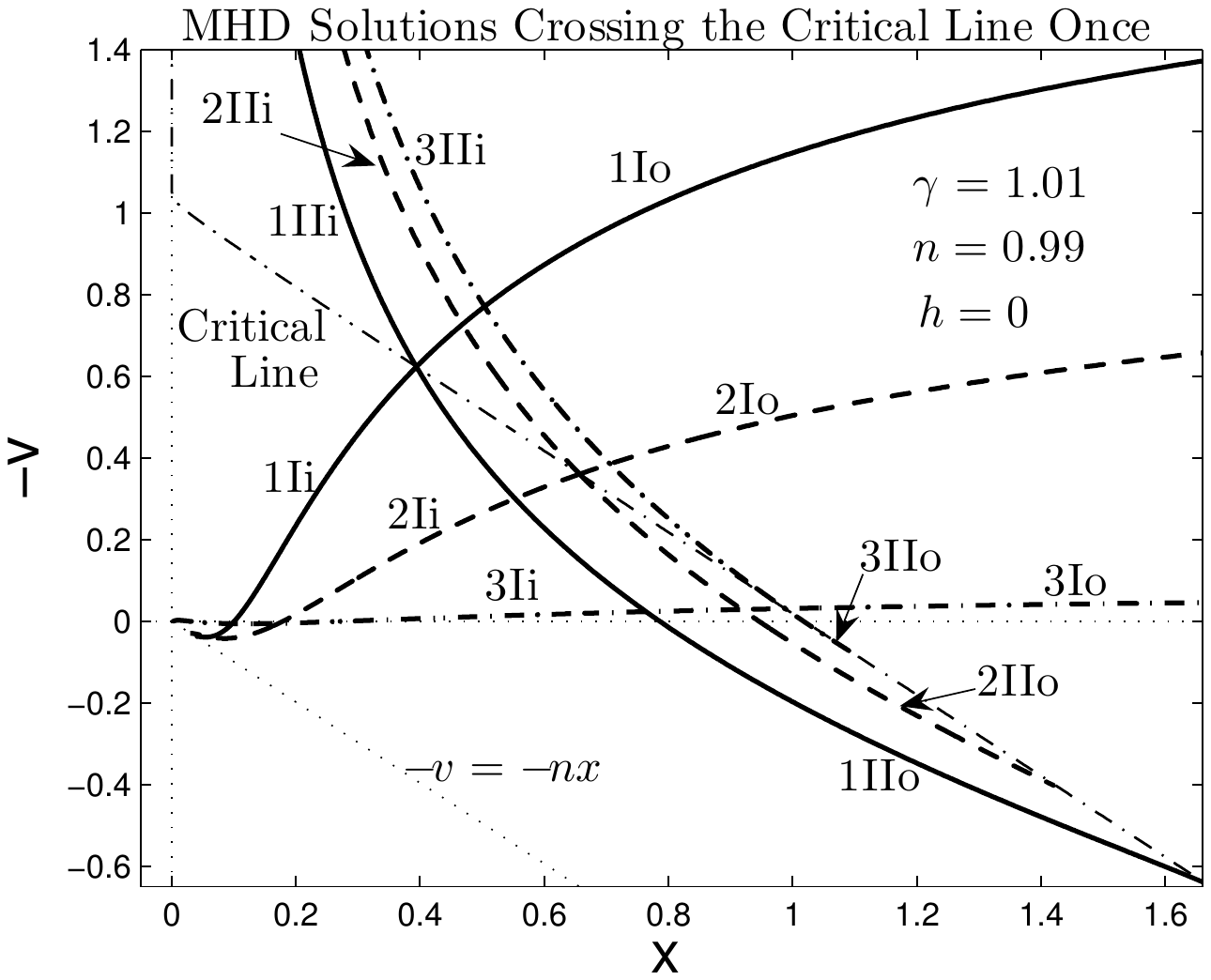}
\caption{Hydrodynamic polytropic solutions crossing the sonic
critical curve once, with $\gamma=1.01$, $n=0.99$ and $h=0$. Point
1 corresponds to $\alpha=5$ and $x=0.3941$, point 2 corresponds to
$\alpha=3$ and $x=0.6570$, and point 3 corresponds to $\alpha=2$
and $x=0.9877$, respectively. The two perpendicular light dotted
straight lines are abscissa and ordinate axes, respectively, while
the light dotted straight line passing through the origin is
$-v=-nx$ to the lower left of which solutions become unphysical
for a negative enclosed mass. The light dash-dotted line is the
sonic critical curve, and the heavy lines are the solutions
crossing the sonic critical curve once.}\label{Once101_0_1}
\end{figure}

First, we show magnetosonic critical curves with different
values of $h$ for given $n=0.99$ and $\gamma=1.01$ (Figs.
\ref{critical101} and \ref{critical101s}). A magnetosonic critical
curve may be divided into two parts as one picks up different
roots of quadratic equation (\ref{wang41}) for $x^2$.

When $h$ increases for stronger magnetic field strengths, the
average slope $\text{d}(-v)/\text{d}x$ of an individual
magnetosonic critical line increases from negative to positive in
our figure displays of $-v$ versus $x$. Meanwhile as the magnetic
field becomes strong enough and as $x$ approaches infinity, $v$
may approach $-\infty$. The critical value $h_c$ in this specific
case is $50.521$ (see subsection \ref{crasy}). This feature is
important in the numerical analysis of similarity MHD flow
solutions not crossing the magnetosonic critical curve. In
contrast to the straight sonic critical line for the isothermal
unmagnetized case (Shu, 1977; Lou and Shen, 2004), the magnetosonic
critical curves here diverge as $x$ approaches zero (see Fig.
\ref{critical101s}).

\begin{figure}
\includegraphics[width=3.3in,bb=100 270 480 570]{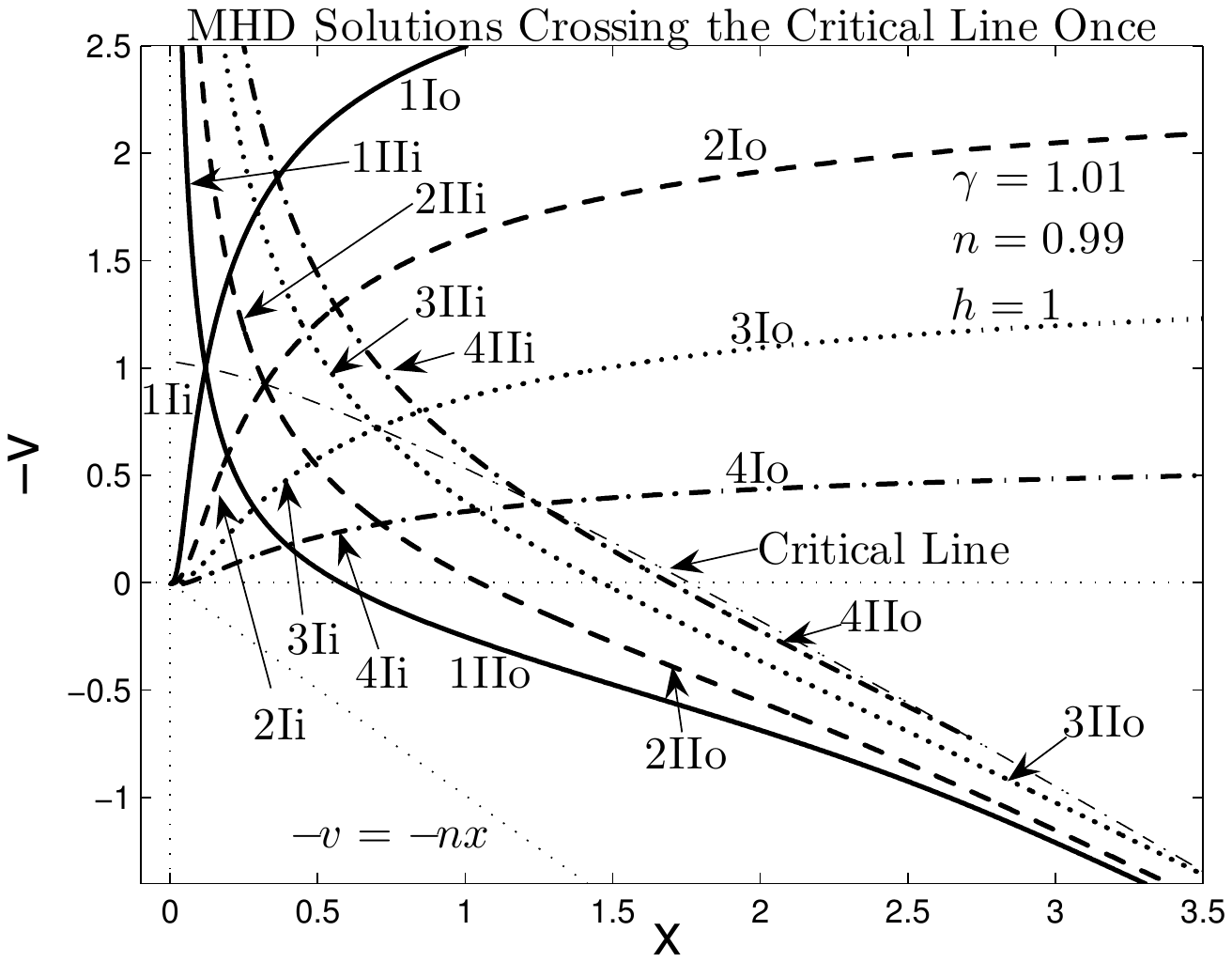}
\caption{Similarity MHD solutions crossing the magnetosonic
critical curve once, with $\gamma=1.01$, $n=0.99$ and $h=1$ being
specified for a conventional polytropic gas. Point 1 corresponds
to $\alpha=15$ and $x=0.1200$, point 2 corresponds to $\alpha=5$
and $x=0.3214$, point 3 corresponds to $\alpha=2$ and $x=0.7021$,
and point 4 corresponds to $\alpha=1$ and $x=1.2456$,
respectively. The two perpendicular light dotted straight lines
are abscissa and ordinate axes, respectively, while the light
dotted straight line passing through the origin is $-v=-nx$ to the
lower left of which solutions become unphysical for a negative
enclosed mass. The light dash-dotted line is the magnetosonic
critical curve, and the heavy lines are the solutions crossing the
magnetosonic critical curve once. }\label{Once101_1_1}
\end{figure}

The magnetosonic critical curves in this nearly isothermal case of
$\gamma\geq 1$ can be compared with those of the isothermal MHD
case of $\gamma=1$ (Yu and Lou, 2005; Yu et al., 2006). Their
asymptotic behaviours are different for both limiting regimes of
$x\rightarrow 0^{+}$ and $x\rightarrow +\infty$. As $x$ approaches
zero, the magnetosonic critical curve in the isothermal case
intersects with the vertical $v-$axis, while in the nearly
isothermal MHD case it diverges as $x\rightarrow 0^{+}$. As $x$
approaches infinity, the magnetosonic critical curve in the
isothermal case remains in the fourth quadrant, while in the
nearly isothermal case it can head up to the first quadrant. These
qualitative differences in asymptotic behaviours in such parallel
cases result from equations $(\ref{wang45})-(\ref{wang47})$.
According to equation (\ref{wang45}), for $\gamma=1$, $v$ remains
finite as $x$ approaches zero, while for $\gamma\geq 1$, even a
small increment in $\gamma$ will lead to a divergence of $v$ as
$x$ goes to zero. In accordance with equation (\ref{wang46}), if
$n=1$ as in the isothermal case, $\alpha$ must approach zero as
$x$ approaching infinity, which means that $v\sim nx$, i.e. the
magnetosonic critical curve remains in the fourth quadrant.
Nonetheless when $n$ is not equal to unity, this constraint on
asymptotic behaviour of the magnetosonic critical curve disappears.
Another perspective is that when $n\rightarrow 1$, we have
$h_c\rightarrow\infty$; thus whatever $h$ values will lead
to asymptotic behaviours such that $v\rightarrow +\infty$ as
$x\rightarrow +\infty$.

The magnetosonic critical curves for different values of $h$ given
$n=0.75$ and $\gamma=1.25$ are shown in Fig. \ref{critical125}.
When $h$ increases, again the average slope
$\text{d}(-v)/\text{d}x$ of an individual magnetosonic critical
line increases from negative to positive in the $-v$ versus $x$
presentation. The value of $h_c$ is $4.5$ in this example. As the
magnetic field becomes extremely strong, one obtains another
branch of the magnetosonic critical curve as shown in Fig.
\ref{critical125} for $h=100$. This new branch is the one
mentioned in equation (\ref{wang40}). The lower branch of the
heavy dotted curve is unphysical for being to the lower left of
the straight line $-v=-nx$.
Also the magnetosonic critical curve diverges as $x$ approaches
zero.

\begin{table}
\center \caption{The corresponding $m(0)$ values for self-similar
MHD flow solutions without crossing the magnetosonic critical
curve for a conventional polytropic gas of $\gamma=1.25$ and
$n=0.75$ with different values of $h$. The two parameters $A_0$
and $B_0$ are specified in asymptotic MHD solutions (\ref{wang22})
and (\ref{wang23}). Parameter $m(0)$ is directly computed from
equation (\ref{wang8}). Here, the $m(0)$ value is computed after
integrating from large $x$ values such as $x=100$ to small $x$
values. }\label{m0_125_1} \vskip 0.3cm
\begin{tabular}{cccccc}\hline
$h$         &$0$     &$0$     &$2$    &$2$     &$10$ \\
$A_0$, $B_0$&$4,\ 0$ &$4,\ -5$&$4,\ 0$&$4,\ -5$&$4,\ -5$\\ \hline
$m(0)$      &$3.8185$&$4.952$ &$3.664$&$4.919$ &$4.747$ \\
\hline
\end{tabular}
\end{table}

\begin{figure}
\subfigure[The smaller $x$ regime. Crossing points 1, 2, and 3
correspond to $\alpha=5$ and $x=0.2170$, $\alpha=0.9$ and
$x=0.8255$, and $\alpha=0.3$ and $x=1.9389$,
respectively.]{\label{Once101_10_1}
\includegraphics[width=3.30in,bb=120 270 480 570]{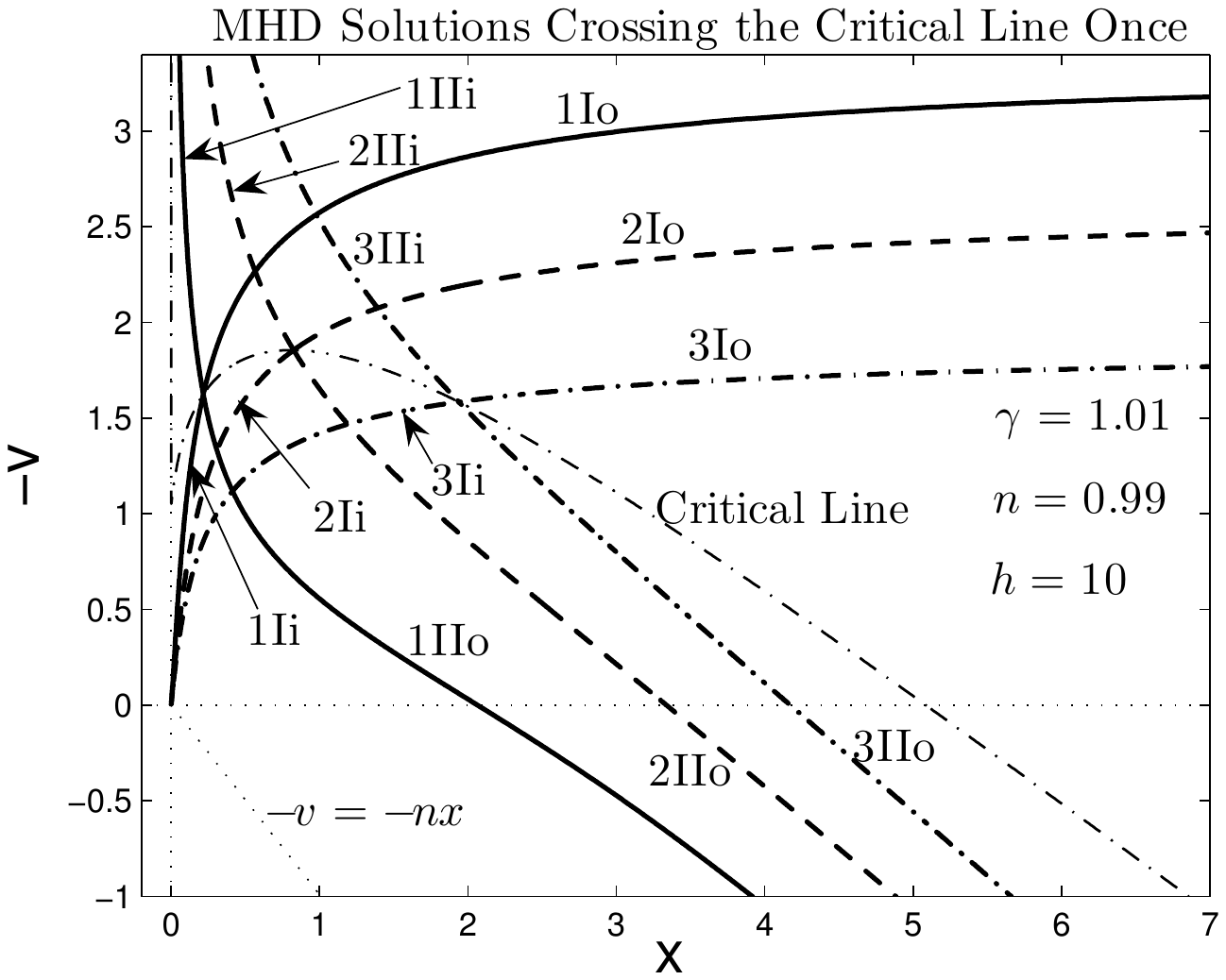} }
\subfigure[The larger $x$ regime. Crossing points 1, 2, and 3
correspond to $\alpha=0.1$ and $x=4.8147$, $\alpha=0.05$ and
$x=9.9587$, and $\alpha=0.04$ and $x=14.0005$,
respectively.]{\label{Once101_10_2}
\includegraphics[width=3.2in,bb=100 270 480 570]{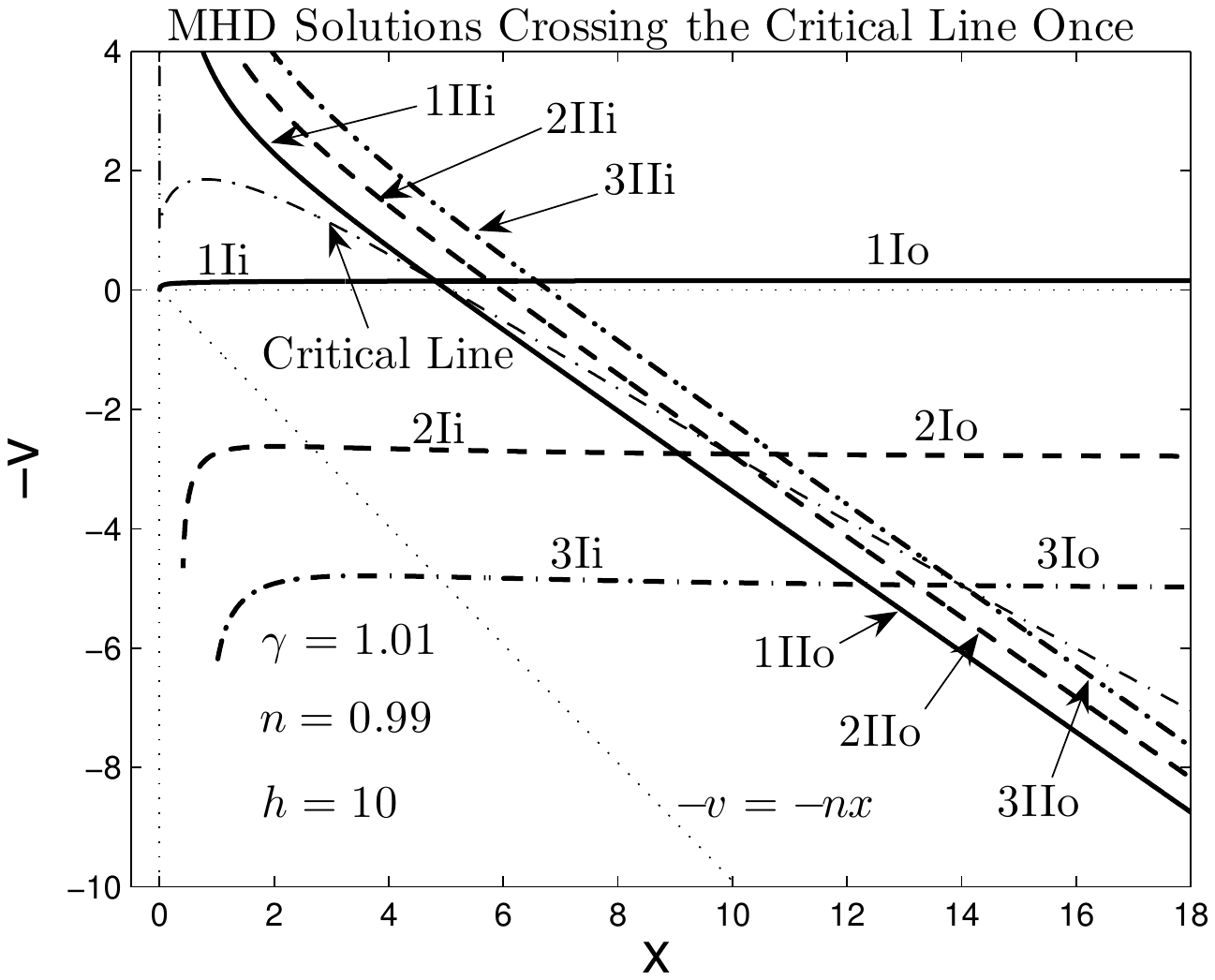} }
\caption{MHD solutions crossing the magnetosonic critical curve
once, with $\gamma=1.01$, $n=0.99$ and $h=10$. Panel (a) displays
the solutions crossing the magnetosonic critical curve at smaller
$x$, while panel (b) displays those crossing the magnetosonic
critical curve at larger $x$. The two perpendicular light dotted
lines are abscissa and ordinate axes, respectively, and the light
dotted straight line is for $-v=-nx$. The light dash-dotted line
is the magnetosonic critical curve, and the heavy curves are the
MHD solutions crossing the magnetosonic critical curve once.
}\label{Once101_10}
\end{figure}

\begin{table}
\center\caption{For Figure \ref{numer101_3}, we summarize the
corresponding $m(0)$ values and the $x$ values of the `kink point'
$x_k$ for MHD expansion wave collapse solutions for $\gamma=1.01$
and $n=0.99$ for a conventional polytropic gas (see Yu \& Lou 2005
and Yu et al. 2006 for an isothermal magnetized gas).
}\label{m0_101_3} \vskip 0.3cm
\begin{tabular}{ccccc}\hline
$h$   & $0$     & $1$     & $10$   & $30$   \\ \hline
$m(0)$& $1.0120$& $1.4931$& $4.346$& $19.26$\\
$x_k$ & $1.019 $& $1.76$  & $5.08$ & $12.11$\\
\hline
\end{tabular}
\end{table}

\begin{figure}
\includegraphics[width=3.3in,bb=100 270 480 570]{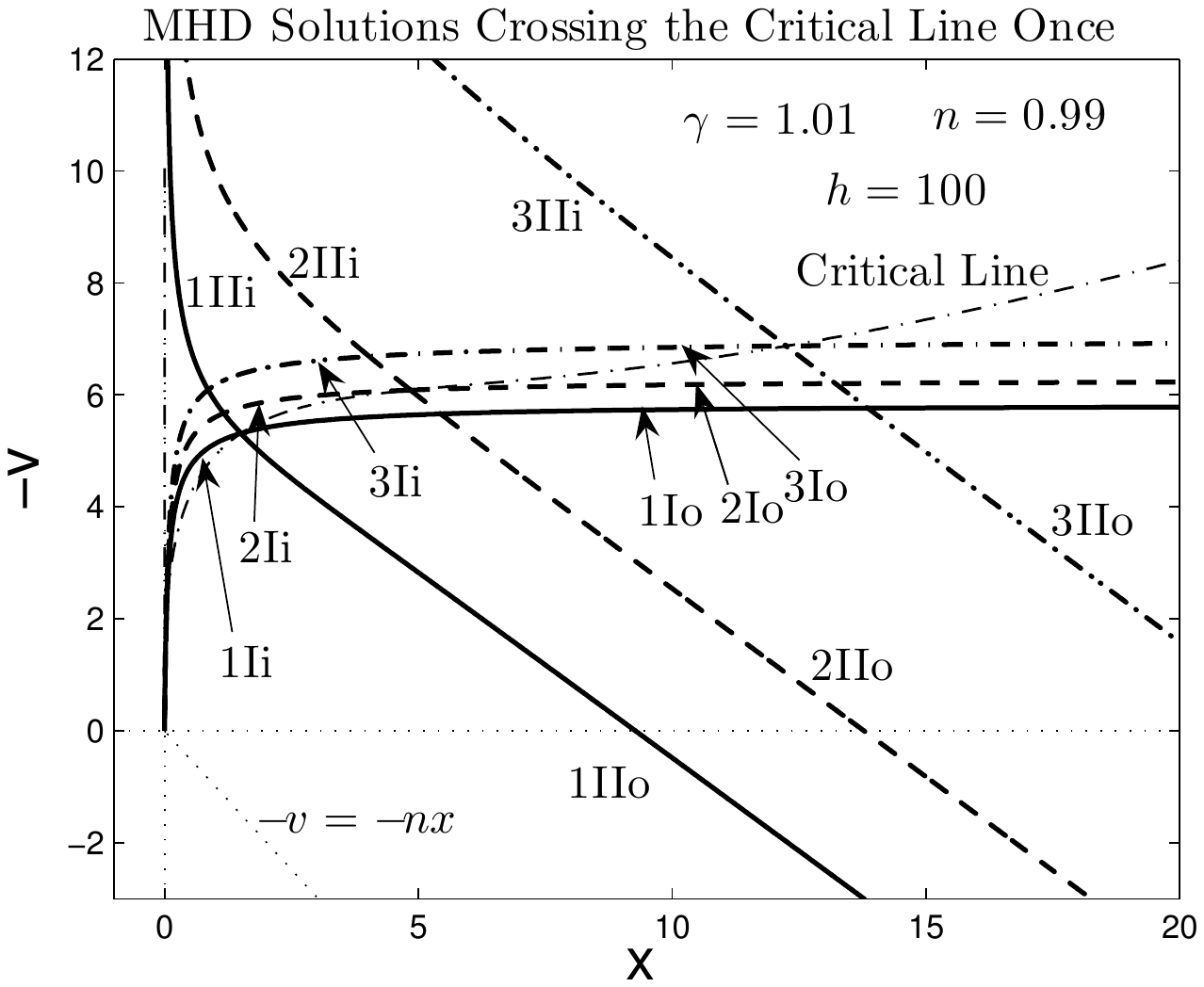}
\caption{MHD solutions crossing the magnetosonic critical curve
once with parameters $\gamma=1.01$, $n=0.99$ and $h=100$ being
specified. Crossing point 1 corresponds to $\alpha=0.2$ and
$x=1.5055$, point 2 corresponds to $\alpha=0.05$ and $x=4.8487$,
and point 3 corresponds to $\alpha=0.024$ and $x=12.2494$,
respectively. The two perpendicular light dotted straight lines
are abscissa and ordinate axes, respectively, and the light dotted
straight line is the demarcation line $-v=-nx$. The light
dash-dotted line is the magnetosonic critical curve, and the heavy
curves are the solutions crossing the magnetosonic critical curve
once. }\label{Once101_100_1}
\end{figure}

\begin{table}
\center \caption{For Figure \ref{numer125_3}, we summarize the
corresponding $m(0)$ and the $x$ values of the `kink point'
values $x_k$ for MHD expansion wave collapse solutions with two
parameters $\gamma=1.25$ and $n=0.75$ specified for a usual
polytropic gas (see Yu and Lou 2005 for an isothermal magnetized
gas).}\label{m0_125_2} \vskip 0.3cm
\begin{tabular}{cccc}\hline
$h$   & $0$     & $2$    & $4$    \\ \hline
$m(0)$& $0.7992$& $2.083$& $24.38$\\
$x_k$ & $1.37$  & $2.52$ & $6.68$ \\
\hline
\end{tabular}
\end{table}

Enlarged features for diverging behaviours of $v(x)$ along
the magnetosonic critical curves for small $x$ in Figs.
\ref{critical101} and \ref{critical125} are shown in Figs.
\ref{critical101s} and \ref{critical125s}, respectively. The
corresponding features of $\alpha$ versus $x$ are displayed in
Figs. \ref{criticalalpha101} and \ref{criticalalpha125},
respectively. The basic facts that for $h=0$, $\alpha$ approaches
infinity both as $x$ approaches zero and infinity, while for
$h>0$, $\alpha$ approaches infinity as $x$ approaches zero and
$\alpha$ approaches a constant as $x$ goes to infinity are all
consistent with the relevant analytical results presented in
subsection \ref{crasy}.

By numerical exploration, we found that the magnetosonic critical
curve has two branches in the $\gamma=1.01$, $n=0.99$ case with
$h=1000$. Also, the critical curve consists of two parts as shown
in Fig. \ref{critical101} in the case of $h=0$ can also be found
in the case of $\gamma=1.01$, $n=0.99$, and $h=0.03$. From the
results for critical curves, one can see that asymptotic analyses
in subsections \ref{crdet} and \ref{crasy} are necessary for
determining the entire magnetosonic critical curve.

\subsection{Solutions without crossing the Magnetosonic
Critical Curve and MHD Expansion Wave Collapse Solutions}

\subsubsection[]{General MHD Solutions without
Crossing the Magnetosonic Critical Curve}

Among the asymptotic MHD solutions derived in subsection
\ref{asym}, the series expansion at large $x$ described by
equations (\ref{wang22}) and (\ref{wang23}) can be readily
integrated from large values of $x$ inward to obtain numerical
solutions without encountering the magnetosonic critical curve.
Specifically, we integrate the solutions from a starting point
of $x=100$.

\begin{figure}
\includegraphics[width=3.3in,bb=100 270 480 570]{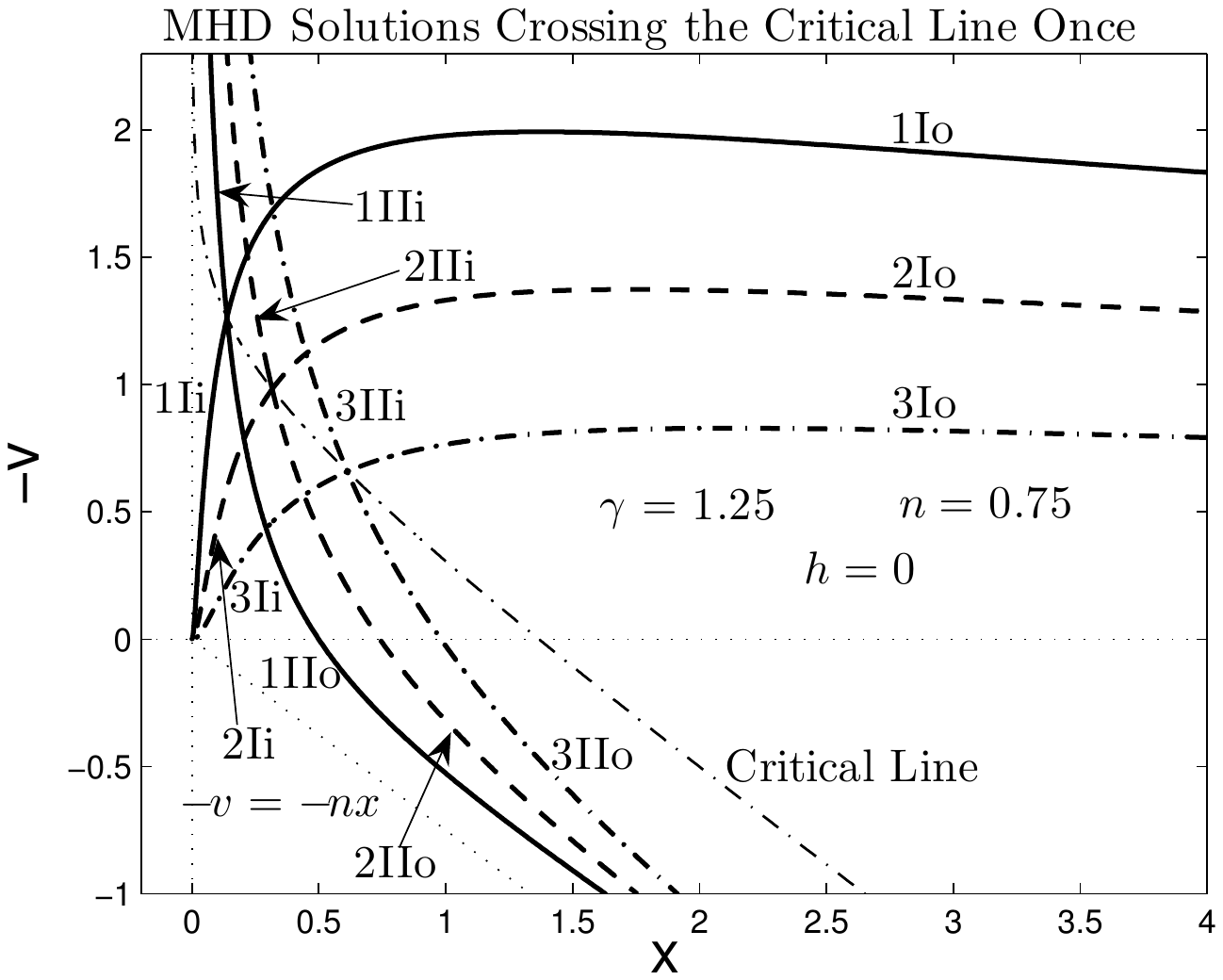}
\caption{Hydrodynamic solutions crossing the sonic critical curve
once, with $\gamma=1.25$, $n=0.75$ and $h=0$. Crossing point 1
corresponds to $\alpha=5$, $x=0.1386$, crossing point 2
corresponds to $\alpha=2$ and $x=0.3166$, and crossing point 3
corresponds to $\alpha=1$ and $x=0.6131$, respectively. The two
perpendicular light dotted lines are the abscissa and ordinate
axes, respectively, and the light dotted straight line $-v=-nx$ is
the demarcation line to the lower left of which solutions become
unphysical. The light dash-dotted line is the sonic critical
curve, and the heavy curves are the solutions crossing the sonic
critical curve once. The oscillatory behaviours of solutions 1Ii,
2Ii and 3Ii as $x$ approaches $0$ are not readily seen and will
be discussed at the end of the main text. }\label{Once125_0_1}
\end{figure}

We present such global semi-complete similarity MHD solutions for
the case of $\gamma=1.01$ and $n=0.99$ in both Figures
\ref{numer101_1} and \ref{numer101_2}. Note that when $x$
approaches zero, the solution approaches the free-fall state as
discussed in subsection \ref{asym} [see equations (\ref{wang27})
and (\ref{wang28})]. Note also that the major difference in $v(x)$
of the two solutions with the same values of $A_0$ and $B_0$ but
with different magnetic field strengths (i.e., different $h$)
manifests mainly at small $x$ about $0\leq x\leq 10$ in both
Figures. Both Figs. \ref{numer101_1} and \ref{numer101_2} show
that the magnetic field mainly accelerates the central collapses,
i.e. when $h$ is larger, the $-v(x)$ becomes larger at the same
$x$, although the asymptotic behaviours of $v(x)$ as $x$ approaches
infinity show that larger $h$ implies smaller $-v(x)$ at the same
large $x$.

\begin{figure}
\includegraphics[width=3.3in,bb=100 270 480 570]{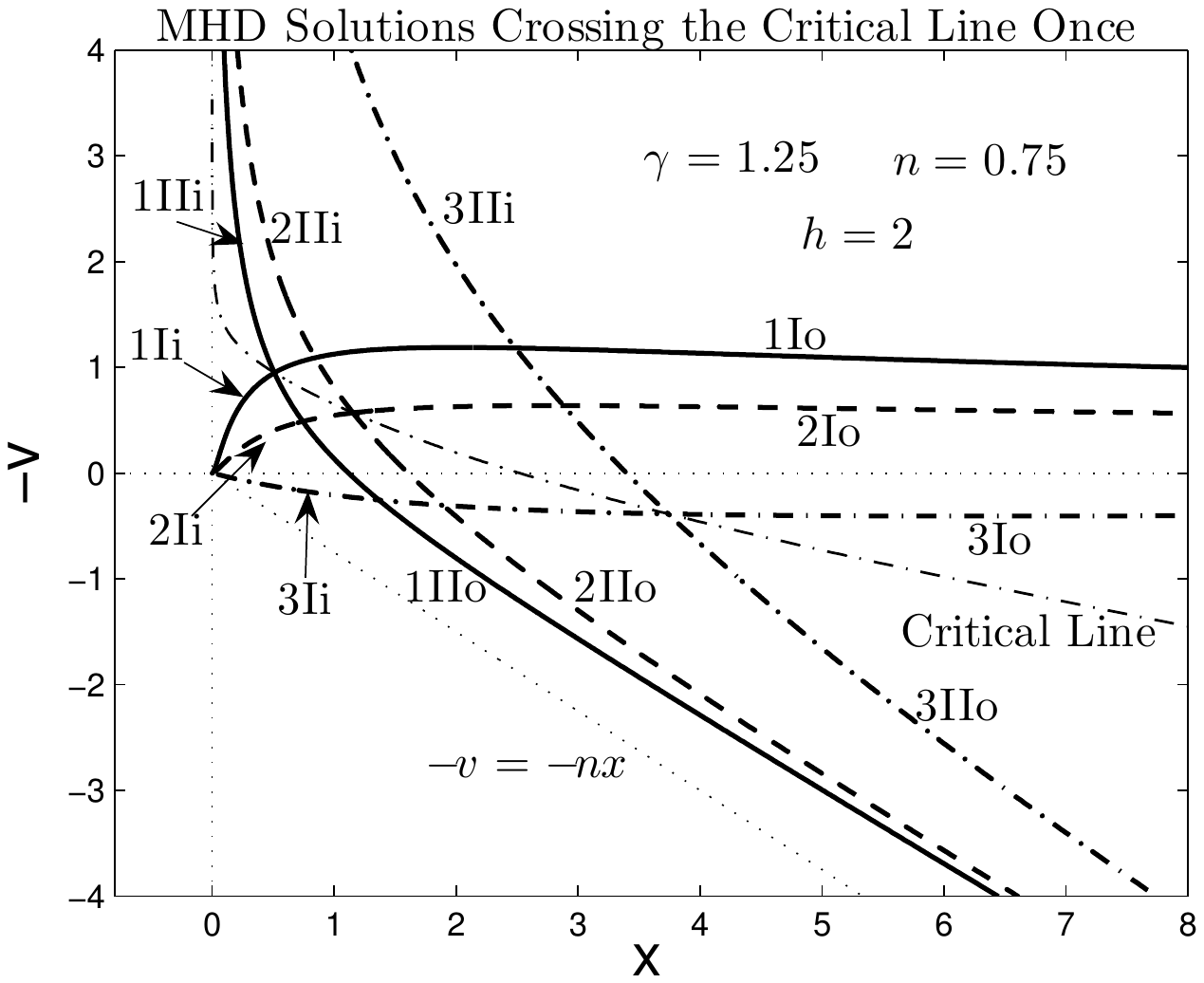}
\caption{MHD solutions crossing the magnetosonic critical curve
once with $\gamma=1.25$, $n=0.75$ and $h=2$ being specified.
Crossing point 1 corresponds to $\alpha=1$ and $x=0.5107$,
crossing point 2 corresponds to $\alpha=0.4$ and $x=1.1549$, and
crossing point 3 corresponds to $\alpha=0.18$ and $x=3.7395$,
respectively. The two perpendicular light dotted lines are the
abscissa and ordinate axes, respectively, and the light dotted
straight line $-v=-nx$ is the demarcation to the lower left of
which solutions become unphysical. The light dash-dotted curve is
the magnetosonic critical curve, and the heavy curves are the
solutions crossing the magnetosonic critical curve once. The
oscillatory behaviours of solutions 1Ii, 2Ii and 3Ii as $x$
approaches $0$ are not readily seen and will be discussed at the
end of the main text. }\label{Once125_2_1}
\end{figure}

Numerical similarity MHD solutions for the case of $\gamma=1.25$
and $n=0.75$ are presented in Figure \ref{numer125_2}. In this
figure the solution with the same values of $A_0$ and $B_0$ but
larger $h$ value cannot catch up with the other solutions with
smaller values of $h$. This does not necessarily mean that the net
magnetic force does not accelerate collapses, because of a smaller
$-v$ in the initial state for the case of a larger $h$.

We briefly note several points here. Firstly, we analyzed in
subsection \ref{asym} the asymptotic behaviour of the MHD free-fall
solutions for small $x$ and inferred from that analysis that these
solutions will not encounter the magnetosonic singular surface at
small $x$, meanwhile the asymptotic MHD solutions as $x$
approaches infinity [equations (\ref{wang22}) and (\ref{wang23})]
also do not encounter the singular surface. Therefore, these
numerical MHD solutions are specific examples of semi-complete
solutions without crossing the magnetosonic critical curve or
encountering the singular surface. Secondly, there exists a
two-dimensional continuum regime of parameters $A_0$ and $B_0$ for
this series expansion of MHD solution, e.g., for a specified $A_0$
parameter, parameter $B_0$ should be larger than a certain
threshold value in order to construct MHD similarity flow
solutions without encountering the magnetosonic critical curve.
Outside such allowed parameter regime, the solutions tend to crash
on to the singular surface but away from the magnetosonic critical
curve so that a global semi-complete MHD solution does not exist.
Thirdly, the MHD similarity flow solutions with $h>h_c$ will cross
the magnetosonic critical curve at the projection to $-v\sim x$
plane, yet they do not actually cross the magnetosonic critical
curve in the $\alpha-v-x$ space because they do not encounter the
singular surface. \vspace{1cm}
\subsubsection[]{Construction of MHD Expansion
\\ \qquad\quad
Wave Collapse Solutions (mEWCSs) }

\begin{figure}
\includegraphics[width=3.3in,bb=100 270 480 570]{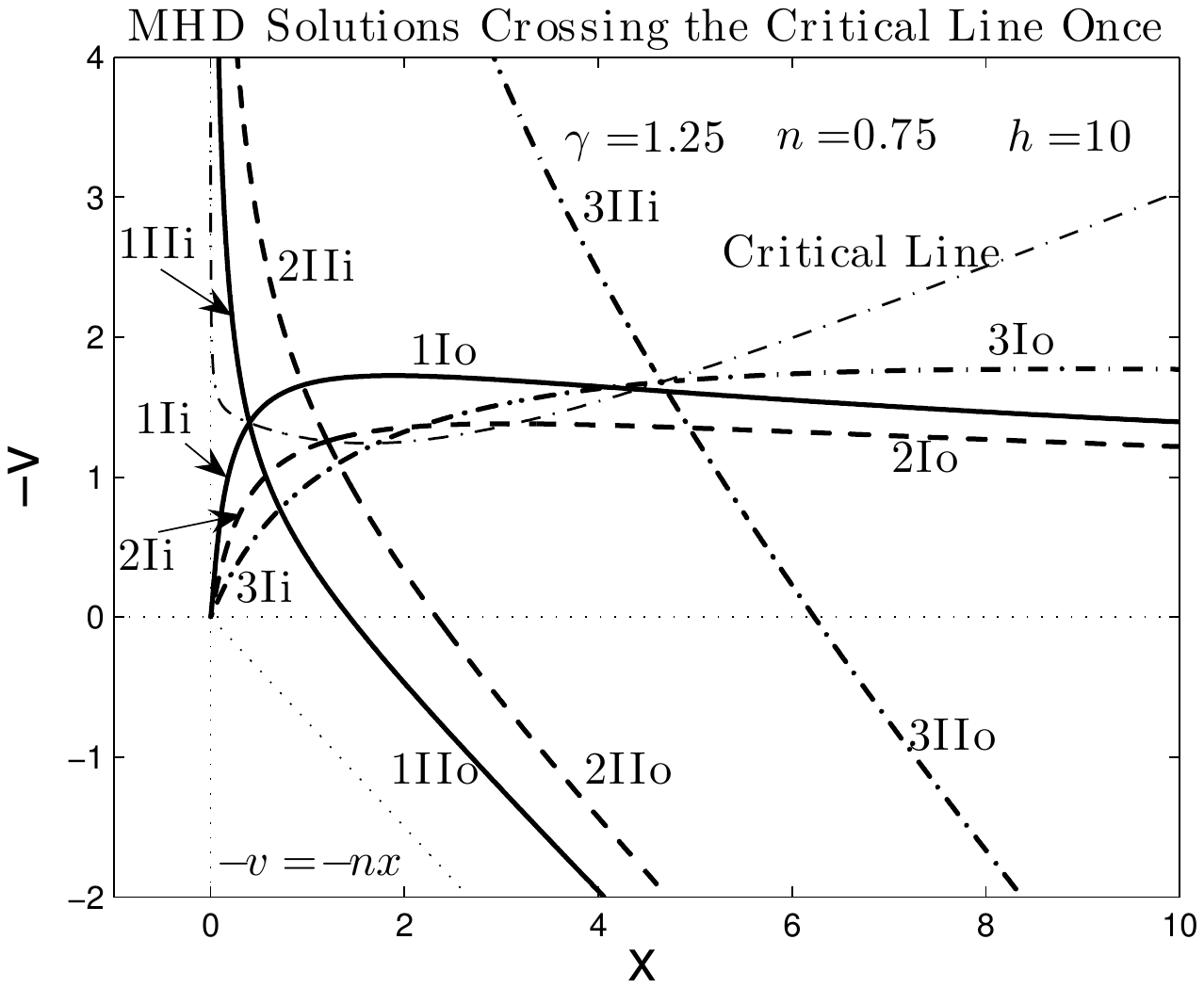}
\caption{MHD solutions crossing the magnetosonic critical curve
once with $\gamma=1.25$, $n=0.75$ and $h=10$. Crossing point 1
corresponds to $\alpha=1$ and $x=0.3998$, crossing point 2
corresponds to $\alpha=0.26$ and $x=1.2053$, and crossing point 3
corresponds to $\alpha=0.12$ and $x=4.6518$, respectively. The two
perpendicular light dotted lines are the abscissa and ordinate
axes, respectively. The straight line $-v=-nx$ is a demarcation
line to the lower left of which solutions become unphysical. The
light dash-dotted curve is the magnetosonic critical curve, and
the heavy curves are the MHD solutions crossing the magnetosonic
critical curve once. }\label{Once125_10_1}
\end{figure}

One interesting solution among global MHD solutions not crossing
the magnetosonic critical curve is the limiting solution
corresponding to MHD expansion wave collapse solutions (mEWCSs)
as a generalization of the isothermal EWCS (Shu, 1977) in two
important aspects, i.e., the polytropic gas and the inclusion of
a random magnetic field. We emphasize here the existence of such
global MHD similarity solutions because magnetic fields do exist
in molecular clouds in general and play important roles in the
evolution of a collapsing cloud. From the perspectives of dynamic
evolution, diffusive processes, radiative signatures, formations
of discs and jets and origin of stellar magnetic fields, one must
take into account of magnetic fields in molecular clouds. By the
model scenario here, our analysis suggests that there exist mEWCSs
for $h<h_c$ (a weaker magnetic field), while no such mEWCS exists
for $h\geq h_c$ (a stronger magnetic field).

\begin{figure}
\includegraphics[width=3.3in,bb=100 270 480 570]{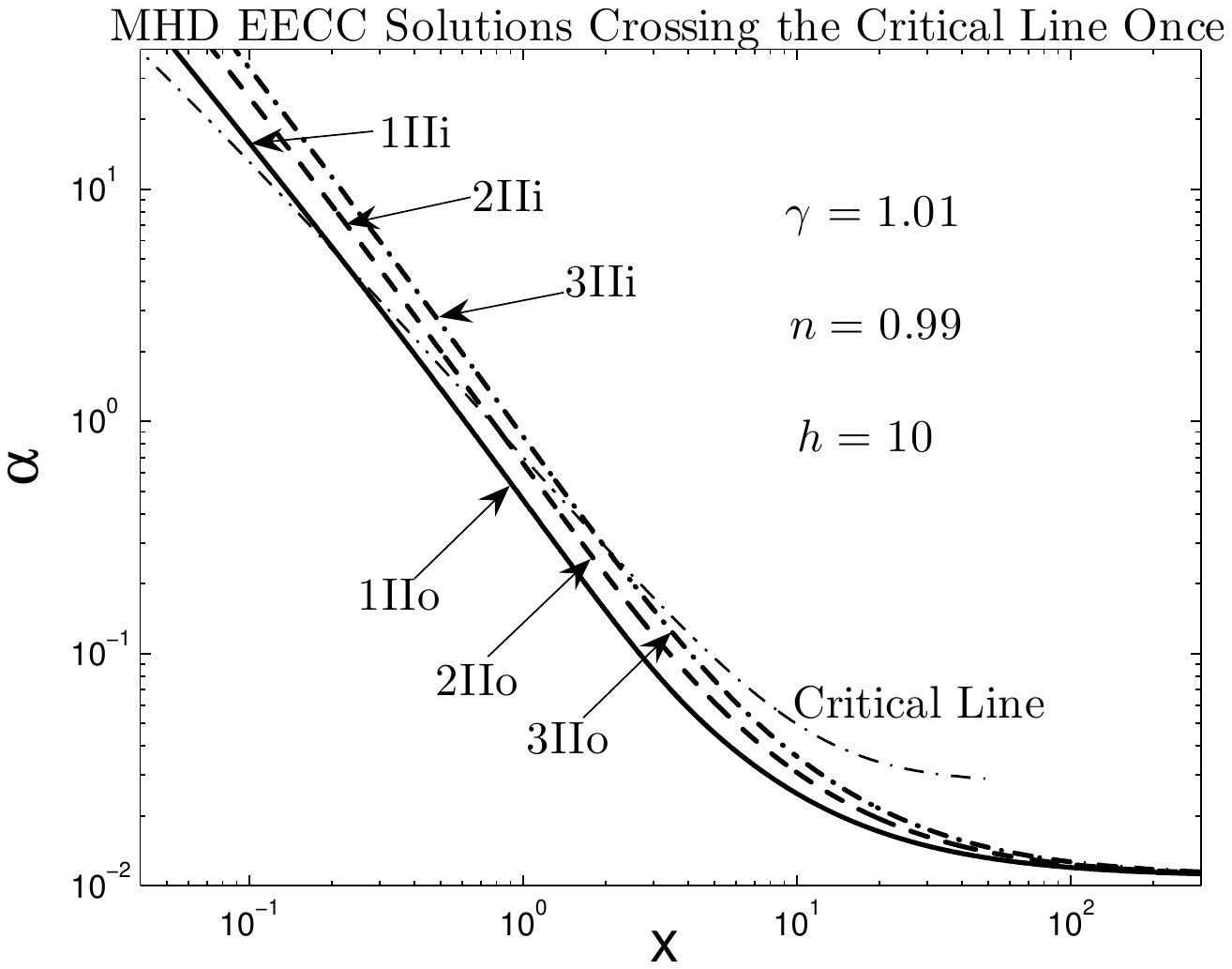}
\caption{Corresponding $\alpha$ behaviours for the MHD solutions
1II, 2II and 3II in Fig. \ref{Once101_10_1}. The light dash-dotted
curve is the magnetosonic critical curve, and the heavy curves are
the solutions crossing the magnetosonic critical curve once.
}\label{Once101_10_3}
\end{figure}

This conclusion can be viewed from the following perspectives.
First, in asymptotic MHD solutions (\ref{wang22}) and
(\ref{wang23}), parameter $B_0$ can be regarded as an external (or
initial) radial flow speed more or less independent of the mass
density profile. For example in the isothermal case, parameter
$B_0$ represents an asymptotic steady flow speed in regions far
from the core (see Lou and Shen 2004 and subsection \ref{asym}
here). Parameter $A_0$ contributes to the radial speed profile due
to gas mass density distribution and the associated self-gravity.
To construct mEWCSs, one should require that $B_0=0$ and the
reduced radial speed $v(x)$ approaches $0^{-}$, i.e., for the
limiting series solution, the radial flow velocity remains
negative and approaches zero far away. In this limiting regime, a
magnetized gas cloud of quasi-spherical symmetry remains at rest
in early times and the core collapse is induced by self-gravity.
In this perspective, the critical $A_0$ for mEWCS is determined by
setting the coefficient of $x^{(n-2)/n}$ term [from the same two
terms $x^{{(n-2)}/{n}}$ and $x^{(2-2\gamma-n)/n}$ for a usual
polytropic gas with $n=2-\gamma$] in asymptotic MHD solution
(\ref{wang23}) to vanish. In contrast, the case of $h\geq h_c$
corresponds to positive coefficients of both terms involving $A_0$
in asymptotic MHD solution (\ref{wang23}) because of
${1}/{(3n-2)}+{2h(n-1)}/{n^2}<0$, indicating outward expansions
for any value of $A_0>0$ when the coefficient of $x^{(n-2)/n}$
term vanishes. Secondly, for a mEWCS, the reduced velocity remains
zero at large $x$ until the solution meets the magnetosonic
critical curve at the point where the magnetosonic critical curve
intersects the $x-$axis. At this intersection point, the slope of
the magnetosonic critical curve is negative for $h<h_c$, which
allows the mEWCS to head up as $x\rightarrow 0^{+}$, while for
$h\geq h_c$, the slope of the magnetosonic critical curve is
positive to force the solution to crash on to the singular surface
without leading to a mEWCS. Thirdly, the physical reason for the
non-existence of a mEWCS when $h>h_c$ is that a sufficiently
strong magnetic field tends to prevent a gravitational collapse
and to drive an outward expansion instead. The stability analysis
of the present MHD problem remains to be examined for the case of
$h\geq h_c$. For $h\geq h_c$ a magnetized gas cloud collapses only
when external (initial) inflows are present. Because of
$h_c\rightarrow\infty$ in the isothermal case, any $h$ may lead to
an mEWCS. Finally, when $h\geq h_c$ or $A_0$ smaller than a
certain critical value representing the mEWCS, parameter $B_0$
should be negative to ensure the core collapse of a magnetized gas
cloud. When $B_0$ becomes sufficiently negative, the corresponding
initial flow speed profile will have a tendency to collapse, and a
core collapse without encountering the magnetosonic critical curve
can happen. Therefore for any $A_0>0$, one can find solutions
without encountering the magnetosonic critical curve.

In Figures \ref{numer101_3} and \ref{numer125_3}, we present the
mEWCSs of a usual polytropic gas cloud for $\gamma=1.01$, $n=0.99$
and $\gamma=1.25$, $n=0.75$, respectively. We conclude from both
Figures \ref{numer101_3} and \ref{numer125_3} that the kink points
$x_k$ of the mEWCSs have larger $x$ values as $h$ increases. The
corresponding values of $m(0)$ and $x_k$ are summarized in Tables
\ref{m0_101_3} and \ref{m0_125_2}, respectively.

\begin{figure}
\includegraphics[width=3.3in,bb=100 270 480 570]{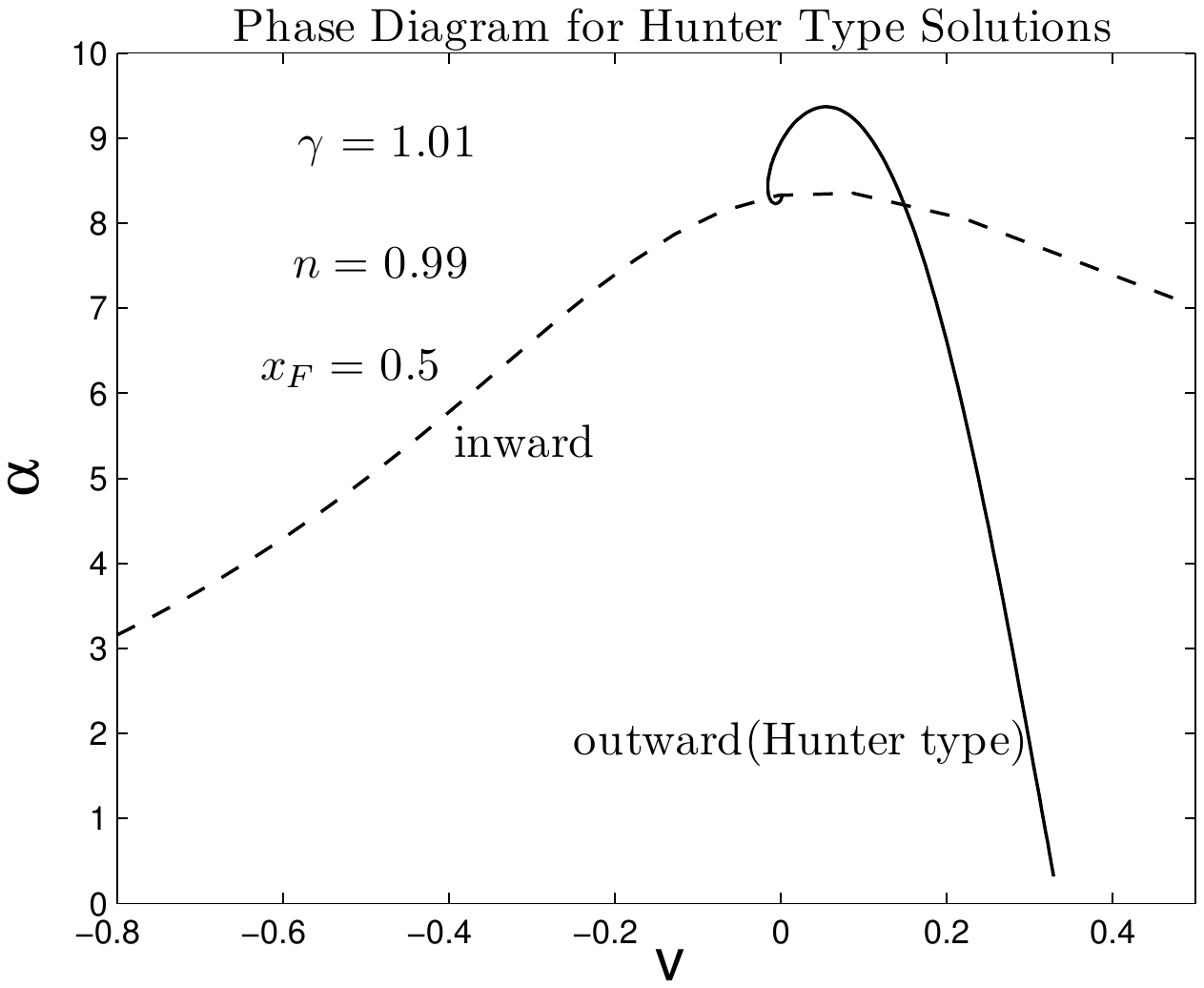}
\caption{Density-speed phase diagram for MHD generalizations of
Hunter type solutions with $\gamma=1.01$, $n=0.99$, $h=1$, and a
chosen meeting point at $x_F=0.5$. The `outward' curve represents
the phase path of Hunter type solutions when the parameter
$\alpha_\ast$ is changed gradually. The `inward' curve represents
the phase path of solutions crossing the magnetosonic critical
curve as the intersection point of the MHD solutions with the
magnetosonic critical curve is gradually adjusted.
} \label{hunter101_1_1}
\end{figure}

We also present the reduced magnetic energy density
$w=h\alpha^2x^2$ associated with a random transverse
magnetic field ${\mathbf B}_t$ versus $x$.
Figure \ref{numer125m} collects a sample of $w$ versus $x$
solution curves for $\gamma=1.25$, $n=0.75$ and $h=2$ (see Fig.
\ref{numer125_2}). Note that the initial magnetic field strengths
are the same $hA_0^2 x^{(2-4/n)}$ as $x$ approaches infinity
(i.e., $t\rightarrow 0^{+}$) for the two upper curves (i.e.,
dashed and dash-dotted linetypes) with $A_0=4$; at first the
magnetic field strength increases faster in the case of $B_0=0$ as
$x$ becomes smaller, and then at some point around $x=0.63549$ the
curve with larger initial velocity ($B_0=-5$) catches up with the
former and grows faster. For the mEWCS with $A_0=2.520$ and
$B_0=0$ and a decreasing $x$, the reduced magnetic energy density
$w$ begins to increase before the kink point $x_k=2.6$ in the
reduced velocity field, and at the kink point the magnetic field
also appears to increase more slowly than at larger $x$ values.
The interpretation of this feature is that at a specific point,
the magnetic field first maintains a constant and then begins to
decrease when the magnetosonic wave front reaches this point.

\begin{figure}
\includegraphics[width=3.3in,bb=100 270 480 570]{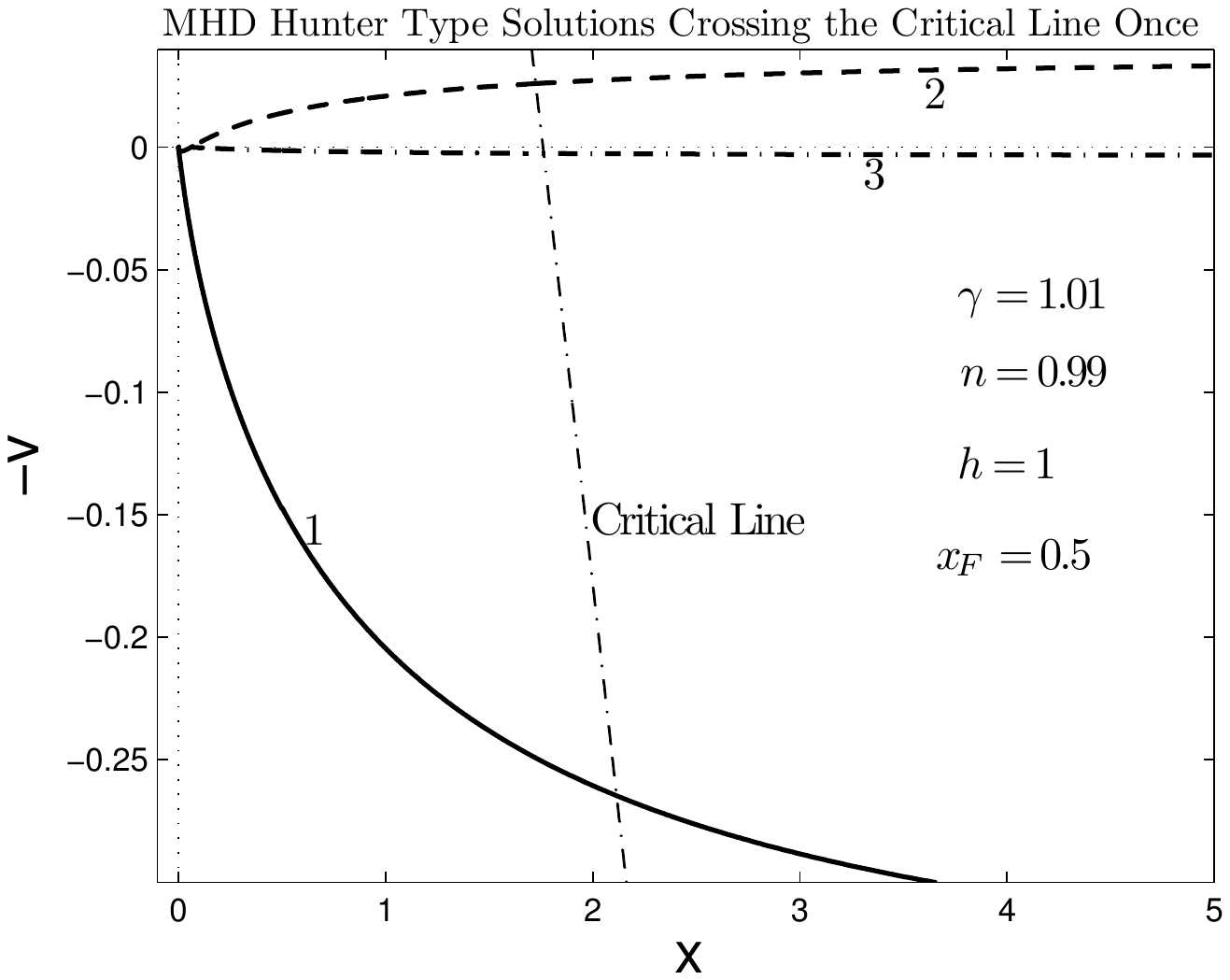}
\caption{The first three discrete MHD Hunter type solutions for
$\gamma=1.01$, $n=0.99$, $h=1$ and $x_F=0.5$ in the semi-complete
space. The two perpendicular light dotted lines are the abscissa
and ordinate axes. The light dash-dotted curve is the magnetosonic
critical curve and the heavy lines are the similarity MHD
solutions curves. Curve 1 corresponds to $\alpha_\ast=359.37$ and
crosses the magnetosonic critical curve at $x=2.1154$, curve 2
corresponds to $\alpha_\ast=8.3941\times 10^6$ and crosses the
magnetosonic critical curve at $x=1.7248$, and curve 3 corresponds
to $\alpha_\ast=1.4769\times 10^{11}$ and crosses the magnetosonic
critical curve at $x=1.7637$, respectively. As $x\rightarrow
+\infty$, the two corresponding $A_0$ and $B_0$ in asymptotic
solutions (\ref{wang22}) and (\ref{wang23}) are $A_0=2.430$ and
$B_0=0.3810$ for curve 1, $A_0=2.016$ and $B_0=-0.03998$ for curve
2, and $A_0=2.058$ and $B_0=0.003869$ for curve 3, respectively.
}\label{hunter101_1_2}
\end{figure}

\begin{table}
\center \caption{Values of the two parameters $A_0$ and $B_0$
in equations (\ref{wang22}) and (\ref{wang23}) at large $x$ of
relevant semi-complete similarity MHD flow solutions in Figures
\ref{Once101_0_1} through \ref{Once125_10_1}.} \label{A0B0Once}
\vskip 0.3cm
\begin{tabular}{cccccc}\hline
\multicolumn{6}{c}{Parameters $A_0$ and $B_0$ in equations
(\ref{wang22}) and (\ref{wang23})} \\ \hline\hline Fig.
&\ref{Once101_10_1} &\ref{Once101_10_1}&\ref{Once101_10_1}
& \ref{Once101_10_2}&\ref{Once101_100_1} \\
Curve&1Io   &2Io   &3Io   &1Io   &1Io \\ \hline
$A_0$&0.145 &0.535 &1.110 &2.392 &0.448\\
$B_0$&-3.468&-2.704&-1.934&-0.166&-6.074\\ \hline
&&&&\\
\hline Fig.
&\ref{Once101_100_1}&\ref{Once101_100_1}&\ref{Once125_10_1}
&\ref{Once125_10_1} & \ref{Once125_10_1}\\
curve&2Io   &3Io    &1Io   &2Io   &3Io \\ \hline
$A_0$&1.222 &3.812  &0.268 &0.767 &9.447 \\
$B_0$&-6.556&-7.340 &-3.268&-2.948&-5.204 \\ \hline
\end{tabular}
\end{table}

\begin{figure}
\includegraphics[width=3.3in,bb=100 270 480 570]{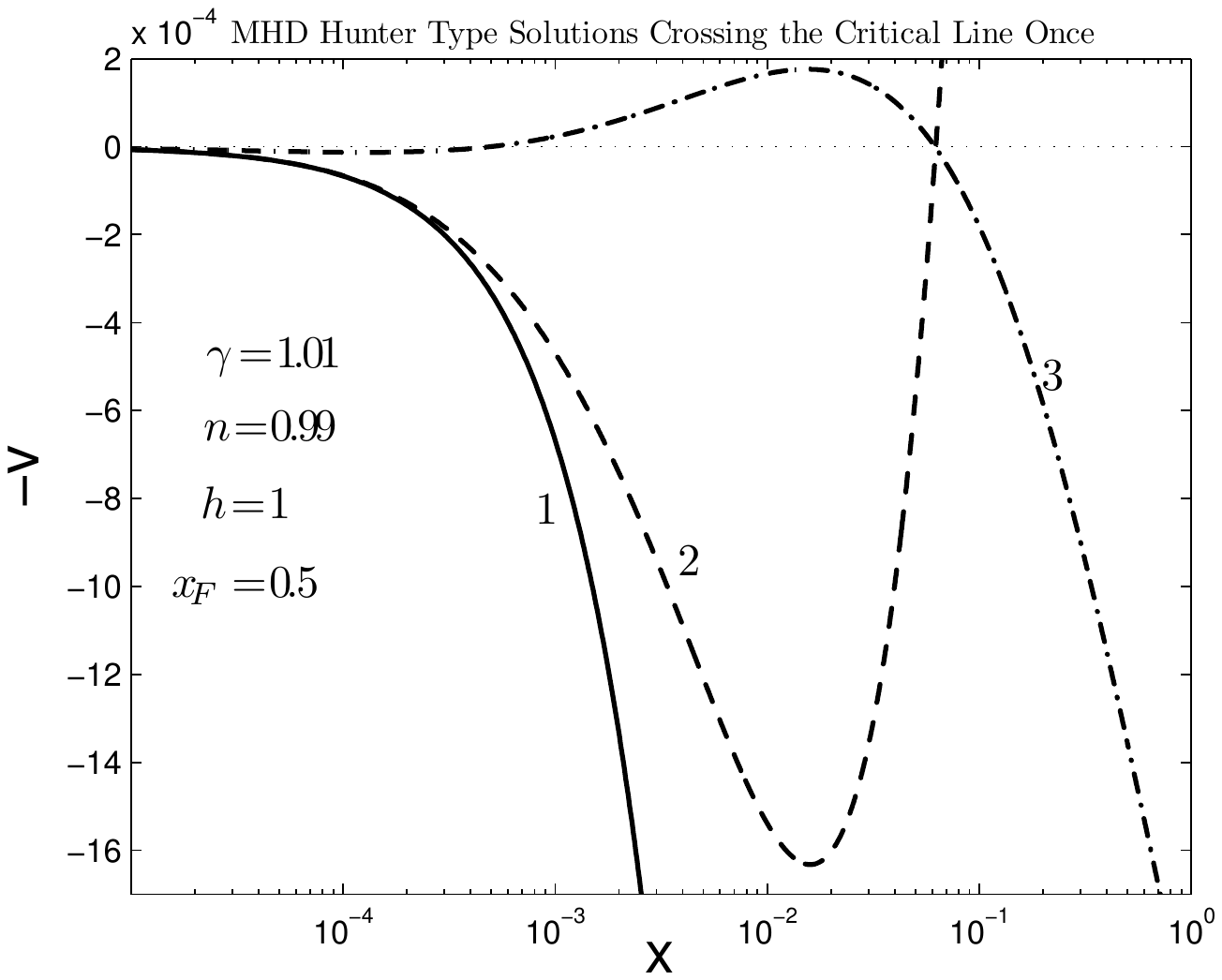}
\caption{Enlarged versions for the MHD generalizations of the
first three Hunter type MHD solutions for $\gamma=1.01$, $n=0.99$,
$h=1$ and $x_F=0.5$ to illustrate that the $i$th solution has $i$
stagnation points ($i=1,\ 2,\ 3$). Other parameters are the same
as those in Figure \ref{hunter101_1_2}. Note that the ordinate
scale has a factor of $10^{-4}$ and the abscissa for $x$ is in
the logarithmic scale. The feature of self-similar magnetosonic
oscillations is shown by these MHD solutions, as a general
extension of the isothermal hydrodynamic feature revealed by
Hunter 1977 and Lou \& Shen 2004. }\label{hunter101_1_3}
\end{figure}

\begin{table}
\center \caption{Corresponding $m(0)$ values of MHD free-fall
solutions at small $x$ for relevant semi-complete solutions in
Figs. \ref{Once101_0_1} through \ref{Once125_10_1}.}\label{m0Once}
\vskip 0.3cm
\begin{tabular}{ccccccc}\hline
\multicolumn{7}{c}{Corresponding values
of $m(0)$ for MHD free-fall solutions} \\
\hline\hline
Fig.&\ref{Once101_10_1}&\ref{Once101_10_1}&\ref{Once101_10_1}
&\ref{Once101_10_2}&\ref{Once101_10_2}&\ref{Once101_10_2} \\
Curve &1IIi &2IIi &3IIi &1IIi &2IIi &3IIi \\ \hline
$m(0)$&0.400&1.291&2.442&4.213&6.933&10.18\\
\hline
&&&&&&\\
\hline Fig.&\ref{Once101_100_1}&\ref{Once101_100_1}&
\ref{Once101_100_1} & \ref{Once125_10_1}&
\ref{Once125_10_1} & \ref{Once125_10_1} \\
Curve &1IIi &2IIi &3IIi &1IIi &2IIi &3IIi \\ \hline
$m(0)$&2.539&8.345&35.59&0.260&0.741&11.44\\ \hline
\end{tabular}
\end{table}

One readily obtains the corresponding value of $m(0)$ for each MHD
free-fall solution using equation (\ref{wang8}). The corresponding
$m(0)$ values for all solutions computed in this section are
summarized in Tables \ref{m0_101_1} to \ref{m0_125_2}. According
to Tables \ref{m0_101_1} and \ref{m0_101_2}, one finds that $m(0)$
increases with either larger $h$ or larger magnitude of $B_0$,
indicating that a stronger magnetic field and a faster inward
initial speed both result in a more rapid core collapse and lead
to an enhanced central mass accretion (n.b. profiles of gas mass
density scalings remain the same). This is related to the fact
that the inward magnetic tension force is twice as large as the
outward magnetic pressure gradient force. In reference to Table
\ref{m0_101_3}, we note the increase of $m(0)$ with a larger $h$
in mEWCSs with the two parameters $\gamma$ and $n$ being
specified. By Table \ref{m0_125_1}, one finds that a larger $B_0$
results in a smaller $m(0)$, but as mentioned above this does not
necessarily mean that the magnetic field does not accelerate the
core collapse because of an initial smaller velocity. By Table
\ref{m0_125_2}, one again gets an increase of $m(0)$ with
increasing $h$. We also provide the $x$ values of the
corresponding `kink points' $x_k$ of the mEWCSs in Tables
\ref{m0_101_3} and \ref{m0_125_2}, showing that for a given
$\gamma>1$, the corresponding $x_k$ increases with an increasing
$h$ value as long as $h<h_c$.

\subsection{Numerical MHD Solutions Crossing the
Magnetosonic Critical Curve Once}\label{once}

\begin{figure}
\includegraphics[width=3.3in,bb=100 270 480 570]{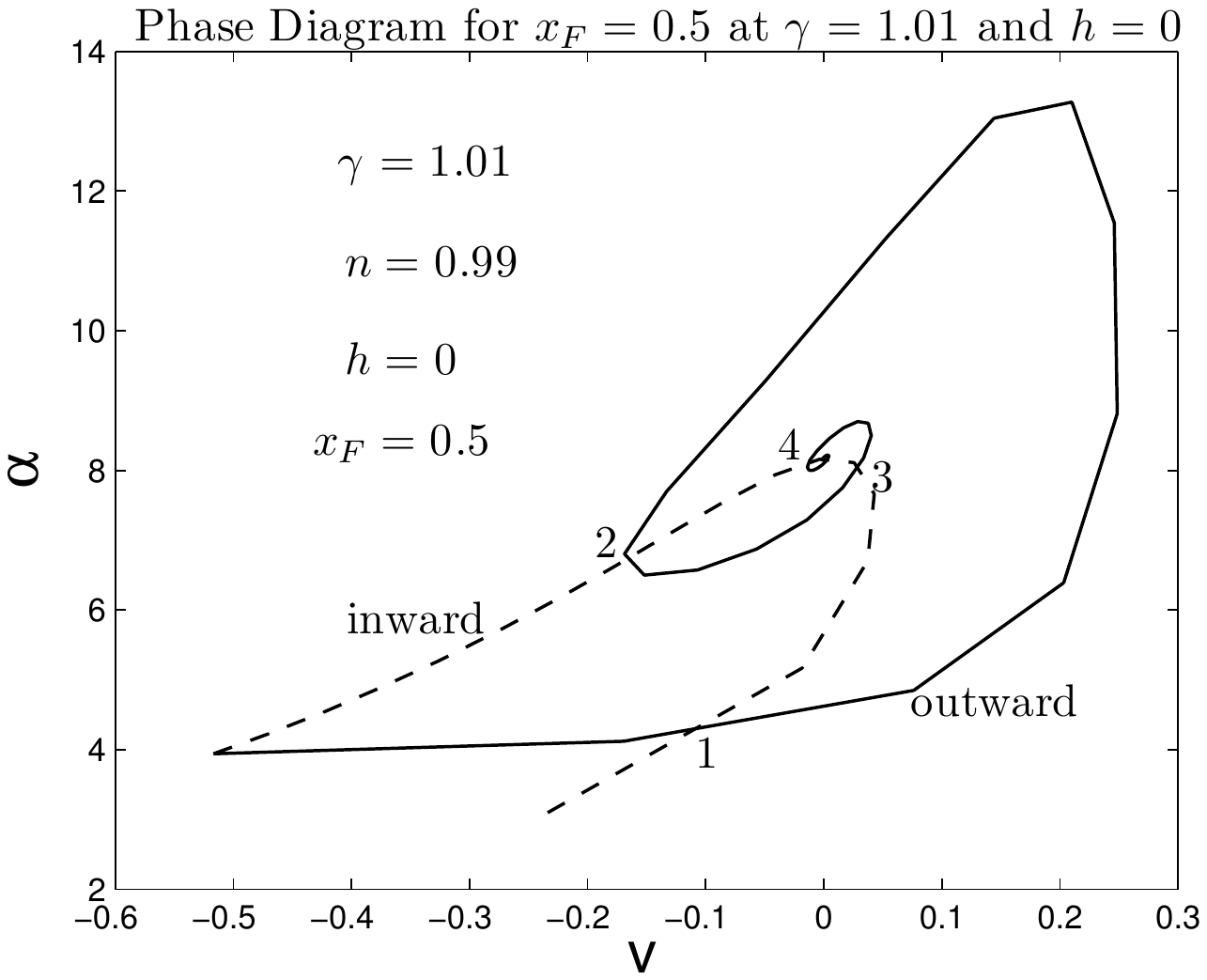}
\caption{Density-speed phase diagram of $\alpha$ versus $v$ for
$\gamma=1.01$, $n=0.99$, $h=0$ and $x_F=0.5$. The `inward curve'
(dashed curve) denotes the phase path by integrating inward from
different $x$ ($x>x_F$) along the sonic critical curve to reach
$x_F$, using the type I eigensolution, while the `outward curve'
denotes the phase path integrating from different $x$ ($x<x_F$)
along the sonic critical curve to reach $x_F$, using the type II
eigensolution. The spiral pattern of the outward curve suggests
the trend of infinitely many matches, leading to infinitely many
semi-complete polytropic EECC solutions (see Lou and Shen
2004).}\label{Match101_0_1}
\end{figure}

Behaviours of similarity MHD flow solutions around the
magnetosonic critical curve are determined by equations
$(\ref{wang53})-(\ref{wang55})$, where
the first derivatives of $v(x)$ and $\alpha(x)$ with respect to
$x$ can be determined along the magnetosonic critical curve.
Numerically, one can integrate from a point in the vicinity of the
magnetosonic critical curve away to obtain a portion of the
eigensolution crossing the magnetosonic critical curve. At one
specific point on the magnetosonic critical curve, there exist two
eigensolutions crossing the magnetosonic critical line. The one of
smaller $v'$ in the vicinity of the magnetosonic critical line is
denoted as type I solutions and the other of larger $v'$ is type
II solutions [for a sufficiently small $x$, these parallel with
type 1 and type 2 derivatives, respectively (Lou and Shen, 2004)].
These notations differ from those used in the isothermal case (Shu,
1977; Lou and Shen, 2004). We have explored MHD solutions crossing
the magnetosonic critical curve once, with several typical results
shown in Figs. \ref{Once101_0_1} to \ref{Once125_10_1}. For
mnemonics, we denote the type Y (Y=I, II) solution outward
(inward) from the $x$th ($x=1,\ 2, \ 3,\ 4$) point by $x$Yo
($x$Yi) solution. We searched similarity MHD solutions for the
cases of $\gamma=1.01$, $n=0.99$ and of $\gamma=1.25$, $n=0.75$
with various $h$ values, i.e., different reduced magnetic energy
density. We mainly focus on semi-complete physical solutions in
$0^{+}<x<+\infty$.

\begin{figure}
\includegraphics[width=3.3in,bb=100 270 480 570]{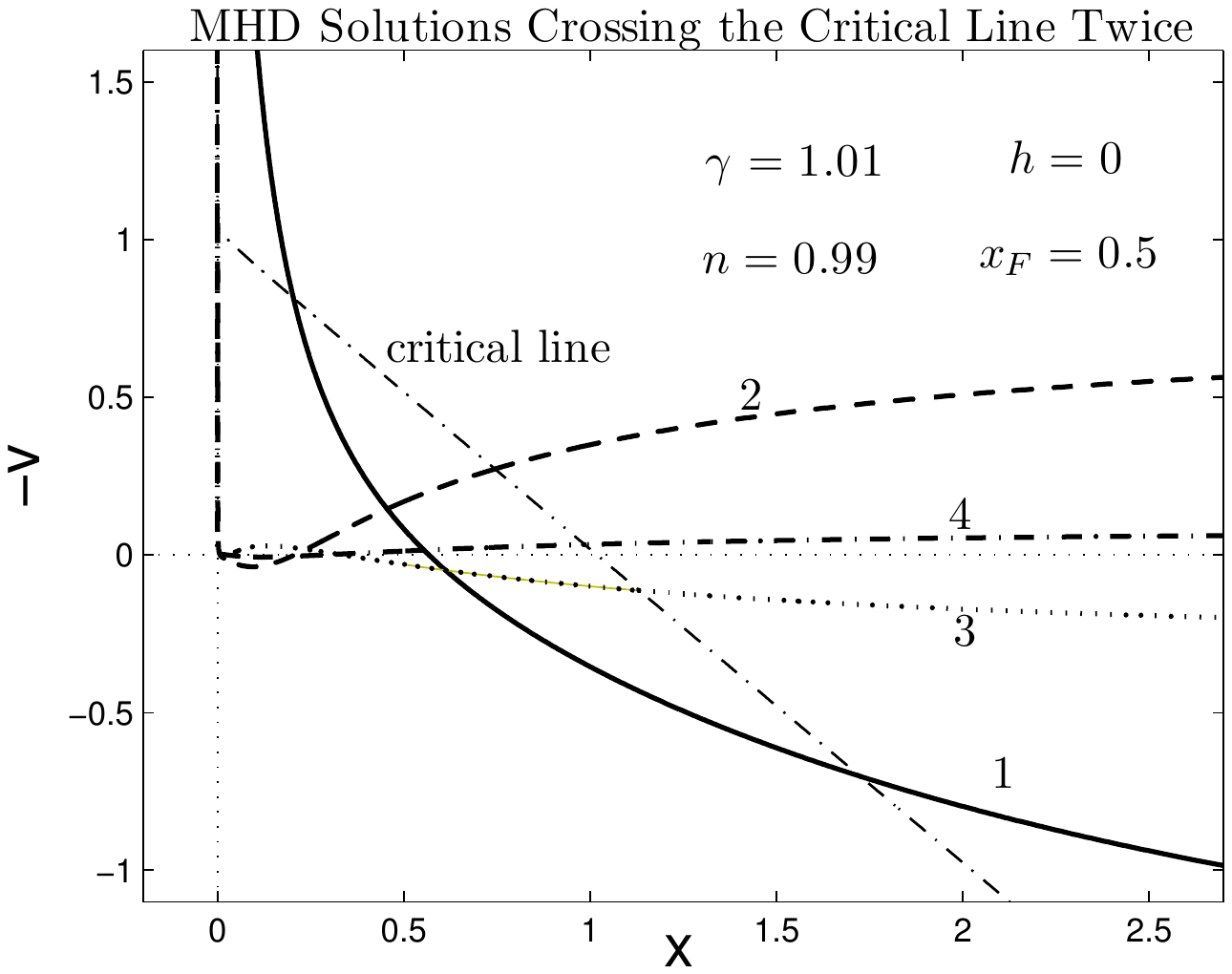}
\caption{Corresponding to the matches in the density-speed phase
diagram of Figure \ref{Match101_0_1}, the first four hydrodynamic
polytropic solutions with $\gamma=1.01$, $n=0.99$, $h=0$ and
$x_F=0.5$ are shown here. Curve 1 (solid line) is obtained from
$x=1.7248$ ($\alpha=1.15739$, $v=0.7018$) and $x=0.2035$
($\alpha=9.7036$, $v=-0.8150$) along the sonic critical curve
(this is a type I$-$type II match), curve 2 (dashed line) is
obtained from $x=0.7448$ ($\alpha=2.64742$, $v=-0.2726$) and
$x=2.2124\times 10^{-4}$ ($\alpha=9224.0$, $v=-1.0517$) along the
sonic critical curve, curve 3 (dotted line) is obtained from
$x=1.1319$ ($\alpha=1.74780$, $v=0.1128$) and $x=4.9114\times
10^{-6}$ ($\alpha=4.2354\times 10^{5}$, $v=-1.0722$) along the
sonic critical curve, and curve 4 (dash-dotted line) is obtained
from $x=0.9855$ ($\alpha=2.00440$, $v=-0.0329$) and
$x=2.8282\times 10^{-8}$ ($\alpha=7.5482\times 10^7$, $v=-1.1004$)
along the sonic critical curve. }\label{Match101_0_2}
\end{figure}

\subsubsection[]{Solutions with $\gamma=1.01$ and Small $h$ Values}

\begin{figure}
\includegraphics[width=3.3in,bb=100 270 480 570]{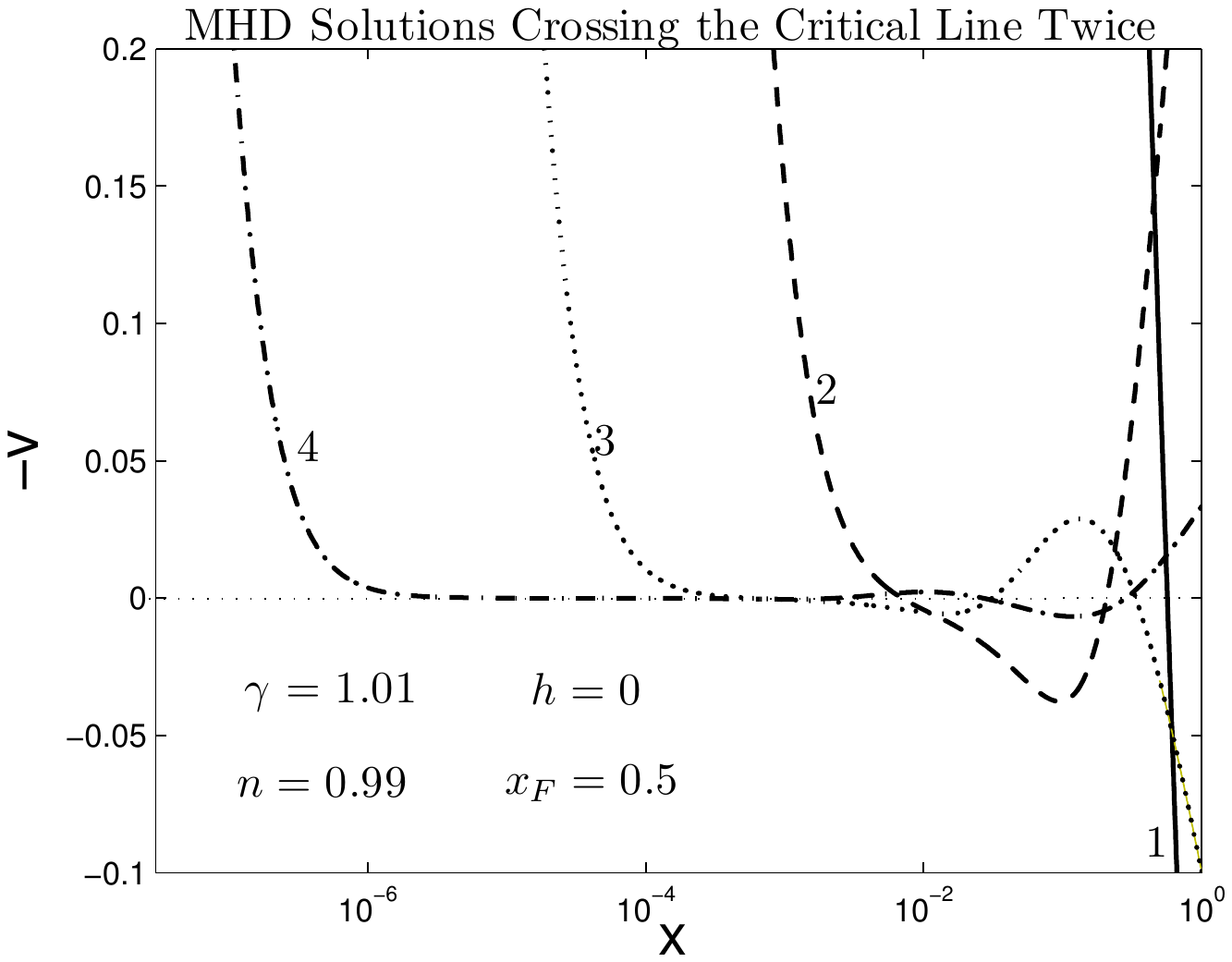}
\caption{Corresponding to Figs. \ref{Match101_0_1} and
\ref{Match101_0_2}, the enlarged versions of the first four
solutions for smaller $x$ in a logarithmic scale with
$\gamma=1.01$, $n=0.99$, $h=0$ and $x_F=0.5$ are shown with
stagnation points and radial oscillations. Curves 1, 2, 3 and 4
are the same as those in Fig. \ref{Match101_0_2}.
}\label{Match101_0_3}
\end{figure}

\begin{figure}
\includegraphics[width=3.3in,bb=100 270 480 570]{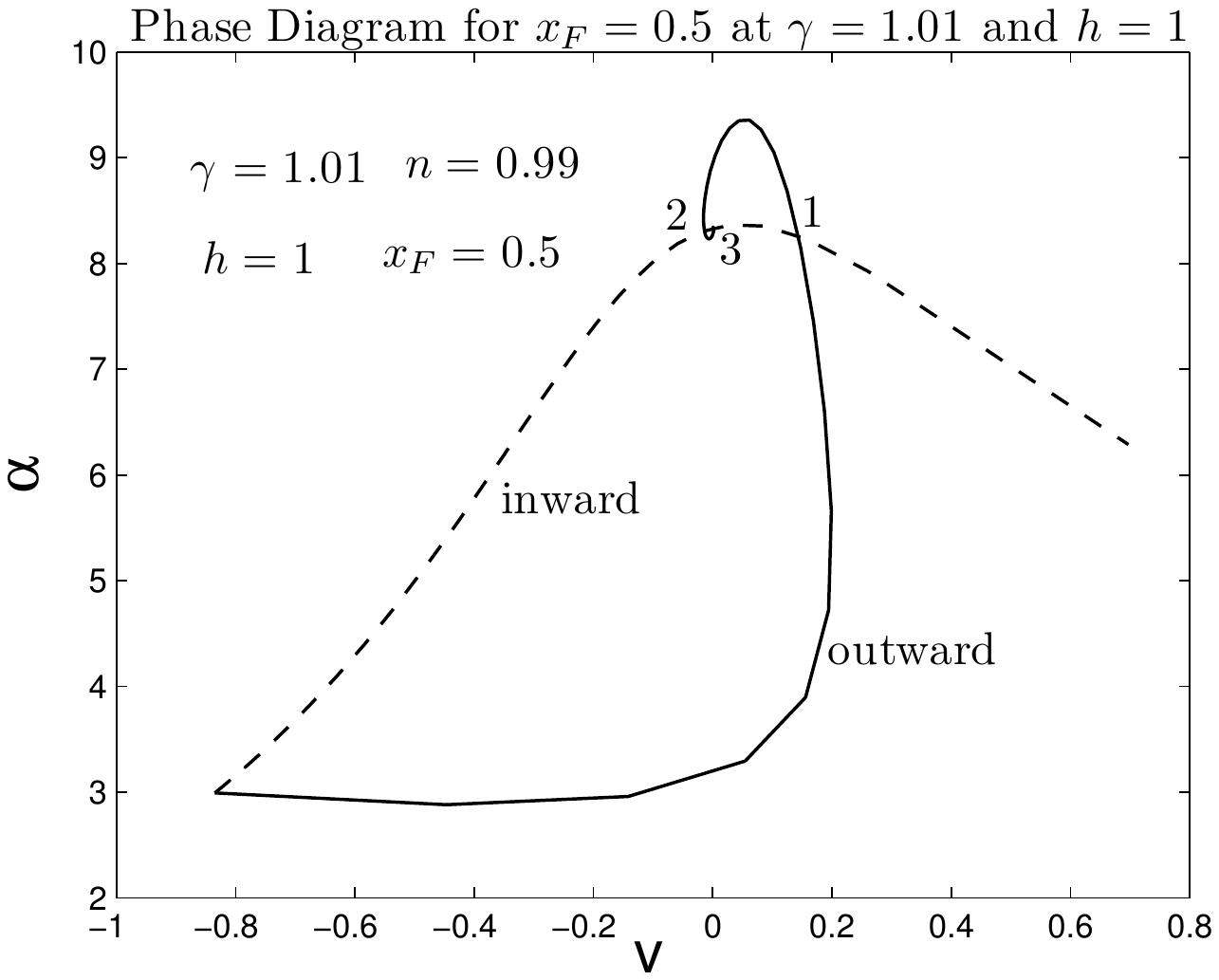}
\caption{Density-speed phase diagram for specified parameters
$\gamma=1.01$, $n=0.99$, $h=1$ and a chosen meeting point
$x_F=0.5$. The `inward curve' is the phase path obtained by
integrating inward from different $x$ ($x>x_F$) along the
magnetosonic critical curve to reach $x_F$, using the type I MHD
eigensolution, while the `outward curve' denotes the phase path
integrating from different $x$ ($x<x_F$) along the magnetosonic
critical curve to reach $x_F$, using the type II MHD eigensolution.
The spiral pattern of the `outward curve' hints at infinitely many
matches, leading to infinitely many semi-complete MHD EECC
similarity solutions. }\label{Match101_1_4}
\end{figure}

\begin{figure}
\includegraphics[width=3.3in,bb=100 270 480 570]{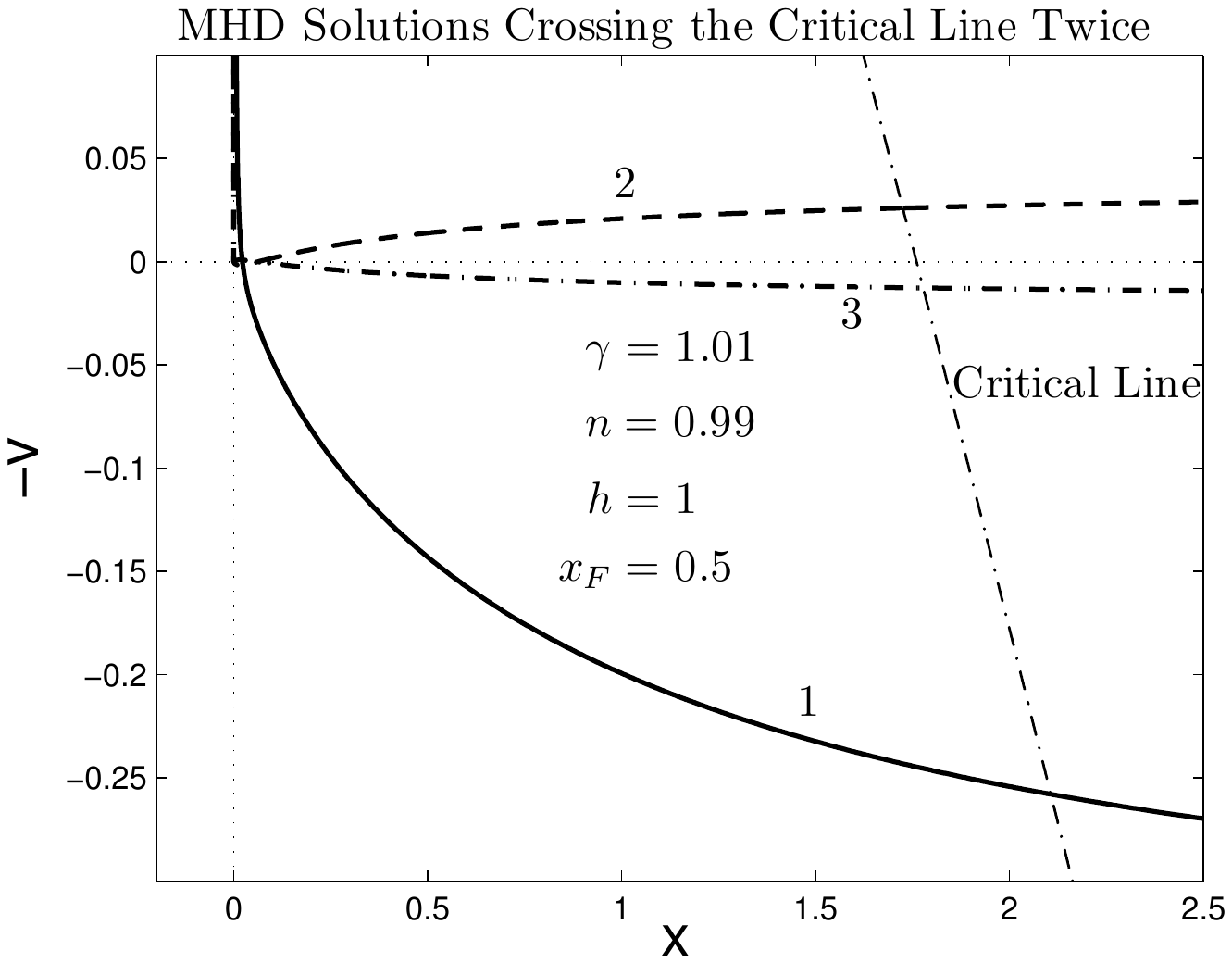}
\caption{Corresponding to the three meeting point matches in the
phase diagram of Figure \ref{Match101_1_4}, we show the first
three MHD similarity solutions with $\gamma=1.01$, $n=0.99$, $h=1$
and $x_F=0.5$. Curve 1 (solid line) is obtained from $x=2.1065$
($\alpha=0.526485$, $v=0.2580$) and $x=0.001157$ ($\alpha=1747.8$,
$v=-1.0432$) along the magnetosonic critical curve, curve 2 (dashed
line) is obtained from $x=1.7248$ ($\alpha=0.671990$, $v=-0.02598$)
and $x=5.5880\times 10^{-8}$ ($\alpha=3.8072\times 10^7$, $v=-1.0966$)
along the magnetosonic critical curve, and curve 3 (dash-dotted line)
is obtained from $x=1.7637$
($\alpha=0.653929$, $v=0.002533$) and $x=3.3391\times 10^{-12}$
($\alpha=6.6905\times 10^{11}$, $v=-1.1516$) along the
magnetosonic critical curve, respectively. For the convenience of
visual inspection, we have multiplied curve 3 by a factor of 5,
i.e., $-5v$ is shown here for curve 3. }\label{Match101_1_2}
\end{figure}

Hydrodynamic solutions with $\gamma=1.01$, $n=0.99$ and $h=0$ are
shown in Figure \ref{Once101_0_1}. Solutions 1Ii, 2Ii and 3Ii for
small $x$ values all run under the straight demarcation line
$-v=-nx$ (to the lower left of which solutions become unphysical)
and then encounter the singular surface (without crossing the
sonic critical curve); they are thus not valid at small $x$
neither mathematically nor physically. Solutions 1IIo, 2IIo and
3IIo crash on to the singular surface and are invalid at large $x$
mathematically, although one may expect special solutions crossing
the critical line twice at some specific points in a discrete
manner (see subsection \ref{twice} and Lou and Shen 2004. Solutions
1IIi, 2IIi and 3IIi approach the free-fall solution [equations
(\ref{wang27}) and (\ref{wang28})] and are valid for small $x$.
Solutions 1Io, 2Io and 3Io approach asymptotic solutions
(\ref{wang22}) and (\ref{wang23}) and are also valid at large $x$.
Solutions shown in Fig. \ref{Once101_0_1} fail to form
semi-complete hydrodynamic polytropic solutions. Likewise, MHD
solutions with the same $\gamma$ and $n$ but $h=1$ shown in Fig.
\ref{Once101_1_1} also fail to form semi-complete MHD solutions,
and the validity of each MHD solution is just like the
corresponding one in Fig. \ref{Once101_0_1}, i.e., MHD solutions
1Ii, 2Ii and 3Ii run to the lower left of the demarcation line
$-v=-nx$ and then crash on to the singular surface, solutions
1IIo, 2IIo and 3IIo crash on to the singular surface, solutions
1IIi, 2IIi and 3IIi approach the MHD free-fall solution and
solutions 1Io, 2Io and 3Io approach asymptotic MHD solutions
(\ref{wang22}) and (\ref{wang23}).

\begin{figure}
\includegraphics[width=3.3in,bb=100 270 480 570]{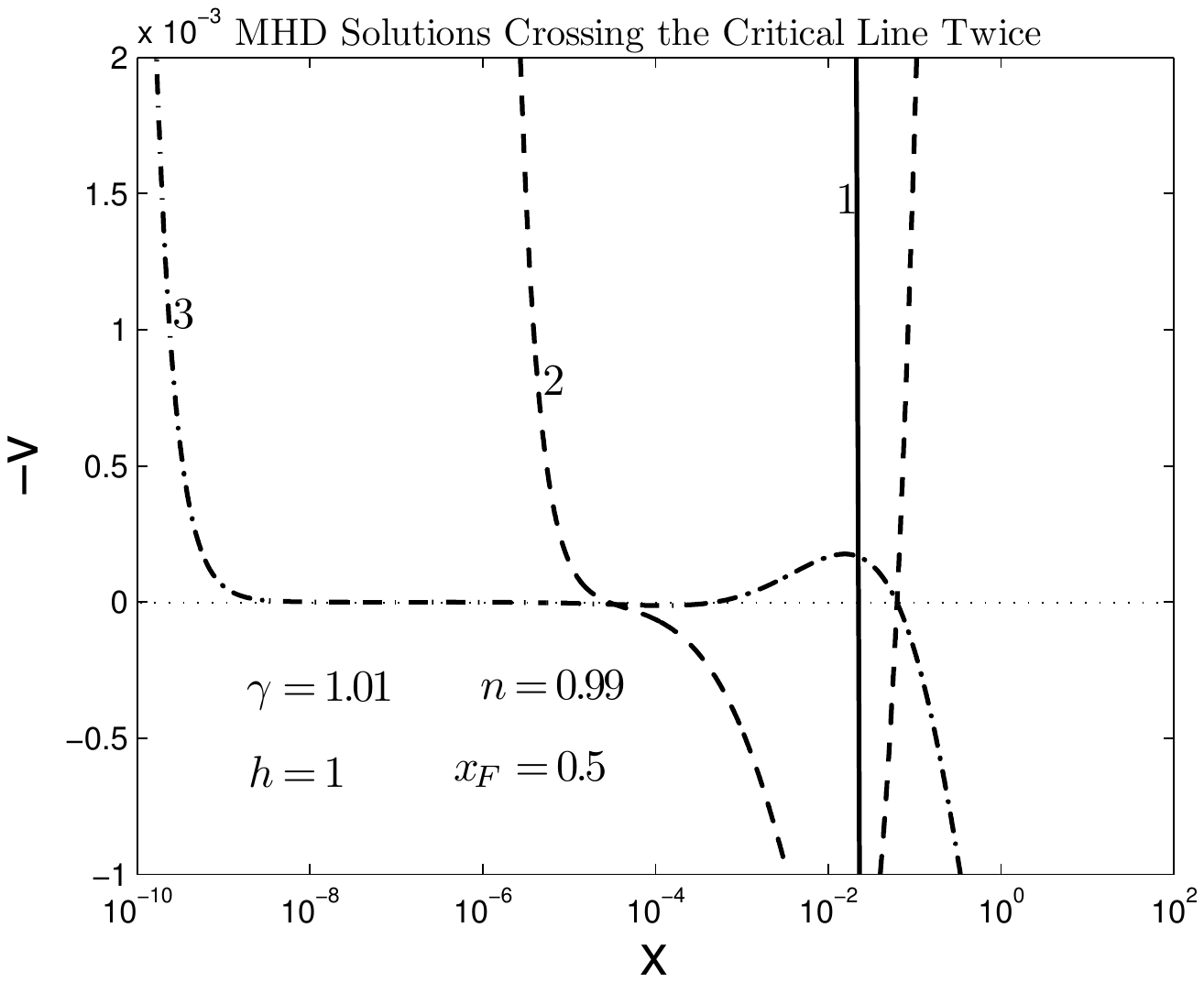}
\caption{The enlarged version of the first three MHD EECC
solutions with $\gamma=1.01$, $n=0.99$, $h=1$ and $x_F=0.5$ in a
logarithmic scale for small $x$ to show the stagnation points with
$v=0$ and self-similar magnetosonic radial oscillations. Curves 1,
2 and 3 are the same as those in Figure
\ref{Match101_1_2}.}\label{Match101_1_3}
\end{figure}

\begin{figure}
\includegraphics[width=3.3in,bb=100 270 480 570]{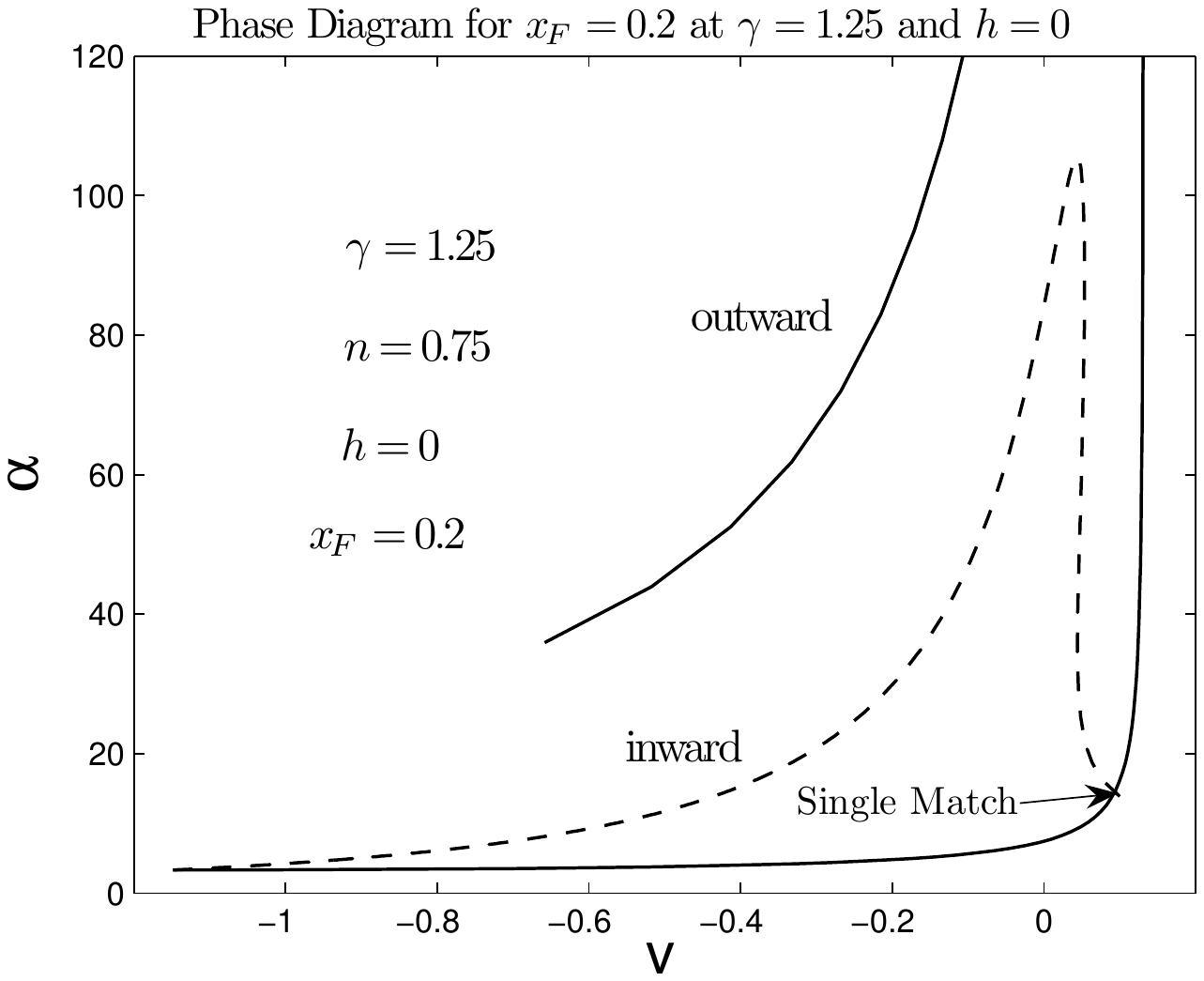}
\caption{Density-speed phase diagram for $\gamma=1.25$, $n=0.75$,
$h=0$ and $x_F=0.2$. The `inward curve' is the phase curve
obtained by integrating inward from different $x$ ($x>x_F$) along
the sonic critical curve to reach the chosen $x_F$ using the type
I eigensolution, and the `outward curve' denotes the phase path
integrating from different $x$ ($x<x_F$) along the sonic critical
curve to reach $x_F$ using the type II eigensolution. There is no
spiral pattern for the `outward curve' and only one single match
can be found at the lower-right corner.}\label{Match125_0_2}
\end{figure}

\begin{figure}
\includegraphics[width=3.3in,bb=100 270 480 570]{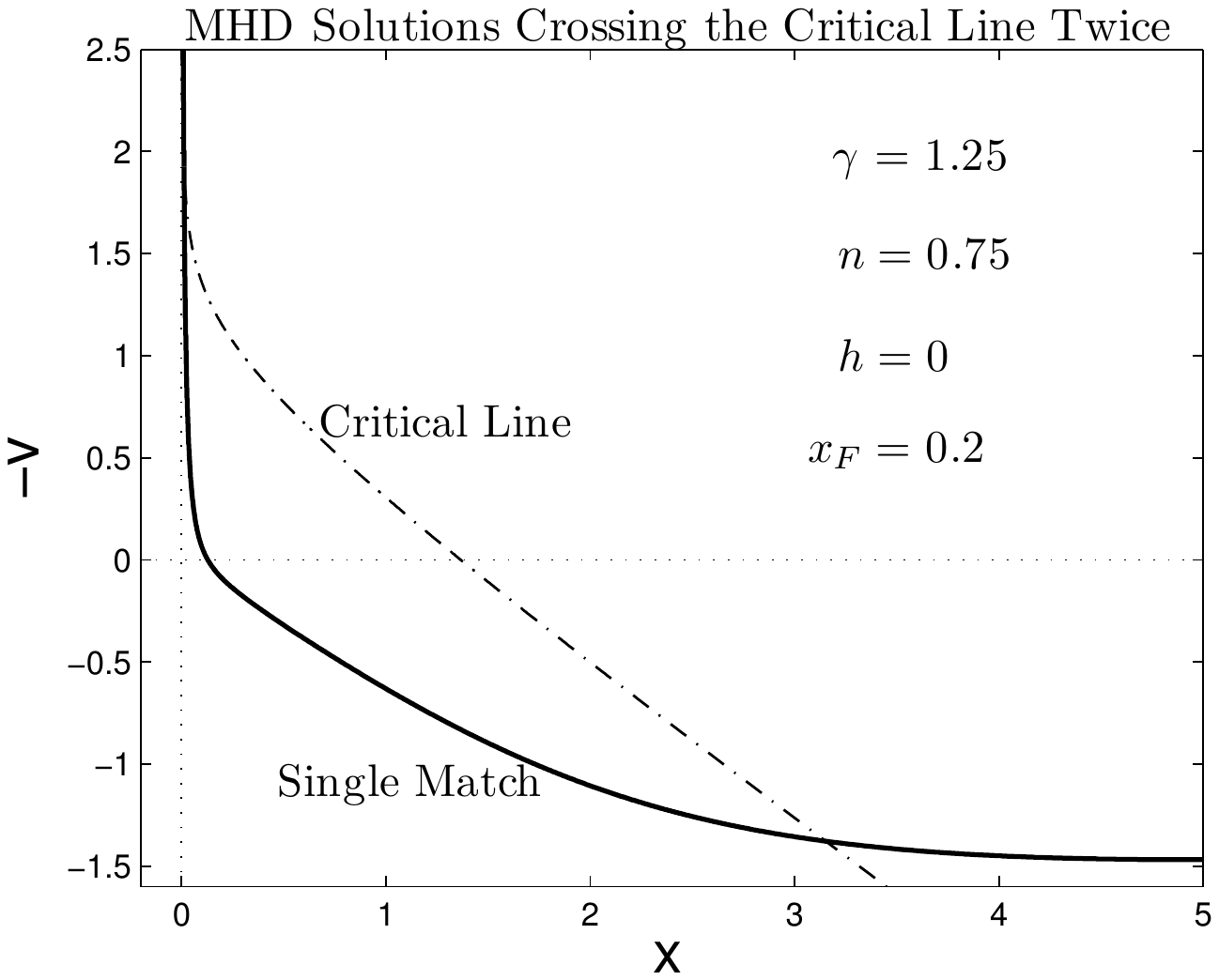}
\caption{Corresponding to the single match in the density-speed
phase diagram of Figure \ref{Match125_0_2}, the relevant
hydrodynamic polytropic solution $-v(x)$ versus $x$ (solid curve)
is shown here for $\gamma=1.25$, $n=0.75$, $h=0$ and $x_F=0.2$,
with the two crossing points given by $x=3.1529$
($\alpha=0.36876104$, $v=1.3777$) and $x=0.01197$
($\alpha=80.978$, $v=-1.9275$) along the sonic critical
curve (dash-dotted curve). }\label{Match125_0_1}
\end{figure}

\subsubsection[]{Solutions with $\gamma=1.01$ and Larger $h$ Values}

For a larger $h$ value of stronger magnetic field, we show
a few more examples of semi-complete MHD flow solutions. Fig.
\ref{Once101_10} shows the solutions with $\gamma=1.01$, $n=0.99$
and $h=10$. The solutions 1Ii, 2Ii, 3Ii in Fig. \ref{Once101_10_1}
and 1Ii in Fig. \ref{Once101_10_2} approach asymptotic MHD
solutions (\ref{wang71}) and (\ref{wang72}) with $C=7.62$ which is
the same for all four solutions here. Solutions 1Io, 2Io and 3Io
in both panels (a) and (b) approach the asymptotic forms of
(\ref{wang22}) and (\ref{wang23}) at large $x$, with the
corresponding parameters $A_0$ and $B_0$ summarized in Table
\ref{A0B0Once}. Solutions 1IIi, 2IIi and 3IIi in both panels (a)
and (b) approach the MHD free-fall solution at smaller $x$ with
$m(0)$ summarized in Table \ref{m0Once}. Solutions 1IIo, 2IIo and
3IIo in both panels (a) and (b) approach asymptotic MHD solution
(\ref{wang133}). Solutions 2Ii and 3Ii in Fig. \ref{Once101_10_2}
run to the lower left of the demarcation line $-v=-nx$ and then
crash on to the singular surface (not crossing the magnetosonic
critical curve) and are thus invalid in the small $x$ regime.
Therefore in Fig. \ref{Once101_10}, among the twelve solution
branches (I1, I2, I3, II1, II2 and II3 in both panels), there
exist in total ten semi-complete solutions, all but the two
solution branches 2I and 3I in Fig. \ref{Once101_10_2}. In Fig.
\ref{Once101_100_1}, the solutions are also all semi-complete
solutions and the corresponding parameters are tabulated in Tables
\ref{A0B0Once} and \ref{m0Once}. The corresponding $C$ parameter
in equations (\ref{wang71}) and (\ref{wang72}) for this case is
$C=95.0$. The novel asymptotic MHD solutions (\ref{wang71}) and
(\ref{wang72}) have been used here and the relevant numerical
MHD results are consistent with the analytical analysis.

\subsubsection[]{Solutions with $\gamma=1.25$ and Small $h$ Values}

For a usual polytropic gas with a different combination of larger
$\gamma$ and smaller $n$, e.g., $\gamma=1.25$ and $n=0.75$, the
situation remains qualitatively similar, but in the smaller $x$
regime with also a small $h$, the situation appears somewhat
different. Fig. \ref{Once125_0_1} presents the unmagnetized case
of $h=0$ where solutions 1Ii, 2Ii and 3Ii show self-similar
oscillation behaviours about the $x$ axis for the regime of small
$x$ (see Fig. \ref{vibrating} and the isothermal case Lou \& Shen
2004). This MHD oscillation behaviour, however, differs from that
described by Lou and Shen (2004), mainly because it is of type II
`quasi static' (i.e., the infall speed remains finite) asymptotic
solution as was recently discovered in the hydrodynamic case (Lou
and Wang, 2006); the presence of such `quasi-static' asymptotic
solutions is intimately related to the polytropic equation of
state. Given other parameters the same but for $h=2$, Fig.
\ref{Once125_2_1} displays semi-complete MHD solutions 1II, 2II
and 3II, with the inner portions approaching the MHD free-fall
state with parameter $m(0)$ summarized in Table \ref{m0Once}
and the outer portions approaching asymptotic MHD solutions
(\ref{wang133}). Meanwhile, MHD solutions 1Ii, 2Ii and 3Ii clearly
display the self-similar magnetosonic oscillation behaviours for
smaller $x$ in an analogous manner. This confirms the existence
of the `quasi-static' asymptotic behaviour in our MHD model. An
analysis of such MHD solutions is contained in Appendix
\ref{asystat}.

\subsubsection[]{MHD Solutions with $\gamma=1.25$ and Larger $h$ Values}

For an even larger $h$ such as $h=10$, one can construct
semi-complete MHD solutions for both types. Besides the similar
type of 1II, 2II and 3II solutions with parameter $m(0)$ also
contained in Table \ref{m0Once}, MHD solutions 1I, 2I and 3I are
also valid both in small and large $x$ regimes. The corresponding
$C$ in solutions (\ref{wang71}) and (\ref{wang72}) for MHD
solutions 1Ii, 2Ii and 3Ii is $C=1.47$, and the parameters $A_0$
and $B_0$ of equations (\ref{wang22}) and (\ref{wang23}) are
summarized in Table \ref{A0B0Once}. The self-similar magnetosonic
oscillation behaviour is described in section \ref{disc}. The novel
asymptotic MHD solutions also exist for $\gamma=1.25$.

\subsubsection[]{Solution Behaviours of $\alpha$ versus $x$}

Behaviours of the corresponding $\alpha(x)$ solution profile of the
MHD EECC solutions in Fig. \ref{Once101_10_1} are presented in
Fig. \ref{Once101_10_3}. Note that the $\alpha(x)$ profiles of all
three MHD solutions approach the same limiting constant
$2/[3(6h+1)]$ at large $x$ as expected. Other MHD EECC solutions
in the figures above are similar to these solutions. This type
of solutions represents a magnetized gas in spherical envelope
expansion with a constant reduced density $\alpha$ at large $x$
(initially) and a free-falling core for small $x$ (finally). For
a specified set of $\gamma$, $n$ and $h$, this series of MHD
solutions may form a one-dimensional continuum, depending on the
point where they cross the magnetosonic critical curve. In other
words, by adjusting the crossing point where such kind of MHD
solutions intersect the magnetosonic critical curve, we are able
to construct infinitely many such solutions and thus the
collection of this kind of MHD solutions forms a one-dimensional
continuum. We emphasize that this series solutions constructed are
closely related to the random magnetic field in the problem.
Complementary to the hydrodynamic EECC solutions (Lou and Shen,
2004), the new MHD EECC solutions in this subsection have
different asymptotic behaviours at large $x$. The fact that
solution (\ref{wang35}) also serves as an asymptotic behaviour at
large $x$ [see equation (\ref{wang133})] should be emphasized
here; we further confirm the validity of this asymptotic MHD
solution by calculating one more term in the series expansion
of this asymptotic MHD solution.

\subsubsection[]{MHD Hunter Type Solutions}

Another type of MHD solutions crossing the magnetosonic critical
curve once is here referred to as the Hunter type solution (Hunter,
1977) generalized to a magnetized polytropic gas and parallel to
isothermal examples shown in figure 6 of Lou and Shen (2004). We
obtain these MHD series solutions by solution matching in the
phase diagram at a chosen meeting point $x=x_F$ where these two
solutions have the same values of $\alpha$, $v$, $\alpha'$ and
$v'$. In reference to EECC solutions in the isothermal and
unmagnetized formulation (Lou and Shen, 2004), we expect that for
$\gamma$ close to unity and $h$ not very large, MHD
generalizations of this Hunter type of solutions should exist. We
explored the case of $\gamma=1.01$, $n=0.99$ and $h=1$, and chose
a meeting point at $x_F=0.5$. The matching in the phase diagram is
displayed in Figure \ref{hunter101_1_1} with a familiar spiral
pattern, indicating that there may exist infinitely many matches,
corresponding to infinitely many discrete semi-complete MHD
solutions of this type. Figure \ref{hunter101_1_2} presents the
first three such MHD solutions and Figure \ref{hunter101_1_3} is
an enlarged version of Figure \ref{hunter101_1_2} revealing that
the number of stagnation points increases for these solutions.
These features show self-similar magnetosonic oscillations in a
magnetized polytropic gas. These similarity solutions are the MHD
counterparts of the isothermal hydrodynamic solutions (Hunter,
1977; Lou and Shen, 2004).

\begin{figure}
\includegraphics[width=3.3in,bb=100 270 480 570]{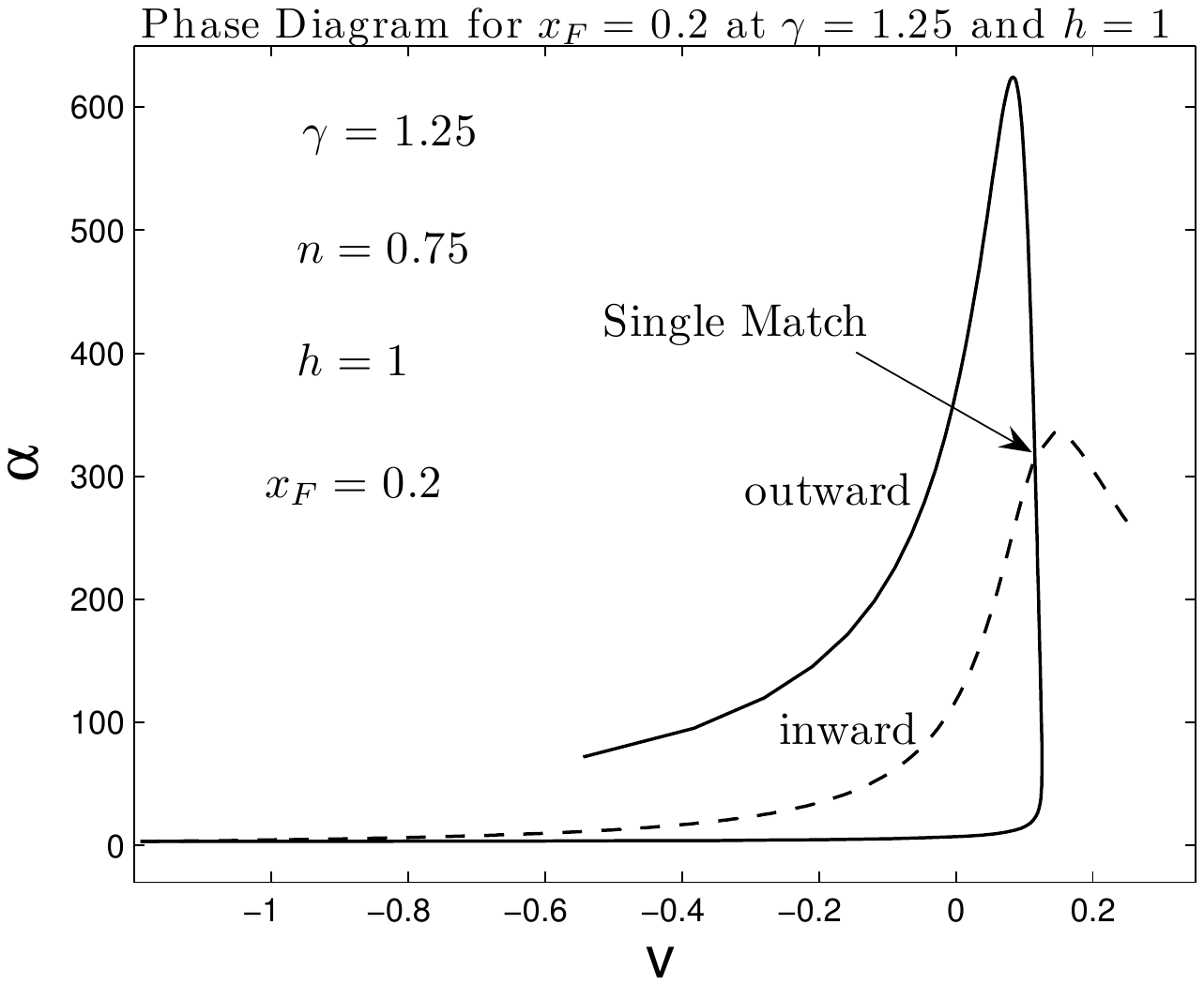}
\caption{Density-speed phase diagram for $\gamma=1.25$, $n=0.75$,
$h=1$ and a chosen meeting point $x_F=0.2$. The `inward curve' is
the phase path by integrating inward from different $x$ ($x>x_F$)
along the magnetosonic critical curve to reach a chosen $x_F=0.2$
using the type I MHD eigensolution, and the `outward curve'
denotes the phase path integrating from different $x$ ($x<x_F$)
along the magnetosonic critical curve to reach $x_F$ using the
type II MHD eigensolution. There is no spiral pattern for the
`outward curve' and only one single match
exists.}\label{Match125_1_1}
\end{figure}

\subsection{Similarity Solutions Crossing the
Magnetosonic Critical Curve Twice}\label{twice}

There exist semi-complete MHD solutions crossing the magnetosonic
critical curve twice. We explore them using the procedure of Lou
and Shen (2004). The advantage of being able to reduce from three
coupled nonlinear MHD ODEs to two will become apparent as we
manage to match solutions in a two-dimensional phase space of
$v(x)$ and $\alpha(x)$.

We chose different meeting points, denoted by $x_F$, in the cases
of $\gamma=1.01$ and $\gamma=1.25$, for a usual polytropic gas
with $n=2-\gamma$. In general, higher values of $h$ do not lead
to matches in the range $0.01-0.8$ for $x_F$ values that we have
systematically searched. For $\gamma=1.01$ and smaller $h$ or
$h=0$, we find that there are likely to be infinitely many matches
as in Lou and Shen (2004) by empirical inferences, while for
$\gamma=1.25$, we could only find a single match for small $h$ or
$h=0$ by a careful numerical search and by observing the variation
trend of phase paths.

\subsubsection[]{Solutions with $\gamma=1.01$ and Small $h$ Values}

Figs. \ref{Match101_0_1} through \ref{Match101_0_3} present the
case of $\gamma=1.01$, $n=0.99$ and $h=0$ for a nearly isothermal
gas without a random magnetic field. At the meeting point
$x_F=0.5$, we find that the `outward curve' obtained by
integrating from small $x$ along the sonic critical curve to
reach a chosen meeting point $x_F$ has a spiral pattern (Fig.
\ref{Match101_0_1}) indicating the trend for discrete yet
infinitely many semi-complete solutions that cross the sonic
critical curve twice. We also find that the number of stagnation
points where a solution intersects the $x$-axis with $v=0$
increases for successive matches along the `outward curve', i.e.,
solution 1 has one stagnation point, and solution 2 has two
stagnation points, and so forth (Fig. \ref{Match101_0_3}). This
feature is very similar to the type 1$-$type 2 match solution in
Lou and Shen (2004). Asymptotic behaviours of the four solutions
obtained here are similar: in the small $x$ regime, they approach
the free-fall solutions (\ref{wang27}) and (\ref{wang28}), while
in the large $x$ regime, they approach asymptotic solutions
(\ref{wang22}) and (\ref{wang23}). The corresponding parameters
$A_0$, $B_0$ and $m(0)$ are summarized in Table \ref{para101_0}.
The number of solutions that one can construct depends upon the
numerical accuracy. With a numerical precision of a relative error
$10^{-14}$, we have found at least five matches in a systematic
and careful numerical exploration.

Figures \ref{Match101_1_4} through \ref{Match101_1_3} show
relevant results for an example of $\gamma=1.01$, $n=0.99$ and
$h=1$, a nearly isothermal and weakly magnetized polytropic gas,
which should be similar to the corresponding unmagnetized
isothermal case (Lou and Shen, 2004). Note that the third curve in
Fig. \ref{Match101_1_2} is multiplied by a factor of 5 for the
compactness and clarity of the presentation. In constructing the
phase diagram, we again find a familiar spiral pattern as in the
$h=0$ case, and properties of stagnation number and asymptotic
behaviours are all qualitatively similar. The corresponding
parameters $A_0$, $B_0$ and $m(0)$ are summarized in Table
\ref{para101_1}. Figures \ref{Match125_0_2} and \ref{Match125_0_1}
present the case of $\gamma=1.25$, $n=0.75$ and $h=0$ for an
unmagnetized polytropic gas. Within our numerical accuracy, we did
not find a spiral phase pattern and there exists only one single
match in this case. This match is very much like solution 1 in the
nearly isothermal case without a random magnetic field.

\subsubsection[]{Solutions with $\gamma=1.25$ and Small $h$ Values}

\begin{figure}
\includegraphics[width=3.3in,bb=100 270 480 570]{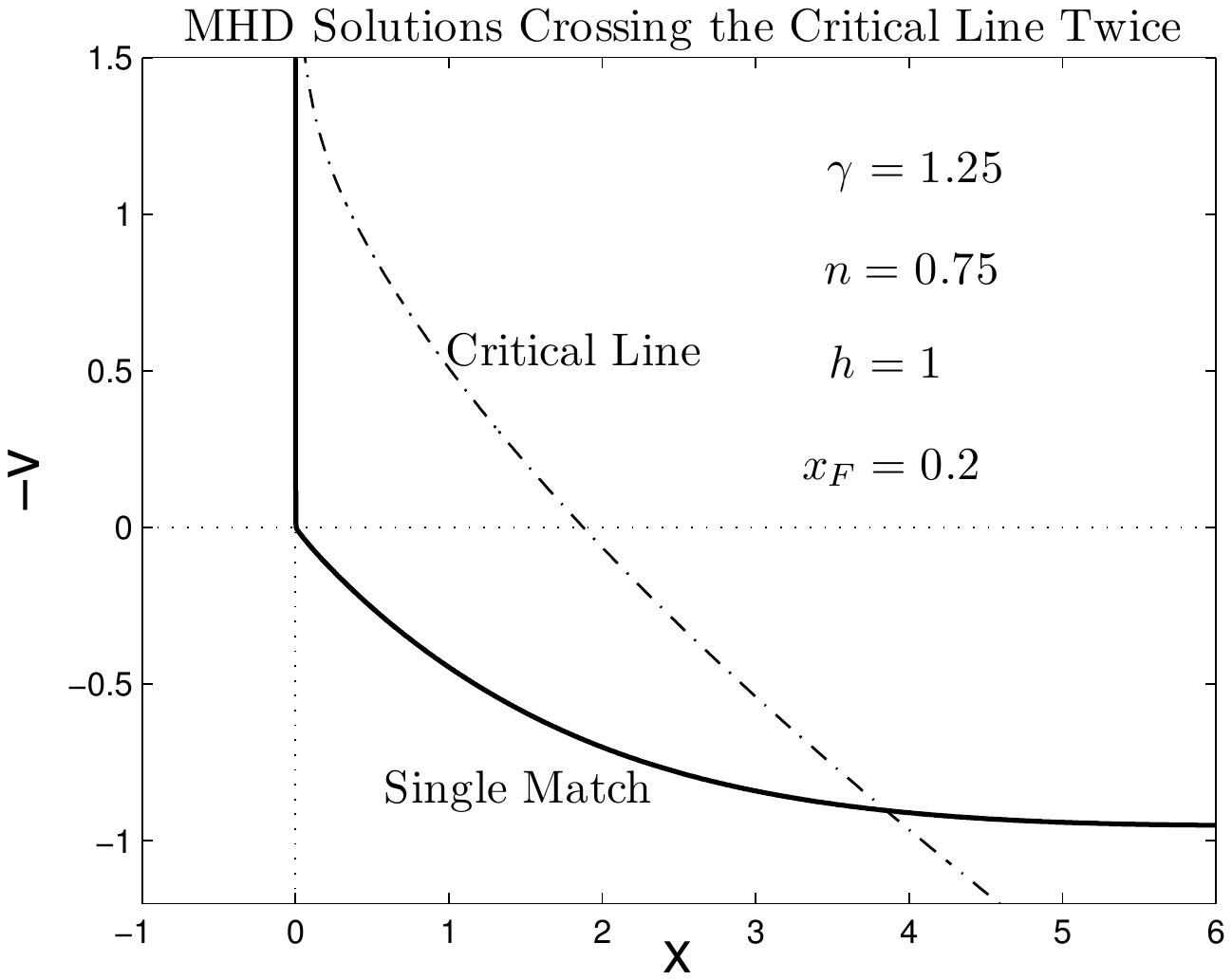}
\caption{The single match of the similarity MHD solution for
$\gamma=1.25$, $n=0.75$, $h=1$ and a chosen meeting point
$x_F=0.2$, obtained with the two magnetosonic crossing points
located at $x=3.8496$ ($\alpha=0.2087287$, $v=0.9027$) and
$x=5.6384\times 10^{-5}$ ($\alpha=36912$, $v=-4.1624$) along the
magnetosonic critical curve. This match is shown in the phase
diagram of Fig. \ref{Match125_1_1}. }\label{Match125_1_2}
\end{figure}

Figures \ref{Match125_1_1} and \ref{Match125_1_2} present the case
of $\gamma=1.25$, $n=0.75$ and $h=1$ for a weakly magnetized
polytropic gas, which is fairly similar to the unmagnetized case
of $h=0$. The asymptotic behaviours in both cases are similar to
previous nearly isothermal cases and the corresponding parameters
$A_0$, $B_0$ and $m(0)$ are contained in Table \ref{para125}.

\subsubsection[]{Solutions with Larger $h$ Values}

For a stronger magnetic field with a larger $h$, no solution
match can be found in the density-speed phase diagram. Figures
\ref{Match101_10_1} and \ref{Match125_10_1} are two typical phase
diagrams showing the two cases of $\gamma=1.01$, $n=0.99$, $h=10$
and of $\gamma=1.25$, $n=0.75$ and $h=10$, respectively. A
possible interpretation of this feature is perhaps that for a
large $h$, one can construct semi-complete solutions crossing
the magnetosonic critical curve once as discussed in Section
\ref{once}, while any outward solution curve will not reach the
magnetosonic critical curve but continue to its lower left to
infinity to match the asymptotic MHD solution (\ref{wang133}).

\section[]{Summary and Discussion}\label{disc}

\begin{table}
\center\caption{The three corresponding parameters $A_0$, $B_0$
and $m(0)$ used in Figure \ref{Match101_0_2} for hydrodynamic
polytropic EECC solutions which are nearly isothermal.
}\label{para101_0} \vskip 0.3cm
\begin{tabular}{cccc}\hline
\multicolumn{4}{c}{Values of $\gamma=1.01$ and $h=0$.} \\
\hline\hline No. & $A_0$ & $B_0$ & $m(0)$ \\ \hline
1&5.200&1.883   &0.3661 \\
2&1.200&-0.7958 &$4.7489\times 10^{-4}$\\
3&2.415&0.3145  &$1.096\times 10^{-5}$ \\
4&1.904&-0.09298&$6.6437\times 10^{-8}$\\ \hline
\end{tabular}
\end{table}

\begin{table}
\center \caption{The three corresponding parameters $A_0$, $B_0$
and $m(0)$ in Fig. \ref{Match101_1_2} for MHD polytropic solutions
which are nearly isothermal.}\label{para101_1}
\vskip 0.3cm
\begin{tabular}{cccc}\hline
\multicolumn{4}{c}{Values of $\gamma=1.01$ and $h=1$.}\\
\hline\hline No. & $A_0$ & $B_0$ & $m(0)$ \\ \hline
1&2.420 &0.3719  &$2.441\times 10^{-3}$ \\
2&2.0155&-0.03997&$1.304\times 10^{-7}$ \\
3&2.058 &$3.868\times 10^{-3}$&$8.590\times 10^{-12}$\\ \hline
\end{tabular}
\end{table}

\begin{table}
\center \caption{The three corresponding parameters $A_0$, $B_0$
and $m(0)$ in Figs. \ref{Match125_0_1} and \ref{Match125_1_2}.
}\label{para125}
\vskip 0.3cm
\begin{tabular}{cccc}\hline
\multicolumn{4}{c}{Specified parameters
are $\gamma=1.25$ and $n=0.75$. }\\
\hline\hline $h$ & $A_0$ & $B_0$ & $m(0)$\\ \hline
0&4.399&3.132&$2.243\times 10^{-2}$ \\
1&5.894&2.267&$4.885\times 10^{-4}$\\ \hline
\end{tabular}
\end{table}
\vskip 0.3cm

We developed an MHD formulation to describe a quasi-spherical
symmetric gas cloud obeying a conventional polytropic equation
of state in the presence of a random magnetic field and searched
for various possible semi-complete MHD similarity solutions
within this model framework. Here, we mainly focus on MHD
effects of a random magnetic field; other effects such as
diffusions and radiative diagnostics will be pursued in
separate papers.

This formulation reveals several key MHD features of a gas
cloud when a polytropic equation of state and a random
magnetic field are combined together.
First, the magnetosonic critical curve diverges when $x$
approaches zero, and may head up to the first quadrant in the $-v$
versus $x$ presentation when $h$ becomes sufficiently large. In
the absence of a random magnetic field, the sonic critical curve
also diverges as $x$ approaches zero. In the presence of a random
magnetic field, the isothermal magnetosonic critical curve remains
finite at $x=0$ (Yu and Lou, 2005; Yu et al., 2006). Therefore,
this small $x$ diverging behaviour is fundamentally related to the
polytropic approximation in contrast to the isothermal
approximation.
Secondly, for a sensible range of magnetic field strengths, the
so-called mEWCSs may describe the large-scale MHD evolution of a
magnetized gas cloud. But for too strong a random magnetic field,
such a mEWCS does not exist. In this regard, our formalism is
naturally applicable to a weakly magnetized gas cloud in
large-scale outward and inward motions. For $B_0=0$, a magnetized
gas cloud may have a tendency to expand as $h$ becomes
sufficiently large.
Thirdly, there exist additional MHD EECC solutions that cross the
magnetosonic critical curve once with asymptotic MHD solution
(\ref{wang133}) at large $x$; these possible asymptotic MHD
expansion solutions at large $x$ were not realized before (Lou and
Shen, 2004; Hunter, 1977). We note that semi-complete or shock
solutions matching with such asymptotic polytropic expansion
solutions may be constructed without a random magnetic field with
$h=0$.
Fourthly, there exists a novel series asymptotic MHD solution
(\ref{wang71}) and (\ref{wang72}) for small $x$ with a very dense
core as well as a strong central magnetic field, and one can
construct semi-complete self-similar MHD solutions using this
asymptotic MHD solution.
Finally, for $\gamma=1.25$ as an example of illustration, we have
not observed the trend of spiral pattern in the density-speed
phase diagram for solution matches. Therefore, there seem to be no
trend of infinitely many semi-complete solutions crossing the
critical curve twice under certain situations.

Our model analysis for a magnetized polytropic gas cloud requires
the following assumptions: the magnetic field is locally random on
small scales with a large-scale quasi-spherical symmetry
Zel'dovich and Novikov (1971), the profiles of mass density, radial
velocity, gas temperature and thermal pressure are all taken to
possess a quasi-spherical symmetry on large scales.
We impose these assumptions as the first approximation for a
quasi-spherical magnetized gas medium, and we try to assess the
role of a random magnetic field under such assumptions
quantitatively and qualitatively. Our model analysis is specific
and in details, while for astrophysical applications, we gain a
physical sense qualitatively and should be careful with the
adopted approximation and thus the model limitations. An important
conceptual issue should be noted here. Even for a hydrodynamic
flow of spherical symmetry in the absence of a random magnetic
field, something must happen around the centre to destroy the
self-similarity and the spherical symmetry except that a central
black hole may accrete gas materials and absorb them in a smooth
and quiet manner (e.g., Cai and Shu, 2005). For any other central
objects such as neutron stars, white dwarfs, or main-sequence
stars, spherical symmetric accretions will unavoidably lead to
central activities. Therefore, a self-similar evolution of
hydrodynamic gas flows is possible outside the influence sphere of
central activities. Parallel to this physical rationale, an MHD
inflow of quasi-spherical symmetry in the presence of a random
magnetic field will eventually give rise to a central sphere of
MHD activities. Again, we expect that a self-similar evolution of
MHD gas flows is possible outside this influence sphere of central
MHD activities.

In this analysis, we have adopted a polytropic equation of state
(\ref{wang132}). We note the existence of two different
definitions for the term `polytropic': one corresponds to an
overall equation of state $p=\kappa\rho^\gamma$ with a constant
$\kappa$ throughout the dynamic evolution in spatial and temporal
domains (Goldreich and Weber, 1980), while the other corresponds
to the entropy conservation along the streamline
\begin{equation}\label{wang130}
\left(\frac{\partial}{\partial t} +u\frac{\partial}{\partial
r}\right) \left[\log\left(\frac{p}{\rho^\gamma}\right)\right]=0\ ,
\end{equation}
indicating that every infinitesimal portion of gas obeys the
polytropic state equation along each streamline, but not
necessarily in the global sense (e.g., Bouquet et al., 1985), i.e.,
the entropy per unit mass assigned to each streamline can be
different. In the present flow problem, all streamlines are radial
and a constant $\kappa$ for all streamlines would certainly meet
the requirement of a quasi-spherical symmetry. In other words, we
have taken the former, yet we recognize that if the latter is
adopted, the parameter $n$ would be a free parameter, which may be
adjusted to fit the requirements of, for examples, initial density
profiles, initial `equations of state' or asymptotic finite radial
speeds at large $x$. By taking the former definition of a
`polytropic' gas, our analysis is primarily for the case of
$n=2-\gamma$. In fact, Suto and Silk (1988) set the equation of
state to be
\begin{equation}\label{wang134}
p=\kappa(t)\rho^{\gamma}\ ,
\end{equation}
where $\kappa(t)$ [corresponding to the notation $K(t)$ of Suto
and Silk (1988)] takes the form of a power law in time $t$ to
accommodate certain unknown energetic processes. This form of
$\kappa(t)$ depending on $t$ constrains the range of index
$\gamma$. If we adopt the usual polytropic equation of state from
the perspective of Suto and Silk (1988), parameter $n$ should take
on the value of $2-\gamma$. If instead, $\kappa$ is allowed to
assume the form of power laws in both $t$ and $r$ and equation
(\ref{wang134}) is adopted, one more free parameter (equivalent to
our parameter $n$ here) would emerge, as has been done for example
by Cheng (1978) and Fatuzzo et al. (2004). For a careful study on
behaviours of `polytropic' gas under the latter definition, we
should allow for the freedom of parameters in this problem and
treat $\kappa$ as a function of both $r$ and $t$. In short, we
have treated here only the case of $\gamma=2-n$ rather than $n=1$
(Suto and Silk, 1988).

Goldreich and Weber (1980) studied the specific case of
$\gamma=4/3$, invoking this model to describe a homologous
evolution of the core collapse of a supernova progenitor. Yahil
(1983) noted that even for a slight departure of $\gamma$ value
from the exact value $4/3$, the core collapse will be no longer
completely homologous. At this stage, the results of Goldreich and
Weber (1980) are beyond the framework of our study, because for
$\gamma=4/3$, our similarity MHD transformation (\ref{wang19})
becomes invalid. With this qualification, we briefly discuss
limiting features of our model analysis as $\gamma$ approaches
$4/3$. In our model framework, especially in reference to
nonlinear MHD ODEs (\ref{wang17}) and (\ref{wang18}), parameter
$n$ will approach $2/3$ as $\gamma\rightarrow 4/3$; it then
follows that the factor $(nx-v)/(3n-2)$ goes to infinity, except
when $v$ approaches $2x/3$ simultaneously, corresponding to a
homologous collapse. Since first derivatives $v'$ and $\alpha'$ in
terms of $x$ should remain finite in both MHD ODEs (\ref{wang17})
and (\ref{wang18}), we conclude that as $n$ approaches $2/3$, we
must require $v=2x/3$ in order to avoid infinite first derivatives
of both $v$ and $\alpha$. Therefore in our MHD case, a homologous
collapse should still exist (see Yahil 1983 for an analysis of
departure from $\gamma=4/3$). This argument serves only as a quick
analysis and detailed calculations will be pursued in a separate
paper.

The consequences of adopting a polytropic equation of state and
incorporating a random magnetic field in our formulation of gas
collapses and outflows lead to several novel features of possible
MHD solutions in comparison to earlier analyses. Firstly, the new
asymptotic MHD solutions (\ref{wang71}) and (\ref{wang72})
describe a gas cloud in which a random magnetic field and gas
materials are both compressed in the core region with a very high
total pressure in the core. This asymptotic MHD solution may
represent a self-similar dynamic evolution course that forms a
compact stellar object that contains extremely intense magnetic
field and high mass density. The formation of this type of objects
requires an embedded random magnetic field to be strong enough,
because for $h=0$ this asymptotic solution for small $x$ does not
exist (see Yu et al. 2006 for this type of asymptotic MHD
solutions in an isothermal gas medium). Secondly, with or without
a random magnetic field, the EECC solutions crossing the critical
curve once provides a possibility that a gas cloud with a
collapsing core has a much faster expanding envelope at a high
radial speed. A more general treatment of flows at relativitistic
speeds under a strong gravity would involve the general theory of
relativity; in fact, the expanding envelope represents a kind of
nonrelativistic Einstein-de Sitter expansion (Shu et al., 2002; Lou
and Shen, 2004; Yu and Lou, 2005). Here, we further speculate the
possibility that a random magnetic field and a polytropic equation
of state might be relevant to the large-scale expansion of the
entire universe involving a weak and random magnetic field.
Finally, although MHD EECC solutions crossing the magnetosonic
critical curve once can be constructed with a strong magnetic
field, it can happen that the kind of EECC solutions crossing the
critical curve twice (Lou and Shen, 2004) does not exist. The point
is that the two MHD solutions having different asymptotic forms
may not connect each other. In other words, it seems to be the
case that for a very strong magnetic field, there are no
semi-complete solutions to approach asymptotic MHD solutions
(\ref{wang22}) and (\ref{wang23}) at large $x$. This is to say
that for these cases, the MHD gas flow system does not have a
non-increasing self-similar radial flow profile.

We now describe and comment on the self-similar oscillatory
behaviour about the abscissa $x-$axis described in subsection
\ref{once}, i.e., of the numerical solutions 1Ii, 2Ii and 3Ii in
Figs. \ref{Once125_0_1} and \ref{Once125_2_1} for the hydrodynamic
and MHD cases, respectively. Typically, the $-v$ versus $x$
solution curve behaves in the following manner. In the range of
small $x$ values, $-v$ appears to cross the $x$-axis repeatedly
and regularly for a number of infinite times, each cycle with a
period decreasing with smaller $x$, and the vibrating amplitude
also decreases as $x$ decreases in an apparent power-law fashion
(see Fig. \ref{vibrating}). We explored this feature down to
$x\geq 10^{-20}$ and did not find the curve to crash on to the
singular surface. An illustrating example of this vibration
behaviour is clearly shown in Fig. \ref{vibrating}. This behaviour,
after careful examination and comparison of numerical and
analytical results, turns out to be the type II `quasi-static'
asymptotic solution derived by Lou and Wang (2006) where the
hydrodynamic problem in the absence of magnetic field was solved.
Appendix \ref{asystat} gives a brief account of the basic results.
This behaviour is a characteristic of the polytropic equation of
state with or without magnetic field. However as shown in Appendix
\ref{asystat}, the presence of a magnetic field modifies the
asymptotic solution behaviour in a non-trivial manner.

\begin{acknowledgements}
This research has been supported in part by the ASCI Center for
Astrophysical Thermonuclear Flashes at the University of Chicago,
by the Tsinghua Center for Astrophysics, by the Collaborative
Research Fund from the National Science Foundation of China (NSFC)
for Young Outstanding Overseas Chinese Scholars (NSFC 10028306)
at the National Astronomical Observatories, Chinese Academy of
Sciences, by the NSFC grants 10373009 and 10533020 at the
Tsinghua University, and by the SRFDP 20050003088
and by the Yangtze Endowment from the Ministry of Education at
the Tsinghua University. The hospitalities of the Mullard Space
Science Laboratory at University College London, U.K., of School
of Physics and Astronomy, University of St Andrews, Scotland,
U.K., and of Centre de Physique des Particules de Marseille
(CPPM/IN2P3/CNRS) et Universit\'e de la M\'editerran\'ee
Aix-Marseille II, France are also gratefully acknowledged.
\end{acknowledgements}

\appendix

\section[]{Magnetic Induction and the Lorentz Force}
\label{QSphere}

In this appendix, we rearrange the magnetic induction equation and
discuss approximations made regarding the magnetic Lorentz force
in our MHD formulation with the simplification of a
quasi-spherical symmetry on large scales (Yu and Lou, 2005; Yu et
al., 2006).

For non-relativistic quasi-neutral MHD flows, the magnetic Lorentz
force density can be split into two terms
\begin{equation}\label{lou1}
\frac{(\nabla\times \overrightarrow{B})\times
\overrightarrow{B}}{4\pi}=-\nabla\bigg(\frac{B^2}{8\pi}\bigg)
+\frac{(\overrightarrow{B}\cdot\nabla)\overrightarrow{B}}{4\pi}\
\end{equation}
by a vector identity and the $\nabla\cdot\vec B=0$ condition. The
first and second terms on the RHS represent the magnetic pressure
and tension force densities.
The radial component of the magnetic tension force on the RHS is
\begin{equation}\label{lou2}
\frac{1}{4\pi}\bigg[B_r\frac{\partial B_r}{\partial r}
+\frac{B_\theta}{r}\frac{\partial B_r}
{\partial\theta}+\frac{B_\phi}{r\sin\theta}\frac{\partial B_r}
{\partial\phi}-\frac{(B_\theta^2+B_\phi^2)}{r}\bigg]\ ,
\end{equation}
where the first term in the square brackets cancels a relevant
term of the magnetic pressure force in expression (\ref{lou1}).
For a large-scale average of a random magnetic field, we assume
$<B_\theta\cdot\partial B_r/\partial r>\cong 0$ and
$<B_\phi\cdot\partial B_r/\partial\theta>\cong 0$. The
nonvanishing radial component of the magnetic Lorentz force
density associated with the mean square of the random transverse
magnetic field is
\begin{equation}\label{lou3}
-\frac{\partial}{\partial r}
\bigg(\frac{<B_t^2>}{8\pi}\bigg)-\frac{<B_t^2>}{4\pi r}\
\end{equation}
as shown in equation (\ref{wang4}).

In the absence of resistivity and ambipolar diffusions etc., the
magnetic induction equation appears as
\begin{equation}\label{lou4}
\frac{\partial\overrightarrow{B}}{\partial t}
=\nabla\times(\overrightarrow v\times \overrightarrow B)\ ,
\end{equation}
which can be written explicitly in three component forms in
spherical polar coordinates $(r,\theta,\phi)$, viz.,
\begin{eqnarray}\label{lou5}
\frac{\partial B_r}{\partial t}
&=&\frac{1}{r\sin\theta}\frac{\partial}
{\partial\theta}\bigg[\sin\theta(v_r B_\theta-v_\theta
B_r)\bigg]\nonumber\\
&&-\frac{1}{r\sin\theta}\frac{\partial}{\partial\phi} (v_\phi
B_r-v_r B_\phi)\ ,
\end{eqnarray}
\begin{eqnarray}\label{lou6}
\frac{\partial B_\theta}{\partial t}
&=&\frac{1}{r\sin\theta}\frac{\partial}{\partial\phi} (v_\theta
B_\phi-v_\phi B_\theta)\nonumber\\
&&-\frac{1}{r}\frac{\partial}{\partial r} \bigg[r(v_r
B_\theta-v_\theta B_r)\bigg]\ ,
\end{eqnarray}
\begin{eqnarray}\label{lou7}
\frac{\partial B_\phi}{\partial t}
&=&\frac{1}{r}\frac{\partial}{\partial r} \bigg[r(v_\phi B_r-v_r
B_\phi)\bigg]\nonumber\\
&&-\frac{1}{r}\frac{\partial}{\partial\theta} (v_\theta
B_\phi-v_\phi B_\theta)\ .
\end{eqnarray}
In the approximation of quasi-spherical symmetry, we take
$<v_\theta>\cong 0$ and $<v_\phi>\cong 0$. The $\theta-$ and
$\phi-$components of the magnetic induction equation can thus be
simplified to the following forms of
\begin{equation}\label{lou8}
\frac{\partial B_\theta}{\partial t}
=-\frac{1}{r}\frac{\partial}{\partial r}(rv_r B_\theta)\ ,
\end{equation}

\begin{equation}\label{lou9}
\frac{\partial B_\phi}{\partial t}
=-\frac{1}{r}\frac{\partial}{\partial r}(rv_r B_\phi)\ ,
\end{equation}
which then immediately lead to
\begin{equation}\label{lou10}
\frac{\partial}{\partial t}\bigg(\frac{r^2 B_t^2}{2}\bigg)
+\frac{v_r}{2}\frac{\partial}{\partial r}(r^2 B_t^2) +r^2
B_t^2\frac{v_r}{r}=0\ ,
\end{equation}
where $B_t^2\equiv B_{\theta}^2+B_{\phi}^2$ is proportional to the
energy density of the random transverse magnetic field. Equation
(\ref{lou10}) is simply the magnetic induction equation
(\ref{wang5}). For the radial component of the magnetic induction
equation (Yu and Lou, 2005; Yu et al., 2006), we have approximately
\begin{equation}\label{lou11}
\frac{\partial (r^2 B_r)}{\partial t} +v_r\frac{\partial
(r^2B_r)}{\partial r} =r^2(\overrightarrow B_t\cdot\nabla)v_r\ .
\end{equation}
Equations (\ref{lou10}) and (\ref{lou11}) here differ from
equations (1') and (1) in the formulation of Chiueh and Chou
(1994).

\section[]{Partial Derivatives of
$A$, $V$ and $X$}\label{partial}

\begin{figure}
\includegraphics[width=3.3in,bb=100 270 480 570]{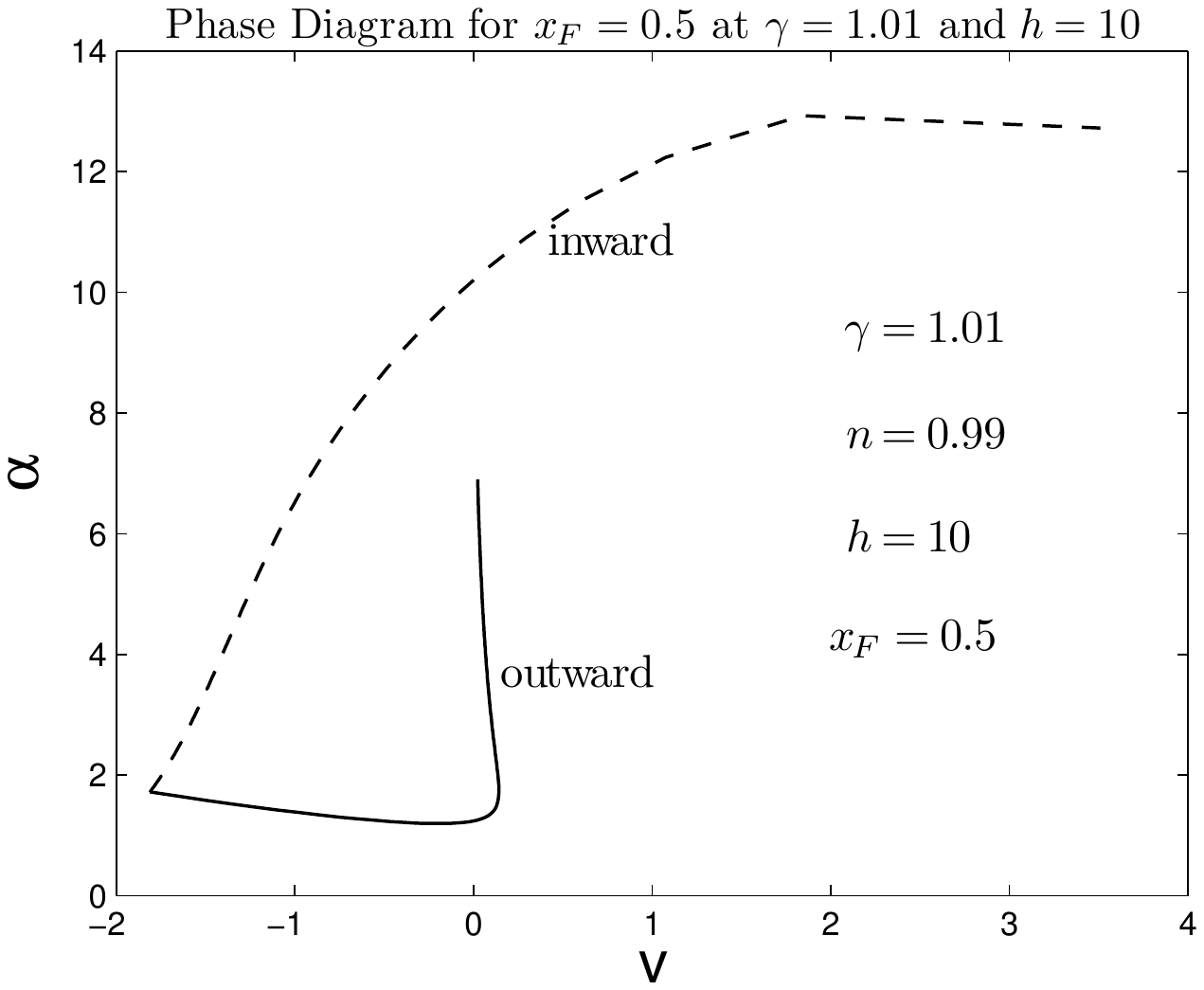}
\caption{Density-speed phase diagram for $\gamma=1.01$, $n=0.99$,
$h=10$ and a chosen meeting point $x_F=0.5$. The `inward curve' is
a phase curve by integrating inward from different $x$ along the
magnetosonic critical curve to $x_F=0.5$ and the `outward curve'
is the phase curve integrating outward from smaller $x$ to reach
the same $x_F$. There is no match in this phase diagram, because
for small $x$ along the magnetosonic critical curve, the solution
curve encounters the magnetosonic singular surface before reaching
$x_F=0.5$.} \label{Match101_10_1}
\end{figure}

The first-order partial derivatives of $A$, $V$ and $X$ as
functions of $\alpha$, $v$, and $x$ are used in subsection
\ref{behav}. They are derived from expressions (\ref{wang77}),
(\ref{wang78}), and (\ref{wang79}) for $A$, $V$ and $X$,
respectively, with straightforward manipulations. We present below
the explicit expressions of these partial derivatives, which are
used in numerical integrations to determine the MHD solution
behaviours in the vicinity of the sonic or magnetosonic critical
curves.
\begin{eqnarray}\label{wang82}
A_\alpha=&&2\alpha\bigg[(n-1)v-\frac{2(x-v)(nx-v)}{x}\bigg]\nonumber\\
&&+3\alpha^2\bigg[2hx+\frac{(nx-v)}{(3n-2)}\bigg]\ ,
\end{eqnarray}

\begin{equation}\label{wang83}
A_v=\alpha^2\bigg[3n+1-\frac{\alpha}{(3n-2)}-\frac{4v}{x}\bigg]\ ,
\end{equation}

\begin{equation}\label{wang84}
A_x=\alpha^2\bigg[\bigg(2h+\frac{n}{3n-2}\bigg)\alpha
-2n+\frac{2v^2}{x^2}\bigg]\ ,
\end{equation}

\begin{eqnarray}\label{wang85}
V_\alpha=&&(n-1)\big[v(nx-v)+4h\alpha x^2\big]\nonumber\\
&&+\frac{2(nx-v)^2}{(3n-2)}\alpha
-2\gamma^2\alpha^{\gamma-1}\frac{(x-v)}{x}\ ,
\end{eqnarray}

\begin{equation}\label{wang86}
V_v=(n-1)\big(nx-2v\big)\alpha -\frac{2\alpha^2(nx-v)}{(3n-2)}
+\frac{2\gamma\alpha^\gamma}{x}\ ,
\end{equation}

\begin{equation}\label{wang87}
V_x=(n-1)\big(nv+4h\alpha x\big)\alpha
+\frac{2n\alpha^2(nx-v)}{(3n-2)} -\frac{2\gamma\alpha^\gamma
v}{x^2}\ ,
\end{equation}

\begin{equation}\label{wang88}
X_\alpha=(nx-v)^2-\gamma^2\alpha^{\gamma-1}-2h\alpha x^2\ ,
\end{equation}

\begin{equation}\label{wang89}
X_v=-2\alpha(nx-v)\ ,
\end{equation}

\begin{equation}\label{wang90}
X_x=2n\alpha(nx-v)-2h\alpha^2x\ .
\end{equation}

\begin{figure}
\includegraphics[width=3.3in,bb=100 270 480 570]{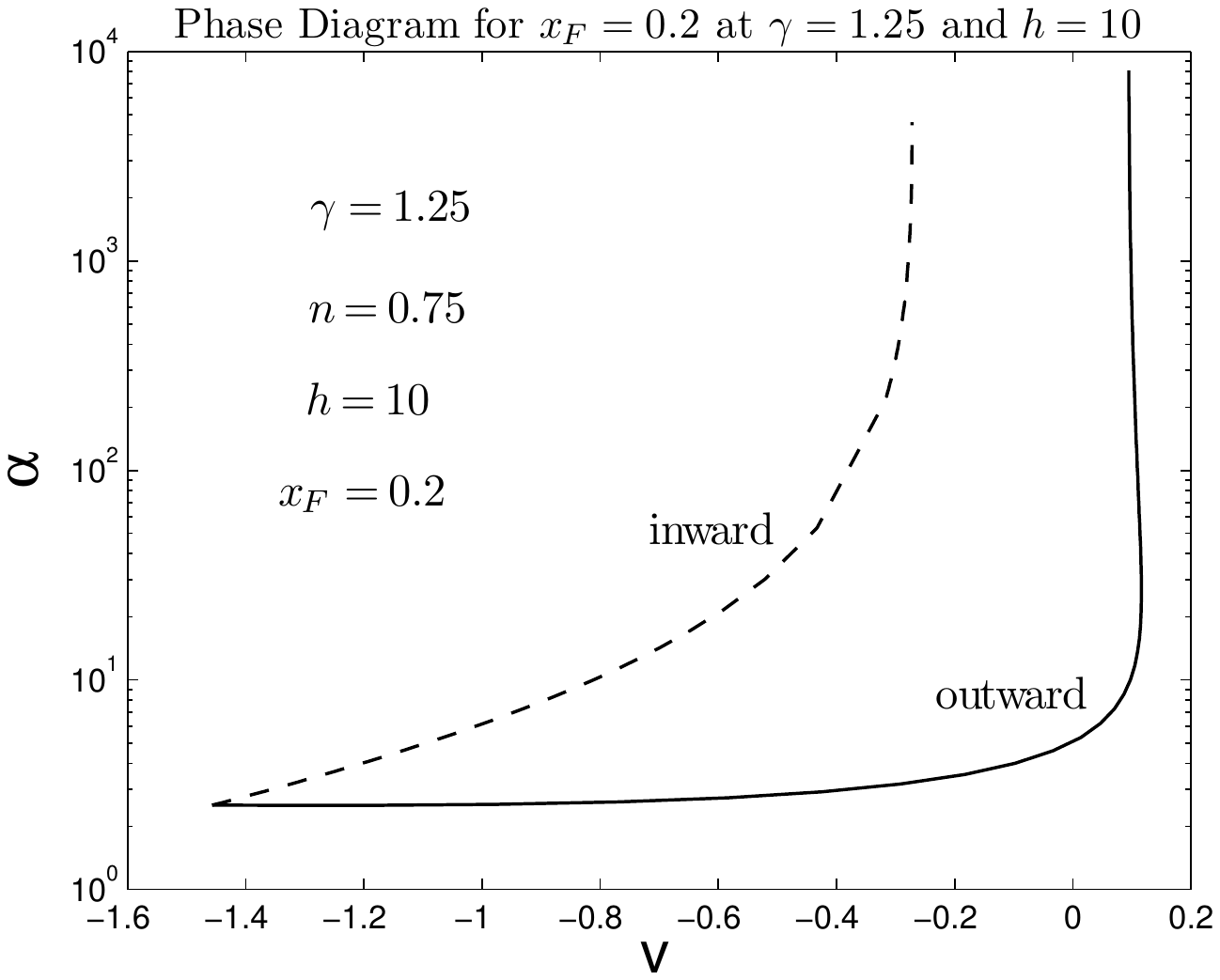}
\caption{Density-speed phase diagram for $\gamma=1.25$, $n=0.75$,
$h=10$ and a chosen meeting point $x_F=0.2$. The `inward curve' is
a phase curve obtained by integrating inward from different $x$
along the magnetosonic critical curve to reach $x_F$ and the
`outward curve' is a phase curve integrating outward from a
smaller $x$. There is no solution match here.
}\label{Match125_10_1}
\end{figure}

\section[]{Second-Order Derivatives of $\alpha(x)$
and \lowercase{$\mathit{v(x)}$}}\label{second}

For nodal points referred to in subsection \ref{behav}, there are
infinitely many solutions crossing the sonic and magnetosonic
critical curves. Only the analytical eigensolutions determined by
equations (\ref{wang53}), (\ref{wang54}) and (\ref{wang55}) are of
main interest here, and a numerical integration is unstable in the
direction away from the critical curve if one uses only the
first-order derivatives of $\alpha(x)$ and $v(x)$ with respect to
$x$. So one should work out the second-order derivatives of
$\alpha(x)$ and $v(x)$ with respect to $x$ in order to pick out
the analytic eigensolutions among many other solutions having weak
discontinuities across the sonic or magnetosonic critical curves
(Lazarus, 1981; Whitworth and Summers, 1985; Hunter, 1986). These
weak discontinuities may be regarded as weak shocks (Boily and
Lynden-Bell, 1995). It is also possible to construct various forms
of shocks across the sonic or magnetosonic critical curve (Tsai and
Hsu, 1995; Shu et al., 2002; Shen and Lou, 2004; Bian and Lou, 2005;
Yu and Lou, 2006; Lou and Wang, 2006, 2007).

We determine the second-order derivatives of $\alpha(x)$ and
$v(x)$ with respect to $x$ by the following procedure. As
$A=V=X=0$ along the sonic or magnetosonic critical curve,
one readily obtains
\begin{equation}\label{wang95}
\alpha''=\frac{X'A''-X''A'}{2(X')^2}=\frac{A''-\alpha'X''}{2X'}\ ,
\end{equation}
\begin{equation}\label{wang96}
v''=\frac{X'V''-X''V'}{2(X')^2}=\frac{V''-v'X''}{2X'}\ .
\end{equation}
We denote the second derivatives of $A$, $V$ and $X$ as
\begin{equation}\label{wang97}
\left( \begin{matrix}A''\cr V''\cr
X''\cr\end{matrix}\right)=\textbf{F} \left(\begin{matrix}\alpha''
\cr v'' \cr 1 \cr\end{matrix} \right)=\left(\begin{matrix}f_{11} &
f_{12} & f_{13} \cr f_{21} & f_{22} & f_{23} \cr f_{31} & f_{32} &
f_{33} \cr
\end{matrix}\right)\left(\begin{matrix}\alpha'' \cr v'' \cr 1 \cr
\end{matrix}\right)\ ,
\end{equation}
where each matrix element $f_{ij}=f_{ij}(\alpha,v,x,\alpha',v')$
of matrix {\bf F} is a function of $\alpha$, $v$, $x$, $\alpha'$
and $v'$ ($i,j=1,2,3$). Substituting equation (\ref{wang97}) into
equations (\ref{wang95}) and (\ref{wang96}), one immediately
obtains
\begin{equation}\label{wang98}
(2X'+\alpha'f_{31}-f_{11})\alpha''
+(\alpha'f_{32}-f_{12})v''=f_{13}-\alpha'f_{33}\ ,
\end{equation}
\begin{equation}\label{wang99}
(v'f_{31}-f_{21})\alpha''
+(2X'+v'f_{32}-f_{22})v''=f_{23}-v'f_{33}\ ,
\end{equation}
from which $\alpha''$ and $v''$ can be solved directly for the
nondegenerate case and be expressed and computed by corresponding
$x$, $\alpha$, $v$ $\alpha'$ and $v'$ at points along the
magnetosonic critical curve. For the convenience of reference and
checking, we summarize the explicit expressions for functions $X'$
and $f_{ij}$ below.
\begin{eqnarray}\label{wang100}
&&X'=(nx-v)^2\alpha'+2\alpha(nx-v)(n-v')\qquad\nonumber\\
&&-\gamma^2\alpha^{\gamma-1}\alpha' -2h\alpha^2x-2h\alpha
x^2\alpha'\ ,
\end{eqnarray}
\begin{equation}\label{wang101}
f_{11}\equiv\alpha^2\bigg[2hx+\frac{(nx-v)}{(3n-2)}\bigg]\ ,
\end{equation}
\begin{equation}\label{wang102}
f_{12}\equiv\alpha^2\bigg[3n+1-\frac{4v}{x}
-\frac{\alpha}{(3n-2)}\bigg]\ ,
\end{equation}
\begin{eqnarray}\label{wang103}
&&f_{13}\equiv 4\alpha\alpha'\bigg[ (3n+1)v'
+\bigg(2h+\frac{n-v'}{3n-2}\bigg)\alpha\qquad\nonumber\\
&&+\bigg(2hx +\frac{nx-v}{3n-2}\bigg)\alpha'-2n
-\frac{2v}{x^2}(2xv'-v)\bigg]\nonumber\\
&&+\alpha^2\bigg[ 2\bigg(2h+\frac{n-v'}{3n-2}\bigg)\alpha'
-\frac{4}{x^3}(v-xv')^2\bigg]\ ,
\end{eqnarray}
\begin{eqnarray}\label{wang104}
&&f_{21}\equiv (n-1)\big[v(nx-v)+4h\alpha x^2\big]\qquad\nonumber\\
&&+2\alpha\frac{(nx-v)^2}{3n-2}
-2\gamma^2\alpha^{\gamma-1}\frac{(x-v)}{x}\ ,\qquad
\end{eqnarray}
\begin{equation}\label{wang105}
f_{22}\equiv (n-1)\alpha(nx-2v)-\frac{2\alpha^2(nx-v)}{3n-2}
+\frac{2\gamma\alpha^\gamma}{x}\ ,
\end{equation}
\begin{eqnarray}\label{wang106}
&&f_{23}\equiv (n-1)\bigg[2(\alpha v'+v\alpha')(n-v')
\qquad\nonumber\\
&&+2(nx-v)\alpha'v'+4h\alpha^2 +16h\alpha
x\alpha'+4hx^2(\alpha')^2\bigg]\nonumber\\
&&+2\frac{(nx-v)^2}{(3n-2)}(\alpha')^2
+\frac{8\alpha(nx-v)(n-v')\alpha'}{(3n-2)}\nonumber\\
&&+\frac{2\alpha^2(n-v')^2}{(3n-2)}
-2\gamma^2(\gamma-1)\alpha^{\gamma-2}
\frac{(x-v)}{x}(\alpha')^2\nonumber\\
&&-4\gamma^2\alpha^{\gamma-1}\alpha'\frac{(v-xv')}{x^2}
-4\gamma\alpha^\gamma\frac{(xv'-v)}{x^3}\ ,
\end{eqnarray}
\begin{equation}\label{wang107}
f_{31}\equiv (nx-v)^2-\gamma^2\alpha^{\gamma-1}-2hx^2\alpha\ ,
\end{equation}
\begin{equation}\label{wang108}
f_{32}\equiv -2\alpha(nx-v)\ ,
\end{equation}
\begin{eqnarray}\label{wang109}
&&f_{33}\equiv
4(nx-v)\alpha'(n-v')+2\alpha(n-v')^2-2h\alpha^2\qquad\nonumber\\
&&-\gamma^2(\gamma-1)\alpha^{\gamma-2}(\alpha')^2 -8h\alpha
x\alpha'-2hx^2(\alpha')^2\ .\nonumber\\
\end{eqnarray}

\begin{figure}
\includegraphics[width=3.3in,bb=100 270 480 570]{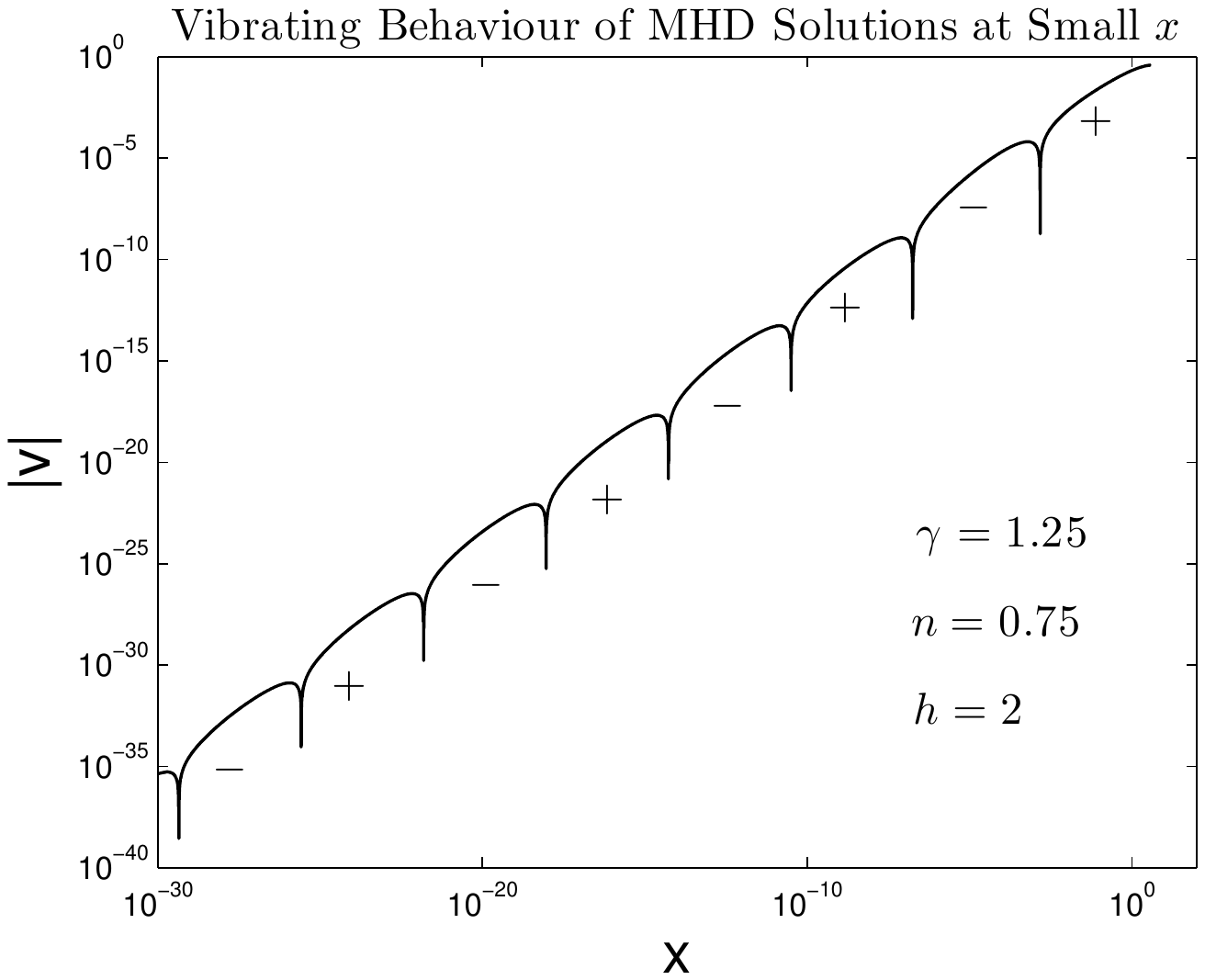}
\caption{An example for the magnetosonic oscillation behaviour
of an MHD similarity solution at small $x$ with both $|v|$ (the
absolute value of $v$) and $x$ displayed in logarithmic scales.
Relevant parameters are $\gamma=1.25$, $n=0.75$ and $h=2$. This
similarity MHD solution is obtained by integrating from the point
$\alpha=0.18$, $x=3.7395$ and $v=0.3863$ on the magnetosonic
critical curve inward to $x=0$. The alternating plus and minus
signs marked along the curve denote the actual sign of $-v$ to
compensate the loss of sign information as we show $|v|$ in a
logarithmic scale here. The plot clearly suggests that the
velocity amplitude scales with $x$ in a power
law.}\label{vibrating}
\end{figure}

\section[]{A Proof of $D_3=0$}\label{proofd3}

We conclude that because the magnetosonic critical curve is a
continuous curve as we have solved it analytically, it must lead
to $D_3=0$ in cubic equation (\ref{wang93}). A brief proof is
presented below. Since functionals $A,\ V,\ X$ as defined by
equations $(\ref{wang77})-(\ref{wang79})$ together determine the
magnetosonic critical curve (i.e., setting $A=V=X=0$
simultaneously), we begin by assuming that the point $(\alpha_1,\
v_1,\ x_1)$ in the $\alpha-v-x$ space
is on the magnetosonic critical curve such that $K(\alpha_1,\
v_1,\ x_1)=0$ where $K$ denotes $A$, $V$ or $X$, respectively. We
further assume that the set $(\alpha_1+\delta\alpha\ ,v_1+\delta
v\ ,x_1+\delta x)$ is also along the magnetosonic critical curve
and is very near to the former point. Then from the requirement
$K(\alpha_1+\delta\alpha\ , v_1+\delta v\ ,x_1+\delta x)=0$, with
$K$ denoting $A$, $V$ and $X$ as appropriate, we infer that
$\delta K\equiv K(\alpha_1 +\delta\alpha,\ v_1+\delta v,\
x_1+\delta x) -K(\alpha_1,\ v_1,\ x_1)=0$. It then follows from
\begin{equation}\label{wang138}
\delta K=\frac{\partial K}
{\partial\alpha}\delta\alpha+\frac{\partial K} {\partial v}\delta
v +\frac{\partial K}{\partial x}\delta x\ ,
\end{equation}
with $K$ denoting $A$, $V$ or $X$ in turn, that
\begin{equation}\label{wang139}
\left(\begin{matrix}A_\alpha & A_v & A_x \cr V_\alpha & V_v & V_x
\cr X_\alpha & X_v & X_x \cr
\end{matrix}\right)\left(\begin{matrix}\delta\alpha \cr \delta v \cr \delta
x\end{matrix}\right)=0
\end{equation}
along the magnetosonic critical curve for the given set of
$(\alpha_1,\ v_1,\ x_1)$. For a nontrivial set of $(\delta\alpha,\
\delta v,\ \delta x)$, we must then require
\begin{equation}\label{wang140}
\left|\begin{matrix}A_\alpha & A_v & A_x \cr V_\alpha & V_v & V_x
\cr X_\alpha & X_v & X_x \cr \end{matrix}\right|=0\
\end{equation}
along the magnetosonic critical curve, which is clearly equivalent
to $D_3=0$ in cubic equation (\ref{wang93}).

\section[]{Determination of Magnetosonic
Critical Curves}\label{infocritical}

\renewcommand{\tabcolsep}{0.13cm}
\begin{table}
\center \caption{Algebraic signs for quantities $A_1$, $B_1$,
$C_1$ and $\Delta$, and the corresponding root numbers for a usual
polytropic gas with $\gamma=1.01$ and $n=0.99$}\label{crinum101}
\vskip 0.3cm
\begin{tabular}{cccc}\hline
\multicolumn{4}{c} {$\gamma=1.01$,\ \ \ \ \
$n=0.99$,\ \ \ \ \ $h=0$}\\
\multicolumn{4}{c}{$\alpha_{11}=0.202728\ ,\ \ \ \alpha_{12}=0.282682$}\\
\hline\hline

$\alpha\in$&$(0\ ,\ \alpha_{11})$&$(\alpha_{11}\ ,
\ \alpha_{12})$&$(\alpha_{12}\ ,\ +\infty)$\\
\hline
signs&           & $B_1>0$     &$B_1,\ \Delta>0$\\
&all$\ <0$&others$\ <0$&$A_1,\ C_1<0$\\
root No.& 0  & 0    & 2     \\ \hline
\end{tabular}
\vskip 0.4cm

\begin{tabular}{ccccc}\hline
\multicolumn{5}{c} {$\gamma=1.01$,\ \ \ \ $n=0.99$,\ \ \ \
$h=0.1$}\\
\multicolumn{5}{c} {$\alpha_{21}=0.048500,\ \alpha_{22}=0.103907,
\ \alpha_{23}=0.202728$}\\
\hline\hline $\alpha\in$&$(0,\ \alpha_{21})$&$(\alpha_{21},\
\alpha_{22})$&$(\alpha_{22},\ \alpha_{23})$ & $(\alpha_{23},\
+\infty)$\\ \hline
signs& &$\Delta>0$&$A_1,\ \Delta>0$&$A_1,\ B_1,\ \Delta>0$\\
&all$\ <0$&others$\ <0$&$B_1,\ C_1<0$&$C_1<0$\\

root No.&0& 0      & 1    & 1\\ \hline
\end{tabular}
\vskip 0.4cm

\begin{tabular}{ccccc}\hline
\multicolumn{5}{c} {$\gamma=1.01$,\ \ \ \ \ $n=0.99$,\ \ \ \ \
$h=1$}\\
\multicolumn{5}{c}{$\alpha_{31}=4.9485\times10^{-3},\
\alpha_{32}=0.051866,\ \alpha_{33}=0.202728$}\\
\hline\hline $\alpha\in$&$(0,\ \alpha_{31})$ &$(\alpha_{31},\
\alpha_{32})$
&$(\alpha_{32},\ \alpha_{33})$&$(\alpha_{33},\ +\infty)$\\
\hline signs& &$\Delta>0$&$\!\!\!A_1,\Delta>0$&
$\!\!\!\!\!A_1,B_1,\Delta>0$\\
&all$\ <0$&others$\ <0$&$B_1, C_1<0$&$C_1<0$\\
root No.&0&$\!\!\!0$&$\!\!\! 1$&$\!\!\!1$\\ \hline
\end{tabular}
\vskip 0.4cm

\begin{tabular}{ccccc}\hline
\multicolumn{5}{c}{$\gamma=1.01$,\ \ \ \ \ $n=0.99$,\ \ \ \ \
$h=10$}
\\
\multicolumn{5}{c} {$\alpha_{41}=4.9440\times10^{-4},\
\alpha_{42}=0.027885,\ \alpha_{43}=0.202728$}\\
\hline\hline $\!\!\!\alpha\in$&$(0,\ \alpha_{41})$
&$(\alpha_{41},\ \alpha_{42})$ &$(\alpha_{42},\ \alpha_{43})$
&$(\alpha_{43},\ +\infty)$\\
\hline
signs& &$\Delta>0$&$\!\!\!A_1,\Delta>0$&$\!\!\!A_1,B_1,\Delta>0$\\
&all$\ <0$&others$\ <0$&$B_1, C_1<0$&$C_1<0$\\
root No.& 0 & 0 & 1 & 1\\ \hline
\end{tabular}
\end{table}

\begin{table}
\center \caption{Algebraic signs for quantities $A_1$, $B_1$,
$C_1$ and $\Delta$, and the corresponding root numbers for a
polytropic gas with $\gamma=1.25$ and $n=0.75$}\label{crinum125}
\vskip 0.3cm
\begin{tabular}{cccc}\hline
\multicolumn{4}{c} {$\gamma=1.25\ ,$\ \ \ $n=0.75\ ,$\ \ \
$h=0$}\\
\multicolumn{4}{c} {$\alpha_{51}=0.279006\ ,\ \ \alpha_{52}=0.368686$}\\
\hline\hline $\alpha\in$& $(0\ ,\ \alpha_{51})$ & $(\alpha_{51}\
,\ \alpha_{52})$ &
$(\alpha_{53}\ ,\ +\infty)$\\
\hline
signs&   & $B_1>0$  & $B_1\ ,\ \Delta>0$ \\
&all$\ <0$&others$\ <0$&$A_1\ ,\ C_1<0$\\
root No.& 0  & 0   & 2 \\ \hline
\end{tabular}
\vskip 0.4cm

\begin{tabular}{ccccc}\hline
\multicolumn{5}{c}{$\gamma=1.25$,\ \ \ \ $n=0.75$,\ \ \ \ $h=1$}\\
\multicolumn{5}{c} {$\alpha_{61}=0.09283\ ,\ \alpha_{62}=0.174661\
,
\ \alpha_{63}=0.279006$}\\
\hline\hline $\alpha\in$&$(0,\ \alpha_{61})$&$(\alpha_{61},\
\alpha_{62})$ &
$(\alpha_{62},\ \alpha_{63})$&$(\alpha_{63},\ +\infty)$\\
\hline
signs&  &$\Delta>0$&$A_1,\Delta>0$&$A_1,B_1,\Delta>0$\\
&all$\ <0$&others$\ <0$&$B_1, C_1<0$&$C_1<0$\\
root No.&0&0&1&1\\ \hline
\end{tabular}
\vskip 0.4cm

\begin{tabular}{ccccc}\hline
\multicolumn{5}{c} {$\gamma=1.25$,\ \ \ \ $n=0.75$,\ \ \ \ $h=10$}\\
\multicolumn{5}{c} {$\alpha_{71}=0.0090897\ ,\
\alpha_{72}=0.107674\ ,\ \alpha_{73}=0.279006$}\\
\hline\hline $\alpha$&$(0\ ,\alpha_{71})$&$(\alpha_{71}\ ,\
\alpha_{72})$ &$(\alpha_{72}\ ,\ \alpha_{73})$&
$(\alpha_{73}\ ,\ +\infty)$\\
\hline signs& &$\!\!\!\!\Delta>0$&$\!\!\!\!A_1,\Delta>0$&
$\!\!\!\!A_1,B_1,\Delta>0$\\
&all$\ <0$&others$\ <0$&$B_1\ ,\ C_1<0$&$C_1<0$\\
root No.&0& 0& 1& 1\\ \hline
\end{tabular}
\vskip 0.4cm

\begin{tabular}{cccc}\hline
\multicolumn{4}{c} {$\gamma=1.25$,\ \ \ \ $n=0.75$,\ \ \ \ $h=100$} \\
\multicolumn{4}{c} {$\alpha_{81}=8.9955\times10^{-4}\ ,\ i
\alpha_{82}=7.1799\times10^{-3}\ ,\ \alpha_{83}=0.038660$}\\
\hline\hline $\alpha\in$&$(0\ ,\ \alpha_{81})$&$(\alpha_{81}\ ,\
\alpha_{82})$ &$(\alpha_{82}\ ,\ \alpha_{83})$\\ \hline
signs&            & $\Delta>0$   &$A_1\ ,\ \Delta>0$\\
&all$\ <0$&others$<0$&$B_1\ ,\ C_1<0$\\
root No.& 0          & 0          & 1   \\ \hline
\multicolumn{4}{c} {$\alpha_{84}=0.079160\ , \ \
\alpha_{85}=0.279006$}\\ \hline $\alpha\in$&$(\alpha_{83}\ ,\
\alpha_{84})$&
$(\alpha_{84}\ ,\ \alpha_{85})$&$(\alpha_{85}\ ,\ +\infty)$\\
\hline
signs&$\Delta>0$& $A_1\ ,\ \Delta>0$& $A_1,\ B_1,\ \Delta>0$\\
&others$\ <0$&$B_1\ ,\ C_1<0$&$C_1<0$\\
root No.&0       & 1     &1     \\ \hline
\end{tabular}
\end{table}
One can numerically determine ranges of $\alpha$ values for which
the signs of $A_1$, $B_1$ and $C_1$ and of the determinant
$\Delta\equiv B_1^2-4A_1C_1$ in equation (\ref{wang42}) do not
change and one can thus decide which root of $x^2$ should be
picked up in equation (\ref{wang41}) and how many roots are
physically relevant once specific values of $\gamma$, $h$ and $n$
are prescribed. This numerical exploration is important as one
needs to know the sonic or magnetosonic critical curve in the
entire range of $0^{+}<\alpha< +\infty$. We have analyzed the
usual polytropic cases of $\gamma=1.01$, $n=0.99$ and
$\gamma=1.25$, $n=0.75$ for different $h$ values. Here, we
summarize the main results. The case that $\gamma=1.01$ and
$n=0.99$ is contained in Table \ref{crinum101} and the case that
$\gamma=1.25$ and $n=0.75$ is contained in Table \ref{crinum125}.

\section[]{MHD Eigensolution Behaviours in the Vicinity
of the Magnetosonic Critical Curve}\label{vicinity}

\begin{table}
\center \caption{Algebraic signs of both determinant $\Delta\equiv
B_3^2-4C_3$ and $C_3$ for equation (\ref{wang93}) and type
identifications for local behaviours of MHD eigensolutions in the
vicinity of magnetosonic critical curves for a conventional
polytropic gas with $\gamma=1.01$ and $n=0.99$, where `s', `n' and
`c' denote a saddle, node, and a centre (or spiral), respectively
(Jordan and Smith, 1977). }\label{vici1} \vskip 0.3cm
\begin{tabular}{cccc}\hline
\multicolumn{4}{c} {$\gamma=1.01$,\ \ \ \ $n=0.99$,\ \ \ \ $h=0$ }
\\
\multicolumn{4}{c} {$x_{11}=1.0455\ ,\ \ \ x_{12}=47.31$}\\
\hline\hline $x\in$ & $(0\ ,\ x_{11})$\ \ & $(x_{11}\ ,\ x_{12})$\
\ & $(x_{12}\ ,\ +\infty)$\
\\ \hline
signs& $\Delta>0$&  $\Delta\ ,\ C_3>0$ & $C_3>0$ \\
&$C_3<0$&&$\Delta<0$\\
type identification\ \ & s\ \  & n\ \ & c\ \ \\ \hline
\end{tabular}
\vskip 0.4cm

\begin{tabular}{ccc}\hline
\multicolumn{3}{c} {$\gamma=1.01$,\ \ $n=0.99$,\ \ $h=0.1$} \\
\multicolumn{3}{c} {$x_{21}=1.1497$}\\
\hline\hline $x\in$ & $(0\ ,\ x_{21})$\  &  $(x_{21}\ ,\ +\infty)$\   \\
\hline
signs& $\Delta>0$ &  $\Delta\ ,\ C_3>0$  \\
&$C_3<0$&\\
type identification\ \ & s\ \  & n\ \  \\ \hline
\end{tabular}
\vskip 0.4cm

\begin{tabular}{ccc}\hline
\multicolumn{3}{c} {$\gamma=1.01$,\ \ \ \ $n=0.99$,\ \ \ \ $h=1$} \\
\multicolumn{3}{c} {$x_{31}=1.9182$}\\
\hline\hline
 $x\in$ & $(0\ ,\ x_{31})$  &  $(x_{31}\ ,\ +\infty)$\ \\ \hline
signs  & $\Delta>0$ & $\Delta\ ,\ C_3>0$   \\
&$C_3<0$&\\
type identification& s  & n  \\ \hline
\end{tabular}
\end{table}

\begin{table}
\center \caption{Algebraic signs of both determinant $\Delta\equiv
B_3^2-4C_3$ and $C_3$ for equation (\ref{wang93}) and type
identifications for local behaviours of MHD eigensolutions for a
conventional polytropic gas in the vicinity of magnetosonic
critical curves as $\gamma=1.25$ and $n=0.75$, where `s', `n', and
`c' denote saddle, node, and centre (or spiral), respectively
(Jordan and Smith, 1977).}\label{vici2} \vskip 0.3cm
\begin{tabular}{cccc}\hline
\multicolumn{4}{c}{$\gamma=1.25$,\ \ \ \ \ $n=0.75$,\ \ \ \ \ $h=0$}\\
\multicolumn{4}{c} {$x_{41}=11.724\ ,\ \ \ \ x_{42}=11.724$}\\
\hline\hline $x$ & $(0\ ,\ x_{41})$\ & $(x_{41}\ ,\ x_{42})$\  &
$(x_{42}\ ,\ +\infty)$ \\ \hline
signs & $\Delta>0$ & $\Delta\ ,\ C_3>0$ & $C_3>0$ \\
&$C_3<0$&&$\Delta<0$\\
type identification& s & n & c \\ \hline
\end{tabular}
\end{table}

Similarity MHD solution behaviours in the vicinity of the
magnetosonic critical curve can be determined by the signs of both
$\Delta\equiv B_3^2-4C_3$ and $C_3$ as defined by equation
(\ref{wang93}) (see subsection \ref{behav} for details). One can
numerically determine the values of $\Delta$ and $C_3$ given
specific $x$ values of different magnetosonic critical curves
corresponding to different parameter sets of $\gamma$, $n$ and
$h$. We describe the main results of this exploration for the
cases of $\gamma=1.01$, $n=0.99$ and $\gamma=1.25$, $n=0.75$ with
different $h$ values. In Table \ref{vici1}, we present the
relevant results for $\gamma=1.01$ and $n=0.99$, while in Table
\ref{vici2}, we show the results for $\gamma=1.25$ and $n=0.75$.
In both Tables \ref{vici1} and \ref{vici2}, the sign `s' indicates
a saddle, `n' indicates a node, and `c' indicates a centre or a
spiral regarded as being unphysical (Jordan and Smith, 1977). Note
in particular that among the cases we have explored and not
displayed in the above Tables, there exist such magnetosonic
critical curves that all points along the lines are saddle points.
These include the cases of $\gamma=1.01$, $n=0.99$, $h=10$ and
$100$, and of $\gamma=1.25$, $n=0.75$, $h=1$, $10$, and $100$.
There are also cases of $h>0$ with the coexistence of nodal,
saddle and spiral or centre points along the magnetosonic curve,
e.g., the case of $\gamma=1.01$, $n=0.99$ and $h=0.0001$. The fact
that there exist spiral or centre points in unmagnetized cases
should be noted because there would be no global semi-complete
smooth solution crossing such points (subsection \ref{behav}).

\section[]{Asymptotic Behaviours Approaching
the Quasi Magnetostatic Solution}\label{asystat}

As mentioned above, global magnetostatic solution (\ref{wang32})
also serves as an asymptotic behaviour for small $x$. This
asymptotic solution turns out to be the only solution
characterized by a $v$ approaching zero faster than ${\cal O}(x)$
and $\alpha\simeq x^{-2/n}$ when $n=2-\gamma$ for a usual
polytropic gas. In order to determine the behaviour of this
asymptotic solution, we assume to the leading order
\begin{equation}\label{wang141}
v=Lx^{K}+\cdots\ ,
\end{equation}
\[
\alpha=\bigg[\frac{(2-\gamma)^{2}}
{2\gamma(4-3\gamma)}+\frac{(1-\gamma)} {\gamma}h\bigg]^{1/
(\gamma-2)}x^{-2/(2-\gamma)}+\Delta\alpha+\cdots\ ,
\]
where the next-order $\alpha$ variation $\Delta\alpha$ is given by
\begin{equation}\label{wang142}
\Delta\alpha=Nx^{K-1-2/n}\ ,
\end{equation}
and $L$, $K$ and $N$ are three constant complex coefficients.
Apparently, $Re(K)>1$ is required for $v$ to approach zero faster
than $\sim {\cal O}(x)$ and $\Delta\alpha$ being the next-order
term of this asymptotic solution as $x$ approaches zero.

Substituting asymptotic expressions (\ref{wang141}) and
(\ref{wang142}) into nonlinear MHD ODEs (\ref{wang9}) and
(\ref{wang13}) and using integral relation (\ref{wang15}), we
obtain the following two relations
\begin{eqnarray*}\label{wang143}
&&\bigg[\frac{n^2(1+K)}{2(3n-2)}+nhK-\gamma h\bigg]N\nonumber\\
&&=\bigg[\frac{n^2}{2\gamma(3n-2)}+\frac{(1-\gamma)}
{\gamma}h\bigg]^{-1/n}\frac{L}{(3n-2)}\ ,
\end{eqnarray*}
\begin{equation}
\frac{n(K-1)N}{(K-2-2/n)} =\bigg[\frac{n^2}{2\gamma(3n-2)}
+\frac{(1-\gamma)}{\gamma}h\bigg]^{-1/n}L\
\end{equation}
among the three complex constants $K$, $L$ and $N$. Clearly, $K$
satisfies the following quadratic equation
\begin{eqnarray}\label{wang144}
&&f(K)\equiv [n^2/2+n(3n-2)h]K^2\nonumber\\
&&-(4-3n)[n/2+(3n-2)h]K\nonumber\\
&&+n^2+\gamma(2/n-2)(3n-2)h=0\ .\qquad
\end{eqnarray}
Once a proper root of $K$ is chosen, the complex ratio of $L$ to
$N$ is determined accordingly. We emphasize that the values of $N$
and $L$ are not determined a priori, i.e., $L$ can take on any
reasonable value, while $N$ is determined by their ratio or vice
versa.

We introduce the handy notation
\begin{eqnarray}\label{wang146}
h_0&=&\frac{\big[(3+2\sqrt{2})n-4\big]
\big[4-(3-2\sqrt{2})n\big]}{2n(3n-2)}\nonumber\\
&=&\frac{(1-n)(-n^2+24n-16)}{n^3}h_c\
\end{eqnarray}
and reach the following conclusions. When $12-8\sqrt 2<n<0.8$ and
$h_0<h<h_c$, or when $2/3<n<12-8\sqrt{2}$ for arbitrary $h$
values, there exist two real roots of $f(K)$, both are larger than
unity; these are referred to as type I `quasi-static' asymptotic
solutions. When $12-8\sqrt{2}<n<0.8$ and $h<h_0$, there exists a
complex root $K$ with its real part larger than unity; this is
referred to as type II `quasi-static' asymptotic solution. For a
complex $K=K_1+iK_2$, we simply have
\begin{equation}\label{wang147}
v=Lx^{K_1}x^{iK_2}=Lx^{K_1}\exp(iK_2\ln x)\ .
\end{equation}
In the limit of $h=0$, the above MHD results reduce to
those of a hydrodynamic analysis (Lou and Wang, 2006).

When $n\neq 2-\gamma$ for an unconventional polytropic gas, an
asymptotic similarity MHD solution in the form of $v\simeq {\cal
O}(x)$ and $\alpha\simeq x^{-2/(2-\gamma)}$ also exists, viz.
\[
v=\frac{2(2-\gamma-n)}{(4-3\gamma)}x+\cdots\ ,
\]
\begin{eqnarray}\label{wang145}
\alpha=\bigg[\frac{(1-\gamma)}{\gamma}h
&+&\frac{(2-\gamma)}{\gamma(3n-2)}\bigg(\frac{n}{2}
-\frac{2-\gamma-n}
{4-3\gamma}\bigg)\bigg]^{-1/(2-\gamma)}\nonumber\\
&\times& x^{-2/(2-\gamma)}+\cdots\ .
\end{eqnarray}


\begin{thebibliography}{}

\bibitem{Bagchietal06}
Bagchi, J., Durret, F., Neto, G. B. L.,
Paul, S.: Science \textbf{314}, 791 (2006)

\bibitem{BarenZel72}
Barenblatt, G. I., Zel'dovich, Ya. B.: AnRFM \textbf{4}, 285 (1972)
%
\bibitem{BianLou05}
Bian F.-Y., Lou, Y.-Q.: MNRAS \textbf{363}, 1315 (2005)
%
\bibitem{bs68}
Bodenheimer, P., Sweigart, A.: ApJ \textbf{152}, 515 (1968)
%
\bibitem{boily95}
Boily, C. M., Lynden-Bell, D.: MNRAS \textbf{276}, 133 (1995)
%
\bibitem{bouquet1985}
Bouquet, S., Feix, M. R., Fualkow, E., Munier, A.: ApJ
\textbf{293}, 494 (1985)
%
\bibitem{caishu05}
Cai, M. J., Shu, F. H.: ApJ \textbf{618}, 438 (2005)
%
\bibitem{cheng78}
Cheng, A. F.: ApJ \textbf{221}, 320 (1978)
%
\bibitem{chiuehchou1994}
Chiueh, T., Chou, J.-K.: ApJ \textbf{431}, 380 (1994)
%
\bibitem{FanLou1999}
Fan, Z.H., Lou, Y.-Q.: MNRAS \textbf{307}, 645 (1999)
%
\bibitem{fatuzzo2004}
Fatuzzo, M., Adams, F. C., Myers, P. C.:
ApJ \textbf{615}, 813 (2004)
%
\bibitem{fillmore84}
Fillmore, J. M., Goldreich, P.: ApJ \textbf{284}, 1 (1984)
%
\bibitem{fosterchevalier93}
Foster, P. N., Chevalier, R. A.: ApJ \textbf{416}, 303 (1993)
%
\bibitem{goldreich80}
Goldreich, P., Weber, S. V.: ApJ \textbf{238}, 991 (1980)
%
\bibitem{hanawa99}
Hanawa, T., Matsumoto, T.: ApJ \textbf{521}, 703 (1999)
%
\bibitem{hanawa00}
Hanawa, T., Matsumoto, T.: PASJ \textbf{52}, 241 (2000)
%
\bibitem{hanawa97}
Hanawa, T., Nakayama, K.: ApJ \textbf{484}, 238 (1997)
%
\bibitem{hennebelle03}
Hennebelle, P.: A\&A \textbf{397}, 381 (2003)
%
\bibitem{HL2004}
Hu, J., Lou, Y.-Q.: ApJ \textbf{606}, L1 (2005)
%
\bibitem{HSLZ2005}
Hu, J., Shen, Y., Lou, Y.-Q., Zhang, S.-N.: MNRAS \textbf{365},
345 (2005)
%
\bibitem{hunter1977}
Hunter, C.: ApJ \textbf{218}, 834 (1977)
%
\bibitem{hunter1986}
Hunter, C.: MNRAS \textbf{223}, 391 (1986)
%
\bibitem{inutsuka92}
Inutsuka, S., Miyama, S. M.: ApJ \textbf{388}, 392 (1992)
%
\bibitem{jordansmith1977}
Jordan, D. W., Smith, P.: Nonlinear Ordinary Differential
Equations, Oxford University Press. Oxford (1977)
%
\bibitem{KC1984a}
Kennel, C. F., Coroniti, F. V.: ApJ \textbf{283}, 694 (1984a)
%
\bibitem{KC1984b}
Kennel, C. F., Coroniti, F. V.: ApJ \textbf{283}, 710 (1984b)
%
%
\bibitem{krasnopolsky02}
Krasnopolsky, R., K\"onigl, A.: ApJ \textbf{580}, 987 (2002)
%
\bibitem{landau1959}
Landau, L. D., Lifshitz, E. M.: Fluid Mechanics,
Pergamon Press, New York (1959)
%
\bibitem{larson69a}
Larson, R. B.: MNRAS \textbf{145}, 271 (1969)
%
\bibitem{Lazarus81}
Lazarus, R. B.: SIAM J. Numer. Anal. \textbf{18}, 316 (1981)
%
\bibitem{lou1993}
Lou, Y.-Q.: ApJ \textbf{414}, 656 (1993)
%
\bibitem{lou1994}
Lou, Y.-Q.: ApJ \textbf{428}, L21 (1994)
%
\bibitem{lou2005}
Lou, Y.-Q.: ChJAA \textbf{5}, 6 (2005)
%
\bibitem{lougao2006}
Lou, Y.-Q., Gao, Y.: MNRAS \textbf{373},
1610 (2006, astro-ph/0609771)
%
\bibitem{loushen2004}
Lou, Y.-Q., Shen, Y.: MNRAS \textbf{348}, 717 (2004)
%
\bibitem{lourosner1986}
Lou, Y.-Q., Rosner, R.: ApJ \textbf{309}, 874 (1986)
%
\bibitem{louwang06}
Lou, Y.-Q., Wang, W.-G.: MNRAS \textbf{372}, 885
(2006, astro-ph/0608043)
%
\bibitem{louwang07}
Lou, Y.-Q., Wang, W.-G.: MNRAS \textbf{378}, L54 (2007,
astro-ph/0704.0223)
%
\bibitem{low1992}
Low, B. C.: ApJ \textbf{390}, 567 (1992)
%
\bibitem{mclaughlin97}
McLaughlin, D. E., Pudritz, R. E.: ApJ \textbf{476}, 750 (1997)
%
\bibitem{myers1998}
Myers, P. C.: ApJL \textbf{496}, L109 (1998)
%
\bibitem{ori88}
Ori, A., Piran, T.: MNRAS \textbf{234}, 821 (1988)
%
\bibitem{penston69a}
Penston, M. V.: MNRAS \textbf{144}, 425 (1969a)
%
\bibitem{penston69b}
Penston, M. V.: MNRAS \textbf{145}, 457 (1969b)
%
\bibitem{press86}
Press, W. H., Flannery, B. P., Teukolsky, S. A., Vetterling, W.:
Numerical Recipes, Cambridge University Press, Cambridge (1986)
%
\bibitem{sedov59}
Sedov, L. I.: Similarity and Dimensional Methods in Mechanics,
Academic Press, New York (1959)
%
\bibitem{semelin01}
Semelin, B., Sanchez, N., de Vega, H. J.: Phys. Rev. D
\textbf{63}, 4005 (2001)
%
\bibitem{shadmehri05}
Shadmehri, M.: MNRAS \textbf{356}, 1429 (2005)
%
\bibitem{shenlou04}
Shen, Y., Lou, Y.-Q.: ApJL \textbf{611}, 117 (2004)
%
\bibitem{shenlou06}
Shen, Y., Lou, Y.-Q.: MNRAS Lett. \textbf{370}, L85
(2006, astro-ph/0605505)
%
\bibitem{shu1977}
Shu, F. H.: ApJ \textbf{214}, 488 (1977)
%
\bibitem{sal87}
Shu, F. H., Adams, F. C., Lizano, S.:
ARA\&A \textbf{25}, 23 (1987)
%
\bibitem{shu02}
Shu, F. H., Lizano, S., Galli, D., Cant\'o, J.,
Laughlin, G.: ApJ \textbf{580}, 969 (2002)
%
\bibitem{sutosilk1988}
Suto, Y., Silk, J.: ApJ \textbf{326}, 527 (1988)
%
\bibitem{tsc}
Terebey, S., Shu, F. H., Cassen, P.:
ApJ \textbf{286}, 529 (1984)
%
\bibitem{th95}
Tsai, J. C., Hsu, J. J. L.: ApJ \textbf{448}, 774 (1995)
%
\bibitem{whitworthsummers1985}
Whitworth, A., Summers, D.: MNRAS \textbf{214}, 1 (1985)
%
\bibitem{wilson1985}
Wilson, A. S., Samarasinha, N. H., Hogg, D. E.:
ApJ \textbf{294}, L121 (1985)
%
\bibitem{wolf03}
Wolf, S., Launhardt, R., Henning, T.:
ApJ \textbf{592}, 233 (2003)
%
\bibitem{woltjer57}
Woltjer, L.: BAN \textbf{13}, 301 (1957)
%
\bibitem{woltjer58a}
Woltjer, L.: BAN \textbf{14}, 39 (1958a)
%
\bibitem{woltjer58b}
Woltjer, L.: ApJ \textbf{128}, 384 (1958b)
%
\bibitem{yahil83}
Yahil, A.: ApJ \textbf{265}, 1047 (1983)
%
\bibitem{YuLou05}
Yu, C., Lou, Y.-Q.: MNRAS \textbf{364}, 1168 (2005)
%
\bibitem{YuLou06}
Yu, C., Lou, Y.-Q., Bian, F. Y., Wu, Y.: MNRAS
\textbf{370} 121 (2006, astro-ph/0604261)
%
\bibitem{magpressure}
Zel'dovich, Ya. B., Novikov, I. D.: Stars and Relativity --
Relativistic Astrophysics, Vol. 1, The University of Chicago
Press, Chicago (1971)
%
\end{thebibliography}
\end{document}